\documentclass[prd,superscriptaddress,nofootinbib,twocolumn,twoside]{revtex4-1}

\usepackage[utf8]{inputenc}

\usepackage{setspace}
\usepackage{footmisc}
\usepackage[colorlinks]{hyperref}
\hypersetup{
    colorlinks = true,
    citecolor  = red,
	linkcolor  = blue
}

\usepackage{bbm}
\usepackage{amsfonts}
\usepackage{mathrsfs}

\usepackage{bbold}
\usepackage{amsmath,amssymb}
\usepackage[english]{babel}

\usepackage{siunitx}
\usepackage{slashed}

\usepackage{epsfig}
\usepackage{graphicx}               
\usepackage{url}
\usepackage{float}
\usepackage{soul}
\usepackage{pstricks}
\usepackage{color}
\usepackage{multirow}

\usepackage{threeparttable}
\usepackage[version=4]{mhchem}
\usepackage{fontawesome}

\newcommand{\be}{\begin{equation}}
\newcommand{\ee}{\end{equation}}
\newcommand{\bea}{\begin{eqnarray}}
\newcommand{\eea}{\end{eqnarray}}
\newcommand{\beq}{\begin{eqnarray}}
\newcommand{\eeq}{\end{eqnarray}}
\newcommand{\LL}{\mathcal{L}}
\newcommand{\MM}{\mathcal{M}}
\newcommand{\OO}{\mathcal{O}}

\def\vect#1{\boldsymbol{#1}}

\newcommand\Tstrut{\rule{0pt}{3.5ex}}         
\newcommand\Bstrut{\rule[-2ex]{0pt}{0pt}}   

\begin{document}

\title{Multi-Channel Direct Detection of Light Dark Matter: Target Comparison}
\author{Sin\'ead M. Griffin}
\author{Katherine Inzani}
\affiliation{Materials Sciences Division, Lawrence Berkeley National Laboratory, Berkeley, CA 94720, USA}
\affiliation{Molecular Foundry, Lawrence Berkeley National Laboratory, Berkeley, CA 94720, USA}

\author{Tanner Trickle}
\author{Zhengkang Zhang}
\affiliation{Department of Physics, University of California, Berkeley, CA 94720, USA}
\affiliation{Theoretical Physics Group, Lawrence Berkeley National Laboratory, Berkeley, CA 94720, USA}

\author{Kathryn M.~Zurek}
\affiliation{Walter Burke Institute for Theoretical Physics, California Institute of Technology, Pasadena, CA 91125,USA}

\begin{abstract}
Direct detection experiments for light dark matter are making enormous leaps in reaching previously unexplored model space. Several recent proposals rely on collective excitations, where the experimental sensitivity is highly dependent on detailed properties of the target material, well beyond just nucleus mass numbers as in conventional searches. It is thus important to optimize the target choice when considering which experiment to build.  We carry out a comparative study of target materials across several detection channels, focusing on electron transitions and single (acoustic or optical) phonon excitations in crystals, as well as the traditional nuclear recoils. We compare materials currently in use in nuclear recoil experiments (Si, Ge, NaI, CsI, CaWO$_4$), a few which have been proposed for light dark matter experiments (GaAs, Al$_2$O$_3$, diamond), as well as 16 other promising polar crystals across all detection channels. We find that target- and dark matter model-dependent reach is largely determined by a small number of material parameters: speed of sound, electronic band gap, mass number, Born effective charge, high frequency dielectric constant, and optical phonon energies. We showcase, for each of the two benchmark models, an exemplary material which has a better reach than in any currently proposed experiment.  
\end{abstract}

\maketitle

\section{Introduction}

Direct detection experiments have traditionally focused on dark matter (DM) with mass near the weak scale.  Cosmologically, however, thermal particle DM may inhabit a much broader mass range between a keV and 10\,TeV.  Recent years have seen bold advances in the efforts to probe DM in the range below 10 GeV, which was less explored previously. Here, despite the existence of well-motivated candidates -- including MeV dark matter~\cite{Boehm2004a,Pospelov:2007mp,Hooper:2008im}, WIMPless miracle DM~\cite{Kumar:2009bw}, GeV hidden sector dark matter~\cite{ArkaniHamed:2008qp,Cheung:2009qd,Morrissey:2009ur}, asymmetric DM~\cite{Kaplan:2009ag,Cohen:2010kn}, freeze-in DM~\cite{Hall:2009bx}, Strongly Interacting Massive Particles~\cite{Hochberg:2014dra}, and many others -- conventional detection techniques based on nuclear recoils lose sensitivity as the energy deposition falls below detector thresholds. This has motivated an extensive exploration of novel detection channels using a variety of target systems. These include electron transitions in atoms and semiconductors~\cite{Essig:2011nj,Graham:2012su,Essig:2012yx,Lee:2015qva,Essig:2015cda,Hochberg:2016ntt,Derenzo:2016fse,Hochberg:2016sqx,Bloch:2016sjj,Essig:2017kqs,Kadribasic:2017obi,Kurinsky:2019pgb,Heikinheimo:2019lwg,Emken:2019tni}, superconductors~\cite{Hochberg:2015pha,Hochberg:2015fth,Hochberg:2016ajh}, Dirac materials~\cite{Hochberg:2017wce,Coskuner:2019odd,Geilhufe:2019ndy}, via the Migdal effect~\cite{Ibe:2017yqa,Dolan:2017xbu,Bell:2019egg,Baxter:2019pnz,Essig:2019xkx}, molecular dissociation or excitation~\cite{Essig:2016crl,Arvanitaki:2017nhi,Essig:2019kfe}, multi-excitation production in superfluid helium~\cite{Schutz:2016tid,Knapen:2016cue,Acanfora:2019con,Caputo:2019cyg}, defect production~\cite{Budnik:2017sbu,Rajendran:2017ynw}, single phonon~\cite{Knapen:2017ekk,Griffin:2018bjn} and magnon~\cite{Trickle:2019ovy} excitations in crystals (see also Refs.~\cite{Riedel:2012ur,Kouvaris:2016afs,Riedel:2016acj,Cavoto:2017otc,Marsh:2018dlj,Alonso:2018dxy,Roberts:2019sfo,Zarei:2019sva} for other recent proposals).

As new experiments are being planned and detection technologies are being discussed and improved, it is important to identify the most promising targets in order to prioritize the experimental program. There are two questions in this respect: {\em (i)} what types of excitations can be utilized as efficient detection paths with current and developing technologies, and {\em (ii)} what materials have the strongest response to DM scattering?

It is the purpose of this paper to initiate a discussion on these questions, and provide theory input to the optimization of experimental strategy. We consider several complementary detection channels:
\begin{itemize}
\item nuclear recoils, sensitive to the heaviest DM masses, down to $\OO(100\,\text{MeV})$ at best;
\item electron transitions across band gaps in crystals, covering DM masses down to $\OO(100\,\text{keV})$;
\item single phonon excitations in crystals, reaching the lightest DM masses, down to $\OO(\text{keV})$.
\end{itemize}
The last two detection channels rely on collective properties of the target, which makes calculating the DM model reach more involved than the standard nuclear recoil calculation. While nuclear recoil was proposed long ago~\cite{Goodman1985a,Drukier1986a}, electron transitions in semiconductors (proposed in Refs.~\cite{Essig:2011nj,Graham:2012su,Essig:2015cda}) and phonon production from sub-MeV DM in crystals (put forth in Refs.~\cite{Hochberg:2016sqx,Knapen:2017ekk,Griffin:2018bjn}) have a much shorter history. Now that all these ideas are available, we hope to find materials which have a strong response in all channels, in order to cover a broad range of DM masses. 

We begin in Sec.~\ref{sec:paths} with a brief review of each detection channel. A common framework to calculate the reach via all three channels is presented in a companion paper \cite{Trickle2019a}, which makes it clear that the detection rate factorizes into the particle-level scattering matrix element squared and a material specific dynamic structure factor that captures the target response. Here we summarize the main results of Ref.~\cite{Trickle2019a}. Our goal is to find materials with strong responses (a large dynamic structure factor) in each channel over the kinematically allowed mass region.

\renewcommand\arraystretch{1.1}
\setlength{\tabcolsep}{0.4em}
\begin{table*}[ht]
\begin{tabular}{|c||c||c|c||c|c|c|c|c|}
\hline
Target & $E_g$ [eV] &$\overline{c}_s^\text{LA} \, [10^{-5}]$ & $A_j$ & $\overline{Z}^*$ & $\overline{\varepsilon}_\infty$ & $ \overline{\omega}_\text{O} \, [\text{meV}]$ & $Q \, [10^{-7}]$ \Tstrut\Bstrut\\
\hline
Si & 1.11 &  2.84 &  28.1 & - & - & 62.3 & - \Tstrut\\
Ge & 0.67  & 1.61 &  72.6 & - & - & 34.8 & -\\
NaI & 5.8 & 0.90 &  23.0, 127  & 1.20  & 3.27  & 12.4 - 20.0 & 23\\
CsI  & 6.14 & 0.46 &  133, 127 & 1.22 & 2.70 & 6.9 - 10.0 & 12\\
\ce{CaWO4} & 5.2 & 1.42 & 40, 184, 16 & 2.84, 4.67 & 3.84 & 8.48 - 106 & 45\Bstrut\\
\hline
\hline
GaAs & 1.42 & 1.57 & 69.7, 74.9 & 2.27 & 10.9 & 31.8 -  34.9 & 2.4\Tstrut\\
\ce{Al2O3} & 8.8 & 3.51 &  27.0, 16.0  & 2.97 & 3.26 & 35.6 - 104 & 130\\
Diamond & 5.47$^*$ & 5.98  &  12.0  & -  & - & 161 & -\Bstrut\\
\hline
\hline
\ce{SiO2}  & 9.2 & 5.76 & 28.1, 16.0 & 3.38 & 2.41 & 13.7 - 149 & 200\Tstrut\\
PbTe & 0.19$^*$ & 1.17 &  207, 128 & 5.69 & 26.3 & 3.91 - 13.5 & 1.3\\
InSb & 0.24$^*$ & 1.13 & 115, 122 & 2.40 & 23.7 & 20.5 -  21.5 & 0.34\Bstrut\\
\hline
\hline
AlN  & 6.20 & 5.70 & 27.0, 14.0 & 2.57 & 4.54 & 29.4 - 109 & 78\Tstrut\\
\ce{CaF2} & 11.81 & 2.15 & 40, 19.0 & 2.36 & 2.26 & 28.4 - 55.6 & 130 \\
GaN & 3.43$^*$ & 4.17 & 69.7, 14.0 & 2.74 & 6.10 & 16.7 - 88.9 & 23 \\
GaSb & 0.720 & 1.32 &  69.7, 122 & 1.92 & 21.6 & 26.4 -  27.3 & 0.33\\
LiF  & 14.2 & 2.17 &  6.9, 19.0 & 1.05 & 2.02 & 33.5 - 77.2 & 270\\
MgF$_2$  & 12.4 & 2.43  & 24.3, 19.0 & 2.00 & 1.97 & 12.1 -  73.7 & 130 \\
MgO & 7.83 & 3.11 &  24.3, 16.0 & 1.97 & 3.38 & 46.3 -  82.6 & 110 \\
NaCl & 8.75 & 1.19 &  23.0, 35.5 & 1.09 & 2.44 & 19.1 - 30.6 & 80 \\
NaF & 11.5 & 1.78 &  23.0, 19.0  & 0.98 & 1.78 & 29.6 - 49.9 & 140\\
PbS  & 0.29$^*$ & 1.41 &  207, 32.1 & 4.45 & 15.0 & 7.27 - 26.9 & 4.9 \\
PbSe  & 0.17$^*$ & 1.27 &  207, 79.0 & 4.86 & 19.5 & 4.86 - 17.1 & 2.2 \\
ZnO & 3.3 & 4.18 &  65.4, 16.0  & 2.17 & 6.13 & 11.1 - 63.4  & 19\\
ZnS & 3.80$^*$ & 1.53 &  65.4, 32.1 & 2.03 & 5.91  & 32.8 -  41.0 & 14\Bstrut\\
\hline
\end{tabular}
\caption{Target materials studied in this work and their key parameters. 
The four blocks contain materials currently in use in nuclear recoil experiments, those considered for proposed near-future experiments, those with superior properties for some specific DM models discussed in this paper, and the remaining ones in alphabetical order, respectively. 
Sensitivity of electron transitions relies heavily on the band gap $E_g$, for which experimental values are shown (those with asterisks are measured at low temperature). Nuclear recoils and acoustic phonon excitations in the nucleon-coupling benchmark model are largely determined by the speed of sound of longitudinal acoustic phonons $\overline{c}_s^\text{LA}$ and atomic mass numbers $A_j$. For optical phonon excitations in the light dark photon mediated model, relevant parameters are the Born effective charges $\overline{Z}^*$, high-frequency dielectric constant $\overline{\varepsilon}_\infty$, optical phonon energies $\overline{\omega}_\text{O}$ as well as $A_j$, all of which combine into a quality factor $Q$, defined in Eq.~\eqref{eq:quality_factor}, which determines the reach at high mass. 
Barred quantities are properly averaged values; see Appendix~\ref{app:table_params_detail} for details. 
}
\label{table:one}
\end{table*}

Toward this goal, in Secs.~\ref{sec:darkphoton} and~\ref{sec:hadrophilic}, we carry out a detailed comparison of target materials, focusing on two benchmark DM scenarios to illustrate how to optimize target choice for the best sensitivity. Our study covers a total of 24 crystal materials, whose key properties that determine sensitivity to DM scattering are summarized in Table~\ref{table:one}. Six of the targets we consider are already used in existing nuclear recoil experiments, including Si (DAMIC~\cite{deMelloNeto:2015mca,AguilarArevalo2019a}, SENSEI~\cite{Abramoff2019a}, SuperCDMS~\cite{Agnese:2014aze,Agnese:2016cpb,Agnese:2015nto,Agnese2018a,Agnese2019a,Agnese2018b}), Ge (SuperCDMS~\cite{Agnese:2014aze,Agnese:2016cpb,Agnese:2015nto,Agnese2018a,Agnese2019a,Agnese2018b}), NaI (DAMA/LIBRA~\cite{Baum2019a}, KIMS~\cite{Kim:2015prm}, ANAIS~\cite{Amare:2019jul}, SABRE~\cite{Shields2015a}, DM-Ice~\cite{Jo:2016qql}), CsI (KIMS~\cite{Kim2008a}), \ce{Al2O3} (CRESST-I~\cite{Cozzini2002}), \ce{CaWO4} (CRESST-II-III~\cite{Petricca:2017zdp,Angloher2016a}), but their responses over all channels have not been studied. Two other targets -- GaAs and diamond -- have been proposed for near-future experiments. We then choose a representative sample of well-known polar semiconductors comprising 16 materials. Our work utilizes state-of-the-art density functional theory (DFT) calculations of material properties. Technical aspects of these calculations are discussed in Appendix~\ref{app:dft}, where we also present our calculated electron band structures and phonon dispersions for the target materials. In the main text, we will highlight a subset of these materials, chosen according to those currently (previously) in use in direct detection (Si, Ge, CsI, \ce{CaWO4}, (\ce{Al2O3})), as well as one or two new materials which demonstrate particularly strong sensitivity to each benchmark model. In particular, for the dark photon mediator, we highlight SiO$_2$ and InSb. Results for the materials not presented in the main text can be found in Appendix~\ref{app:additional_plots}, along with other parameters assumed when calculating the reach.

\section{Detection Channels}
\label{sec:paths}

We begin by briefly reviewing the detection channels, which are discussed thoroughly in our companion paper~\cite{Trickle2019a}. Generally, for a DM particle $\chi$, the event rate per unit target mass is given by
\be
R = \frac{1}{\rho_T} \frac{\rho_{\chi}}{m_\chi} \int d^3v\, f_\chi(\vect{v})\, \Gamma(\vect{v}),
\ee
where $\rho_T$ is the target mass density, $\rho_{\chi}$ is the local DM energy density, $m_\chi$ is the DM mass, and $f_\chi(\vect{v})$ is the incoming DM's velocity distribution in the target rest frame. The event rate $\Gamma(\vect{v})$ for an incoming DM particle with velocity $\vect{v}$ is usually normalized against a reference cross section, defined from the particle-level scattering matrix element $\MM$ (in the nonrelativistic normalization) evaluated at a reference momentum transfer $q_0$. Here we adopt the following definitions,
\beq
\overline{\sigma}_n &\equiv& \frac{\mu^2_{\chi n}}{\pi} \overline{|\MM_{\chi n}(q_0)|^2}_{q_0=m_\chi v_0}\,,\\
\overline{\sigma}_e &\equiv& \frac{\mu^2_{\chi e}}{\pi} \overline{|\MM_{\chi e}(q_0)|^2}_{q_0=\alpha m_e}\,,
\eeq
for DM-nucleon and DM-electron interactions, respectively, where $\mu_{\chi n}$, $\mu_{\chi e}$ are the reduced masses, and $v_0$ is the dispersion of the DM's velocity distribution. They coincide with the total particle-level scattering cross sections in the case of a heavy mediator. As we show in Ref.~\cite{Trickle2019a}, for spin-independent (SI) scattering off a target material via tree-level exchange of a mediator, the matrix element factorizes into a DM component that is universal, and a target response component captured by a dynamic structure factor $S(\vect{q},\omega)$ that is target and excitation specific, such that
\be
\Gamma (\vect{v}) = \frac{\pi\overline\sigma}{\mu^2} \int\frac{d^3q}{(2\pi)^3} \,{\cal F}_\text{med}^2(q)\, S\bigl(\vect{q},\omega_{\vect{q}} \bigr)\,.
\ee
Here $\overline{\sigma}$, $\mu$ represent either $\overline\sigma_n$, $\mu_{\chi n}$ or $\overline\sigma_e$, $\mu_{\chi e}$, $\vect{q}$ is the momentum transfer from the DM to the target, and
\be
\omega_{\vect{q}}  = \frac{1}{2}m_\chi v^2 -\frac{(m_\chi\vect{v}-\vect{q})^2}{2m_\chi} = \vect{q}\cdot\vect{v} -\frac{q^2}{2m_\chi}
\ee
is the corresponding energy deposition. The mediator form factor is given by\footnote{When present, in-medium screening effects are incorporated in the dynamic structure factor $S(\vect{q},\omega)$ instead of the mediator form factor ${\cal F}_\text{med}(q)$.}
\be
{\cal F}_\text{med}(q) = 
\begin{cases}
1 & \text{(heavy mediator)}\,, \\
\left( q_0/q \right)^2 & \text{(light mediator)}\,.
\end{cases}
\ee
The dynamic structure factor, which captures the target's response to a general energy-momentum transfer $\omega,\vect{q}$, is given by
\be
S(\vect{q},\omega) \equiv \frac{1}{V} \sum_f \bigl|\langle f| {\cal F}_T(\vect{q}) |i\rangle \bigr|^2\, 2\pi\, \delta\bigl(E_f-E_i-\omega\bigr)\,, 
\label{eq:S}
\ee
where $V$ is the total volume, $|i\rangle$, $|f\rangle$ are the initial and final states of the target system, and ${\cal F}_T$ is the quantum mechanical operator acting on the target Hilbert space that the DM couples to.

For an isotropic target, the dynamic structure factor depends only on the magnitude but not the direction of $\vect{q}$, so the velocity integral can be evaluated independently, giving
\begin{align}
\eta(v_\text{min}) & \equiv \int d^3v  \frac{f_\chi(\vect{v})}{v} \Theta(v-v_\text{min})\,, \\
v_\text{min} & = \frac{q}{2 m_\chi} + \frac{\Delta E}{q} \,,
\end{align}
for which analytic expressions can be obtained assuming a boosted truncated Maxwell-Boltzmann (MB) distribution. On the other hand, for the more general case of anisotropic target response, the dynamic structure factor depends on the direction of $\vect{q}$, and we can utilize the delta function in Eq.~\eqref{eq:S} to evaluate the velocity integral first, giving 
\be
g(\vect{q},\omega)\equiv \int d^3 v f_\chi(\vect{v})\,2\pi\,\delta(\omega-\omega_{\vect{q}}),
\label{eq:g}
\ee
which can be computed analytically for the usually assumed boosted truncated MB distribution. 

In the following subsections, we consider each detection channel in turn, summarizing the formalism presented in Ref.~\cite{Trickle2019a} on the dynamic structure factors and detection rates, building on the discussion in previous works (particularly \cite{Essig:2015cda,Griffin:2018bjn,Coskuner:2018are}).

\subsection{Nuclear Recoils}
\label{sec:paths-nuclear}

For each nucleus species,
\be
S\bigl(\vect{q},\omega\bigr) = 2\pi \frac{\rho_T}{m_N} \frac{f_N^2}{f_n^2} \,F_N^2(q)\,\delta \biggl( \frac{q^2}{2m_N} -\omega \biggr) \,,
\ee
where $m_N$ is the nucleus mass, $f_n$, $f_p$ and $f_N=f_p Z+f_n(A-Z)$ are the DM-neutron, DM-proton and DM-nucleus couplings respectively, and $F_N(q)$ is the Helm form factor 
\begin{align}
&F_N(q) = \frac{3\,j_1(qr_n)}{qr_n}\,e^{-(qs)^2/2}\,, \\
&r_n \simeq \, 1.14\, A_n^{1/3} \,\text{fm} \;, \, s \simeq \, 0.9 \,\text{fm}\,,
\label{eq:helm}
\end{align}
which approaches 1 in the $q\to0$ limit. The differential rate with respect to energy deposition, generalized to the case of multiple nucleus species, is
\beq
\frac{dR}{d\omega} &=& \frac{\rho_\chi}{m_\chi}\frac{\overline\sigma_n}{2\mu_{\chi n}^2} \frac{1}{\sum_N A_N} \nonumber\\
&&\;\; \biggl[\sum_N A_N\, \frac{f_N^2}{f_n^2}\,F_N^2\,{\cal F}_\text{med}^2 \,\eta (v_{\text{min}})\biggr]_{q^2=2m_N \omega},\qquad
\label{eq:diff_rate_NR}
\eeq
where $v_\text{min}=\frac{q}{2\mu_{\chi N}}$. 

The conventional nuclear recoil calculation is valid when each nucleus can be considered independent of the other nuclei. In a crystal target, this is true if the scattering happens at a timescale $1/\omega$ much shorter than the inverse phonon frequencies $1/\omega_\text{ph}$, {\em i.e.}\ if the energy deposition $\omega \gg \omega_\text{ph}\sim\OO(100\,\text{meV})$, or equivalently, $q\gg\sqrt{m_N \omega_{\text{ph}}}$ (note that this momentum cutoff is essentially the inverse of the spatial extent of nucleus wavefunctions in a harmonic potential). For lower energy depositions, the scattering event proceeds by direct production of (single or multiple) phonons. We discuss single phonon excitations in Sec.~\ref{sec:paths-phonon}. We will see that single phonon excitation rates are suppressed by the Debye-Waller factor for $q \gtrsim \sqrt{m_N \omega_{\text{ph}}}$, which shows the complementarity between the two channels.

\subsection{Electron Transitions}

In solids, electrons form band structures with energy eigenstates labeled by a band index $i$ and a wave vector $\vect{k}$ within the first Brillouin zone (1BZ). In an insulator or semiconductor, all electrons occupy the valence bands at low temperatures, and can be excited across the band gap to conduction bands. The dynamic structure factor encapsulates all such transitions from $i_1, \vect{k}_1$ to $i_2,\vect{k}_2$:
\beq
&& S\bigl(\vect{q},\omega\bigr) 
= 2  \sum_{i_1,i_2} \int_\text{1BZ} \frac{d^3k_1d^3k_2}{(2\pi)^6} \,2\pi\,\delta\bigl(E_{i_2,\vect{k}_2}-E_{i_1,\vect{k}_1}-\omega\bigr) \nonumber\\
&&\qquad\qquad \times \sum_{\vect{G}}\, (2\pi)^3\delta^3(\vect{k}_2-\vect{k}_1+\vect{G}-\vect{q}) \,\bigl|f_{[i_1\vect{k}_1,i_2\vect{k}_2,\vect{G}]}\bigr|^2 ,\nonumber\\
\label{eq:structure_electron}
\eeq
up to screening effects. Here $\vect{G} = n_1 \vect{b}_1 + n_2 \vect{b}_2 + n_3 \vect{b}_3$, with $n_1, n_2, n_3 \in \mathbb{Z}$ and $\vect{b}_{1,2,3}$ are reciprocal primitive vectors. The crystal form factor is defined by
\beq
f_{[i_1\vect{k}_1,i_2\vect{k}_2,\vect{G}]} &\equiv& \sum_{\vect{G}_1,\vect{G}_2} \delta_{\vect{G}_2-\vect{G}_1,\vect{G}} \nonumber\\
&&\qquad u_{i_2}^* \bigl(\vect{k}_2+\vect{G}_2 \bigr) u_{i_1} \bigl(\vect{k}_1+\vect{G}_1 \bigr) \,,
\eeq
where $u_i(\vect{k}+\vect{G})$ are Bloch wavefunction coefficients computed from DFT (see Appendix~\ref{app:dft-electron}). We neglect possible spin dependence of the electron band structures, and simply sum over contributions from the degenerate spin states. The total rate is given by
\beq
R &=& \frac{2}{\rho_T} \frac{\rho_\chi}{m_\chi} \frac{\pi\overline\sigma_e}{\mu_{\chi e}^2} 
\sum_{i_1,i_2} \int_\text{1BZ} \frac{d^3k_1d^3k_2}{(2\pi)^6} \nonumber\\
&&\quad  \sum_{\vect{G}} g(\vect{q},\omega)\,{\cal F}_\text{med}^2(q)\,\bigl|f_{[i_1\vect{k}_1,i_2\vect{k}_2,\vect{G}]}\bigr|^2 \,,
\label{eq:electron_rate}
\eeq
where $\vect{q} = \vect{k}_2-\vect{k}_1 +\vect{G}$ and $\omega = E_{i_2,\vect{k}_2} -E_{i_1,\vect{k}_1}$ are assumed.  Note that unlike in nuclear recoils, the dynamic structure factor for electron transitions is generally not isotropic in $\vect{q}$ for all energy-momentum depositions. When anisotropies are significant, the rate cannot be expressed in terms of $\eta(v_\text{min})$, and the $g$ function in Eq.~\eqref{eq:g} should be used instead. The physical implication is that the rate depends on the direction of the DM wind and exhibits daily modulation. An example of this is discussed in Ref.~\cite{Trickle2019a}.

\subsection{Single Phonon Excitations}
\label{sec:paths-phonon}

Phonons are quanta of lattice vibrations in crystals. For a three-dimensional crystal with $n$ atoms/ions in the primitive cell, there are $3n$ phonon branches, with dispersions $\omega_{\nu,\vect{k}}$ ($\nu = 1,\dots, 3n$), where the wave vector $\vect{k}$ is in the 1BZ. The dynamic structure factor has the general form
\beq
S\bigl(\vect{q},\omega\bigr) &=& \frac{\pi}{\Omega}\sum_\nu \,\delta\bigl(\omega-\omega_{\nu,\vect{k}}\bigr) \times \nonumber\\
&& \frac{1}{\omega_{\nu,\vect{k}}}\, \biggl|\sum_{j} \frac{e^{-W_j(\vect{q})}}{\sqrt{m_j}}\,e^{i\vect{G}\cdot \vect{x}_j^0} \bigl(\vect{Y}_j\cdot\vect{\epsilon}_{\nu,\vect{k},j}^*\bigr) \biggr|^2 ,\qquad
\label{eq:structure_phonon}
\eeq
where $\Omega$ is the volume of the primitive cell, $j=1,\dots,n$ runs over the atoms/ions in the primitive cell, $\vect{x}_j^0$ are their equilibrium positions, and $m_j$ are their masses. $\vect{Y}_j$ contains the DM-atom/ion couplings, whose general definition is given in Ref.~\cite{Trickle2019a}. We explicitly state the expression of $\vect{Y}_j$ for each benchmark model below. $\vect{\epsilon}_{\nu,\vect{k},j}$ are the phonon polarization vectors. $\vect{k}$ is the momentum within the 1BZ that satisfies $\vect{q}=\vect{k}+\vect{G}$ for some reciprocal lattice vector $\vect{G}$ --- only those phonon modes that match the momentum transfer up to reciprocal lattice vectors can be excited, as a result of lattice momentum conservation. At large $q$, the dynamic structure factor is suppressed by the Debye-Waller factor, given by
\be
W_j (\vect{q}) = \frac{\Omega}{4m_j}\sum_\nu \int_\text{1BZ}\frac{d^3k}{(2\pi)^3} \frac{|\vect{q}\cdot\vect{\epsilon}_{\nu,\vect{k},j}|^2}{\omega_{\nu,\vect{k}}}\,.
\ee
We obtain the total rate
\beq
R &=& \frac{1}{m_\text{cell}} \frac{\rho_\chi}{m_\chi} \frac{\pi\overline\sigma}{2\mu^2} \int\frac{d^3q}{(2\pi)^3} \,{\cal F}_\text{med}^2(q) \,\sum_\nu\, g(\vect{q},\omega_{\nu,\vect{k}}) \nonumber\\
&& \frac{1}{\omega_{\nu,\vect{k}}}\, \biggl|\sum_{j} \frac{e^{-W_j(\vect{q})}}{\sqrt{m_j}} \,e^{i\vect{G}\cdot\vect{x}_j^0} \bigl(\vect{Y}_j\cdot\vect{\epsilon}_{\nu,\vect{k},j}^*\bigr) \biggr|^2 \,,
\label{eq:single_phonon_rate}
\eeq
where $m_\text{cell}= \rho_T\Omega$ is the mass contained in a primitive cell. The phonon dispersions $\omega_{\nu,\vect{k}}$ and polarization vectors $\vect{\epsilon}_{\nu,\vect{k},j}$ that enter this equation are obtained from DFT calculations (see Appendix~\ref{app:dft-phonon}).

\subsubsection{Acoustic vs.\ Optical Phonons}
\label{sec:acoustic_v_optical}

It is useful to distinguish acoustic and optical phonons, as they are sensitive to different types of DM interactions. Among the $3n$ phonon branches, three are gapless with linear dispersions $\omega_{\nu,\vect{k}} \sim c_s |\vect{k}|$ near $|\vect{k}|=0$ (with $c_s$ the sound speed), as a result of spontaneous breaking of translation symmetries; these are acoustic phonons that, in the long wavelength limit, correspond to in-phase oscillations of atoms/ions in the same primitive cell. The remaining $3(n-1)$ branches are gapped ``optical'' phonons, corresponding to out-of-phase oscillations.  

Due to the nature of in-phase oscillations, acoustic phonons can be efficiently excited if DM couples to different atoms/ions in a correlated way. An example is a DM particle coupling to nucleons via a scalar or vector mediator. In this case, $\vect{Y}_j$ is proportional to a linear combination of $A_j$ and $Z_j$, and can have the same sign and similar magnitudes for all $j$.

By contrast, the out-of-phase oscillations associated with gapped phonon modes have enhanced sensitivity to DM coupling to the atoms/ions in the same primitive cell differently. This is the case for dark-photon-mediated DM scattering with polar materials.  The dark photon mediator kinetically mixes with the SM photon, and as a result, $\vect{Y}_j$ point in opposite directions for oppositely charged ions. We follow convention and call all gapped phonon modes ``optical,'' though only in polar materials where there are both positively and negatively charged ions in the primitive cell ({\em e.g.}\ GaAs) do these modes couple strongly to the (dark) photon via the oscillating dipole. Diamond, Si and Ge, for example, all have gapped phonon modes, but none of these materials has a strong coupling to the dark photon as the primitive cell does not contain oppositely charged ions.

\section{Target Comparison: Kinetically Mixed Light Dark Photon Mediator}
\label{sec:darkphoton}

\begin{figure*}[t]
\centering
\includegraphics[width=0.95\textwidth]{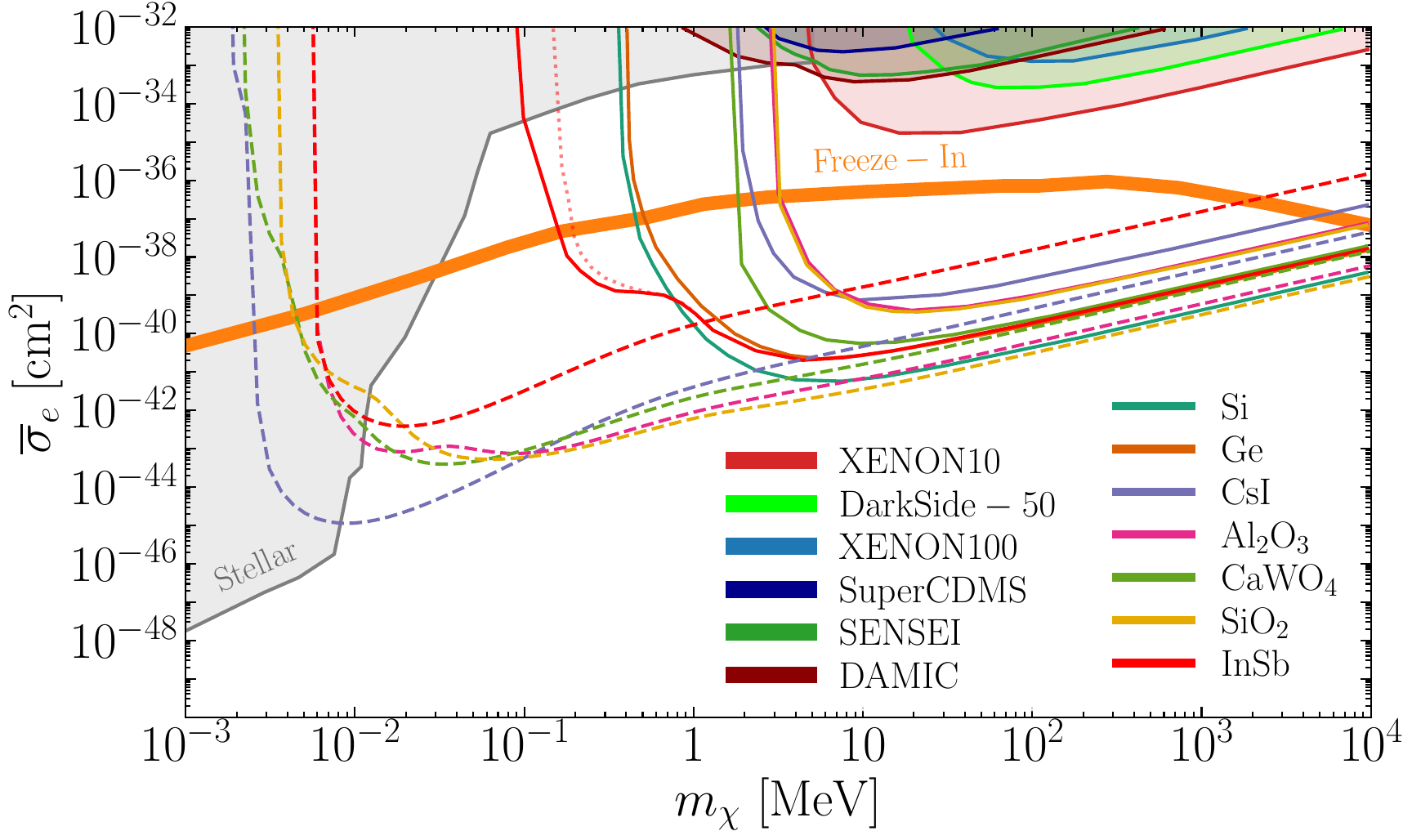}
\caption{Projected reach from single phonon excitations (dashed) and electron transitions (solid) for DM scattering mediated by a kinetically mixed light dark photon (the smallest-gap target InSb suffers from slow convergence in the electronic transition calculation at $m_\chi<1\,$MeV, for which we show results of the two most accurate runs with solid and dotted curves, see Appendix~\ref{app:dft-electron} for details). Nuclear recoils (not shown) can also probe this model, but the conclusion on which targets are superior is the same as for the light hadrophilic mediator model. A detector threshold of 1\,meV is used for the phonon calculations, and all transitions with energy deposition greater than the band gaps are included in electron excitations. The freeze-in benchmark is taken from Refs.~\cite{Chu:2011be,Essig:2011nj}, corrected by including plasmon decay for sub-MeV DM~\cite{Dvorkin:2019zdi}. Stellar constraints are from Ref.~\cite{Vogel:2013raa} and direct detection constraints are from DAMIC~\cite{AguilarArevalo2019a}, DarkSide-50~\cite{Agnes2018a}, SENSEI~\cite{Abramoff2019a}, SuperCDMS~\cite{Agnese2018b}, XENON10~\cite{Essig:2017kqs,Essig:2012yx}, and XENON100~\cite{Agnes2018a,Aprile2014a}.\footref{foot:one}}
\label{fig:ultralight_sigmae}
\end{figure*}

A well motivated model of light dark matter involves interaction with the SM via a light dark photon $A'$ that kinetically mixes with the photon:
\beq
\LL &=& - \frac{1}{4} F'^{\mu \nu}F'_{\mu \nu} + \frac{1}{2} \kappa \, F^{\mu \nu}F'_{\mu \nu} + \frac{1}{2} m_{A'}^2 A'^2 \nonumber\\
&& +\, \bigl(|D_\mu \chi|^2 -m_\chi^2|\chi|^2\bigr) \;\;\text{or}\;\; \bigl( i\overline\chi \slashed{D} \chi - m_\chi \overline\chi \chi\bigr) \,,\quad
\eeq
where $D_\mu = \partial_\mu-ie'A'_\mu$, and the DM $\chi$ can be either a complex scalar or a Dirac fermion. The gauge boson kinetic terms can be diagonalized by redefining $A_\mu \rightarrow A_\mu + \kappa A'_\mu$, which gives $J_\text{EM}^\mu$ a charge under the dark $U(1)$ of $\kappa e$. The reference cross section, utilized in present results for this model, is given by
\be
\overline{\sigma}_e = \frac{\mu_{\chi e}^2}{\pi} \frac{\kappa^2 e'^2 e^2}{\left( \alpha^2 m_e^2 + m_{A'}^2 \right)^2}.
\ee
The projected 95\% condifence level (C.L.) exclusion reach on $\overline{\sigma}_e$ assuming zero background (\textit{i.e.}, the cross section needed to obtain three events) from electron transitions and single phonon excitations are shown in Fig.~\ref{fig:ultralight_sigmae}, for $m_{A'}\to 0$ and an exposure of one kg-yr.\footnote{\label{foot:one} Changes in the constraint projections via single phonon excitations in Figs.~\ref{fig:ultralight_sigmae},~\ref{fig:fig10}-\ref{fig:fig12}, relative to previous versions, are due to a bug fix when using the Born effective charges, $\mathbf{Z}^*_j$. Targets with significant changes, at low $m_\chi$, are \ce{SiO2}, \ce{CaWO4} and \ce{MgF2}. The calculations have been updated using \textsf{PhonoDark} v1.1.0~\cite{phonodark}.} In the rest of this section, we describe in detail the features in this plot, and also discuss nuclear recoils.

\subsection{Single Phonon Excitations}

Optical phonon excitation is the dominant detection mode for dark photon mediated scattering. As shown in Ref.~\cite{Trickle2019a}, in the low $q$ limit (which dominates the momentum integral for a light mediator since ${\cal F}_\text{med}^2\propto q^{-4}$), the interaction is described via the Born effective charges of the ions, $\vect{Z}_j^*$ (which are generally $3\times3$ matrices),
\be
\vect{Y}_j = -  \frac{q^2}{\vect{q} \cdot \vect{\varepsilon}_\infty \cdot \vect{q}} \,\bigl(\vect{q}\cdot\vect{Z}_j^* \bigr) +\OO(q^2)\,,
\ee
where $\vect{\varepsilon}_\infty$ is the high-frequency dielectric matrix. The total rate is given by Eq.~\eqref{eq:single_phonon_rate}. Only \textit{polar} materials, or those which have differently charged ions in the primitive cell, can couple phonon modes to the dark photon, which explains the absence of phonon reach curves for Si and Ge in Fig.~\ref{fig:ultralight_sigmae}.

As explained in the previous section, optical phonon modes involve out-of-phase oscillations and are gapped. Because the optical modes are the dominant contribution to the rate, the properties of the optical modes determine the shape of the phonon excitation curves in Fig.~\ref{fig:ultralight_sigmae}: when there are sharp changes in the reach as a function of mass, it is because there is a transition in the dominance of a particular optical mode. For low momentum transfer, the dispersion of the gapped modes is approximately a constant, such that the lowest DM mass reachable is determined by setting the maximum kinetic energy of the incoming DM, $m_\chi v_\text{max}^2/2$, equal to the energy of the lowest optical mode,\footnote{One has to be careful with this estimate, as the lowest optical mode is generally not the dominant mode, rather it is the mode which is most ``longitudinal,'' or maximizes $\vect{q} \cdot \vect{\epsilon}$. For simple diatomic materials, there is one precisely longitudinal mode in the low $q$ limit, but the same is not true for more complex materials such as \ce{Al2O3}, as many gapped modes have a longitudinal component. A general rule of thumb is that the highest energy optical mode is the most longitudinal.}
\be
m_{\chi,\text{min}} \sim 3\,\text{keV} \biggl(\frac{\omega_\text{O}}{10\,\text{meV}}\biggr)\,.
\label{eq:mxminopt}
\ee
Thus materials having low energy optical phonon modes are desirable to search for light dark matter; CsI, for example, has particularly low-lying optical phonon excitations, and its sensitivity to the lightest DM masses is seen in Fig.~\ref{fig:ultralight_sigmae}. 

We can also see that at higher masses, single optical phonon production rates vary widely between materials. This can be understood analytically.  Consider first the simplest case of a diatomic polar crystal ({\em e.g.}\ GaAs). The dominant contribution to the $q$ integral in Eq.~\eqref{eq:single_phonon_rate} is well within the 1BZ and therefore we can set $\vect{G}=\vect{0}$, $W_j\simeq 0$, and $g(\vect{q}, \omega) \propto q^{-1}$. Approximating $\vect{Z}_j^*\simeq Z_j^* \,\mathbb{1}$, and noting that $Z_1^*=-Z_2^*\equiv Z^*$, we see that the rate is dominated by the longitudinal optical (LO) mode, for which one can show $\vect{\epsilon}_{\text{LO},\vect{k},1}$ and $\vect{\epsilon}_{\text{LO},\vect{k},2}$ are anti-parallel, and $|\vect{\epsilon}_{\text{LO},\vect{k},j}|=\sqrt{\mu_{12}/m_j}$ in the limit $k\to 0$, where $\mu_{12}$ is the reduced mass of the two ions. Further approximating the phonon dispersion as constant and $\vect{\varepsilon}_\infty\simeq \varepsilon_\infty\,\mathbb{1}$, the rate simplifies to
\beq
R &\propto& \frac{q_0^4}{m_\text{cell}} \frac{\rho_\chi}{m_\chi} \frac{\overline{\sigma}_e}{\varepsilon_\infty^2 \omega_\text{LO}} \frac{Z^{*2}}{ \mu_{\chi e}^2 \mu_{12}} \log{\left( \frac{m_\chi v_0^2 }{\omega_\text{LO}}\right)} \nonumber\\
&\propto& \frac{Z^{*2}}{A_1A_2\varepsilon_\infty^2} \left( \frac{\text{meV}}{\omega_\text{LO}} \right) \equiv Q\,.
\eeq
We call $Q$ a quality factor, since it is the combination of material-specific quantities that determines the direct detection rate. A higher-$Q$ material has a better reach in the high mass regime. More concretely, we find
\begin{align}
R \simeq \frac{1}{\text{kg yr}} & \left( \frac{Q}{10^{-7}} \right)  \left( \frac{m_e}{m_\chi} \right) \left( \frac{m_e^2}{\mu_{\chi e}^2}\right) \left( \frac{\overline{\sigma}_e}{10^{-39} \, \text{cm}^2}\right) \nonumber  \\
& \times \log{\left( \frac{q_\text{max}}{q_\text{min}} \right)} \, .
\end{align} 

Note that although we have focused on the special case of diatomic polar crystals in order to derive analytic estimates, similar considerations apply for more complicated crystals. For example, it is not surprising that larger Born effective charges and lighter ions are helpful. When comparing the targets, we adopt the following prescription for the quality factor,
\be
Q \equiv \frac{1}{\overline{\varepsilon}_\infty^2 \overline\omega_O} \prod_{j = 1}^n \left( \frac{|Z_j^*|}{A_j} \right)^\frac{2}{n}\,,
\label{eq:quality_factor}
\ee
where $n$ is the total number of ions in the primitive cell, and $\overline\omega_O$ is the directionally averaged optical phonon energy of the highest mode near $\vect{k}=\vect{0}$, given in Table~\ref{table:one}. In our list of materials LiF has the largest quality factor, with \ce{SiO2} second. We choose to highlight \ce{SiO2} in Fig.~\ref{fig:ultralight_sigmae} because LiF is a less desirable experimental target due to large backgrounds~\cite{Miuchi:2002zp}. 

A further consideration for optimizing $Q$ given a fixed chemistry (atomic species) is to maximize the Born effective charges. For example, cubic tungsten trioxide (WO$_{3}$) has been reported to have anomalously high Born effective charges of up to $+12.5$ and $-9.1$ on W and O respectively~\cite{Detraux_et_al:1997}. Materials with such high Born effective charges, a manifestation of highly covalent bonding character, provide a further route for maximizing $Q$.\footnote{Cubic WO$_{3}$ is dynamically unstable giving imaginary frequencies in the phonon band structure. Therefore we do not include it in phonon comparison plots, and leave a study of other stable isomorphs for future work.}

We comment in passing that also in the case of a heavy dark photon mediator, the rate is largely determined by the quality factor defined in Eq.~\eqref{eq:quality_factor} for sub-MeV DM; for heavier DM, couplings to ions cannot be simply captured by the Born effective charges at high momentum transfer, and the total rate is more challenging to compute~\cite{Trickle2019a}.

\subsection{Electron Transitions}

The typical band gaps between valence and conduction bands, $E_\text{g}$, range from a fraction of an eV (InSb and Ge) to as high as 10 eV ({\em e.g.}\ \ce{SiO2}). This gap sets the lightest DM mass to which the experiment is sensitive, as kinematics requires that $m_\chi v_\text{max}^2/2 > E_\text{g}$, implying
\be
m_{\chi,\text{min}} \sim 0.3\,\text{MeV} \biggl(\frac{E_g}{\text{eV}}\biggr)\,.
\label{eq:mxmin_electron}
\ee
Thus, small gap materials will generally have better reach. For example, InSb is superior to Si for $m_\chi\lesssim$ MeV, as seen in Fig.~\ref{fig:ultralight_sigmae}; in fact, the sub-eV band gap of InSb allows for a significant $\vect{G}=\vect{0}$ contribution that is absent for larger gap materials, and this contribution dominates at $m_\chi\lesssim$ MeV, greatly extending the reach. However, note that Ge, which has a smaller band gap than Si, does not have a better reach. The difference here is due to a direct vs indirect band gap.\footnote{The HSE06 exchange-correlation functional used in our DFT calculations slightly underestimates the direct band gap of Ge whilst being a close match to the indirect band gap~\cite{Peralta2006a}. This leads to the prediction of a direct band gap when optimized lattice parameters are used, contrary to experiment.} When depositing energy via a scattering process, there must be some momentum transfer, and therefore, strictly speaking, $E_g$ in Eq.~\eqref{eq:mxmin_electron} should be replaced by the minimum kinematically allowed energy difference. For direct gap materials this means that $m_{\chi, \text{min}}$ will increase, as it does in Ge, which is why Ge has worse reach than Si. Note that there is a complementarity between single phonon excitations and electron transitions. In the phonon case, materials with the best sensitivity tend to be insulators, as they have small values of $\varepsilon_\infty$. However, for electron transitions, one prefers materials with smaller band gaps, which generally have larger values of $\varepsilon_\infty$. This is because loop corrections to the in-medium photon propagator are larger for a smaller band-gap: virtual electrons can be more easily created because of the smaller energy difference. 

For higher masses an analytic comparison is not tractable. The wavefunction coefficients in Eq.~\eqref{eq:electron_rate} cannot be modeled well analytically, and hence the reach must be computed numerically. Note that for Si, Ge, NaI, CsI, GaAs, and diamond, our results are roughly consistent with previous calculations in Refs.~\cite{Essig:2015cda,Derenzo:2016fse,Kurinsky:2019pgb}, where the DFT calculation is implemented differently. However, we find discrepancies in the semi-core electron contributions, which are subdominant for our light mediator benchmark, but become important for a heavy mediator. We will investigate this issue in detail in an upcoming publication. Another improvement of the calculation that we plan to address is the treatment of in-medium screening effects (see Ref.~\cite{Trickle2019a} for further discussion), which we have neglected in the present calculation. Such effects are expected to be weak for materials with band gaps larger than about 1 eV. However, for sub-eV gap targets such as InSb, for masses below $\sim 1$ MeV, the result here should be taken with caution, as the effects may not be negligible.

\subsection{Nuclear Recoils}
\label{sec:darkphoton-nr}

The dark photon mediator coupling in a target system is momentum dependent. At very small momentum transfers $q \to 0$, the coupling is negligible as the total target is assumed to have no net charge. For $q \lesssim r_\text{ion}^{-1}$, where $r_\text{ion}$ is the size of an atom without the binding electrons, ionic charges, if present, can be coupled to. As the momentum transfer increases further, outer-shell electrons will respond incoherently, possibly transitioning to conduction bands independent of proton and inner-shell electron responses. On the other hand, in a nuclear recoil event, $q\gg\sqrt{m_N\omega_\text{ph}} \gg r_\text{ion}^{-1}$. In this regime, protons respond coherently as long as $F_N(q)\simeq 1$, since they are bound in the nucleus, whereas electron couplings are irrelevant since even the core electron wavefunctions do not have such high momentum components. Therefore, nuclear recoils can happen in an overall neutral crystal via coupling to the proton number of each nucleus without any atomic form factor suppression.

In order to compare against phonon and electron excitations, we express the reach in terms of $\overline\sigma_e$ instead of $\overline\sigma_n$. This corresponds to replacing $(f_N/f_n)^2\to Z_N^2$ for each nucleus species, and $\mu_{\chi n} \rightarrow \mu_{\chi e}$, $q_0 \rightarrow \alpha m_e$, and lastly, $\overline\sigma_n \rightarrow \overline\sigma_e$ in Eq.~\eqref{eq:diff_rate_NR}. While we discuss material comparison in this subsection, nuclear recoil reach curves have been omitted in Fig.~\ref{fig:ultralight_sigmae} in order not to further complicate the plot; they can be approximately rescaled from the reach curves in Fig.~\ref{fig:ultralight_sigman} below, and are straightforward to compute from Eq.~\eqref{eq:diff_rate_NR}.\footnote{\label{foot:two} Changes in the constraint projections via single phonon excitations in Figs.~\ref{fig:ultralight_sigman}-\ref{fig:massive_sigman},~\ref{fig:fig13}-\ref{fig:fig18}, relative to previous versions, are due to a bug fix in computing $F_{N_j}$ which altered the constraints by a factor of, approximately, 2.25. The calculations have been updated using \textsf{PhonoDark} v1.1.0~\cite{phonodark}.}

The low mass reach of nuclear recoils is material and threshold dependent, and can be understood from kinematics. The maximum momentum transfer is given by $q_\text{max} = 2 \mu_{\chi N} v_\text{max} $, and therefore the maximum energy deposited is given by $\omega_\text{max} = 2 \mu_{\chi N}^2 v_\text{max}^2/m_N$. Requiring that this be larger than the threshold sets the minimum DM mass. For a threshold around $500$ meV (which almost saturates the validity bound for some of the crystal targets as discussed in Sec.~\ref{sec:paths-nuclear}), and $v_\text{max} = 10^{-3}$, $m_\chi \ll m_N$, the minimum DM mass within reach is
\begin{equation}
m_{\chi, \text{min}} \sim 100 \,\text{MeV} \, \left( \frac{\omega_\text{min}}{500 \text{ meV}} \right)^\frac{1}{2} \left( \frac{m_N}{10 \text{ GeV}} \right)^\frac{1}{2} \, ,
\label{eq:mXmin_nuclear}
\end{equation}
Therefore, materials with lighter nuclei are more favorable for kinematic matching.

At higher masses, kinematics is not a limiting factor, and we can obtain an analytic approximation for the rate. Assuming a singular nuclear species, $A_N = A$, $Z_N = Z$ simplifies Eq.~\eqref{eq:diff_rate_NR} to
\begin{align}
\frac{dR}{d\omega} \propto \frac{\overline{\sigma}_e}{m_\chi\mu_{\chi e}^2}\,\frac{Z^2}{A^2} \frac{1}{\omega^2} \eta(v_\text{min})\, ,
\label{eq:dRdED-dp-nr}
\end{align}
and we see that the rate is dominated by small $\omega$. At masses above a few hundred MeV and small $\omega$, $\eta(v_\text{min})$ approaches $\eta(0)$. The total rate then becomes
\begin{align}
R \propto \frac{\overline{\sigma}_e}{m_\chi\mu_{\chi e}^2}\,\frac{Z^2}{A^2} \frac{1}{\omega_\text{min}} \, ,
\label{eq:R-dp-nr}
\end{align}
and is approximately material independent. Note that if the dark photon mediator is heavy, the factor $A^2 \omega^2$ in the denominator of Eq.~\eqref{eq:dRdED-dp-nr} would be absent, and heavier (larger $Z$) elements are advantageous.

\section{Target Comparison: Hadrophilic Scalar Mediator}
\label{sec:hadrophilic}

\begin{figure*}[t]
\centering
\includegraphics[width=0.95\textwidth]{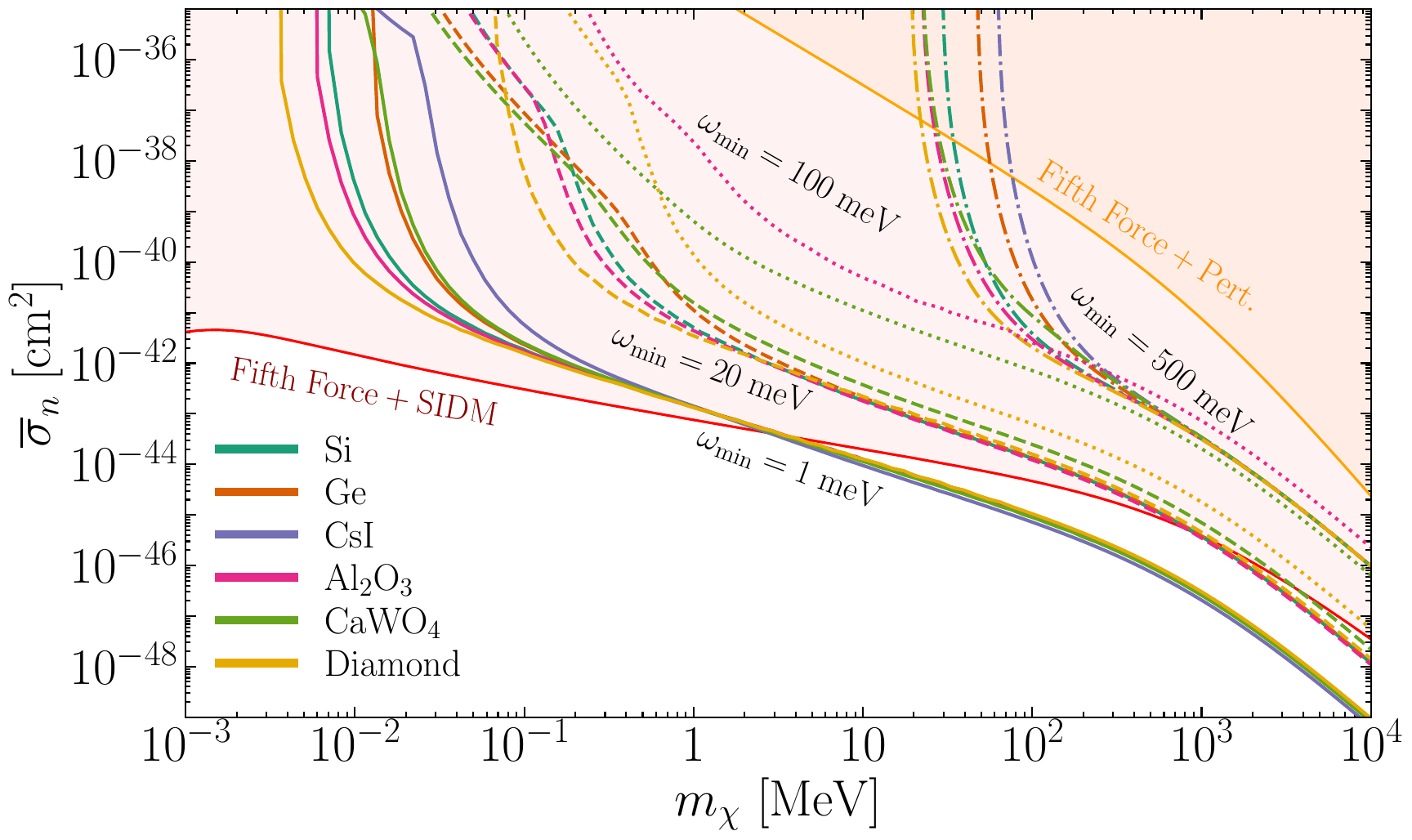}
\caption{Single phonon and nuclear recoil reach for a light ($m_\phi = 1$ eV) hadrophilic scalar mediator. 1, 20, and 100 meV thresholds are shown for the single phonon reach (solid, dashed, and dotted respectively), and 500 meV threshold is assumed for the nuclear recoil reach (dot-dashed). For $m_\phi = 1$ eV the dominant constraint on $f_n$ is from fifth force experiments~\cite{Knapen:2017xzo}. If $m_\chi$ makes up all the DM then the dominant constraint on $y_\chi$ is from DM self-interactions (SIDM)~\cite{Knapen:2017xzo}. If $m_\chi$ is only a subcomponent, we only require perturbativity $y_\chi<1$ (Pert.); in this case the reach curves can be easily rescaled.\footref{foot:two}}
\label{fig:ultralight_sigman}
\end{figure*}

\begin{figure*}[t]
\centering
\includegraphics[width=0.95\textwidth]{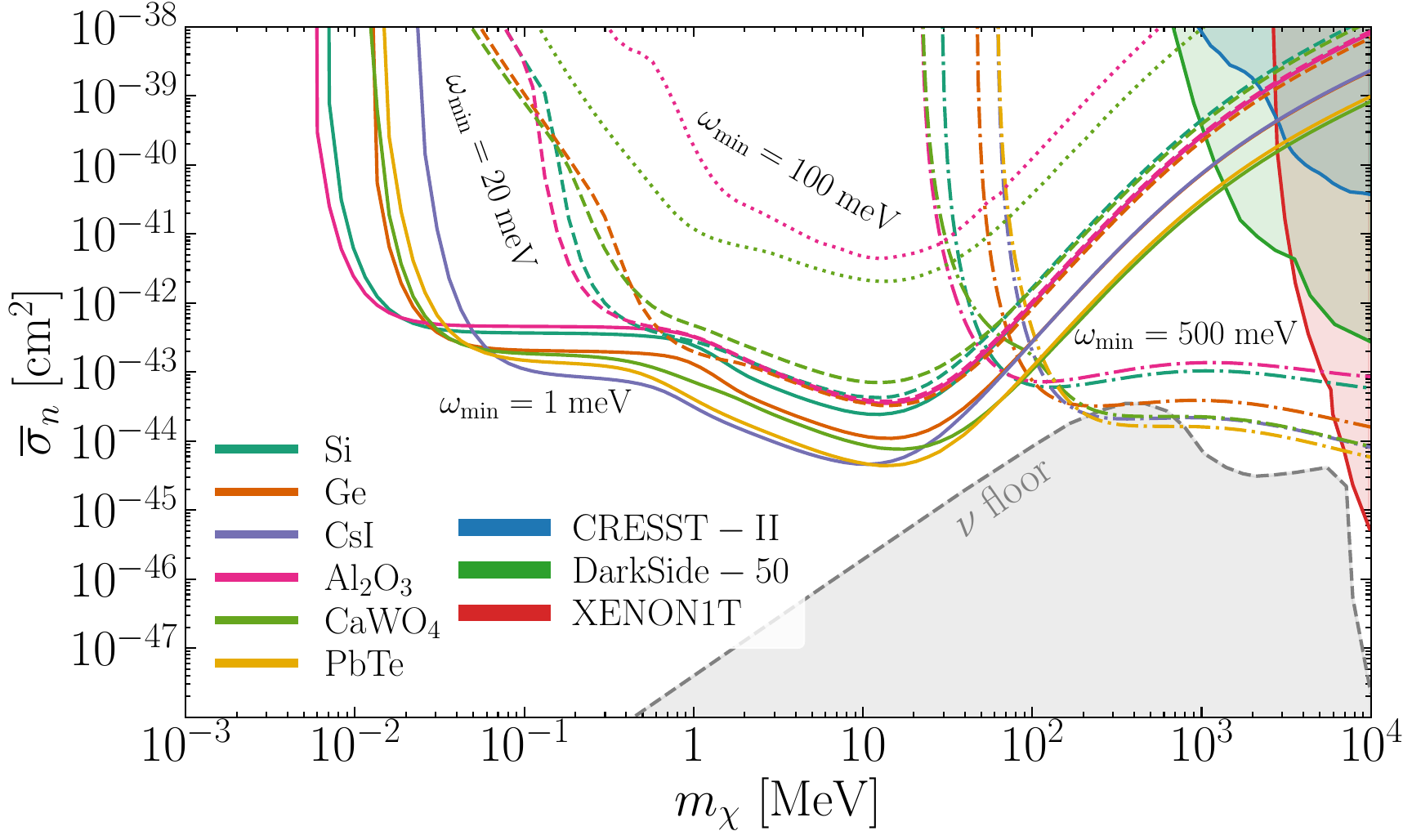}
\caption{Single phonon and nuclear recoil reach for a massive ($m_\phi \gtrsim 400$ MeV) hadrophilic scalar mediator. 1, 20, and 100 meV thresholds are shown for the single phonon reach (solid, dashed, and dotted respectively), and 500 meV threshold is assumed for the nuclear recoil reach (dot-dashed). There are no stellar constraints for $m_\phi \gtrsim 400$ MeV~\cite{Knapen:2017xzo}. Currently, the best experimental nuclear recoil constraints in this region of parameter space are from DarkSide-50~\cite{Agnes2018b} (assuming binomial fluctuations), and XENON1T (combined limits from~\cite{Aprile2018a, Aprile2019a}). We also show the constraint from CRESST-II~\cite{Angloher2016a}, which is stronger than the DarkSide-50 constraint at low masses assuming no fluctuation in energy quenching. A more complete collection of nuclear recoil constraints can be found in Refs.~\cite{Agnes2018b,Akerib2019b,Aprile2019a}. The neutrino floor is taken from Ref.~\cite{Battaglieri:2017aum}.\footref{foot:two}}
\label{fig:massive_sigman}
\end{figure*}

As a second benchmark model, we consider a real scalar mediator $\phi$ coupling to the proton and neutron,
\beq
{\cal L} &=& \frac{1}{2} (\partial_\mu\phi)^2 -\frac{1}{2} m_\phi^2\phi^2  + f_p \, \phi \, \overline pp + f_n \, \phi \, \overline nn \nonumber\\
&& +\,\biggl(\frac{1}{2}(\partial_\mu\chi)^2 -\frac{1}{2}m_\chi^2\chi^2 + \frac{1}{2} y_\chi m_\chi \phi\chi^2\biggr) \nonumber\\
&&\quad\; \text{or}\;\; \bigl( i\overline\chi \slashed{\partial} \chi - m_\chi \overline\chi \chi +y_\chi \phi \overline\chi \chi\bigr)\,,
\eeq
where the DM $\chi$ is taken to be either a real scalar or a Dirac fermion. In the absence of electron couplings, the relevant search channels are single phonon excitations and nuclear recoils. We will quote the reach in terms of $\overline{\sigma}_n$, given by
\be
\overline{\sigma}_n = \frac{\mu_{\chi n}^2}{4\pi} \frac{y_\chi^2 f_n^2}{\bigl( m_\chi^2 v_0^2 + m_\phi^2 \bigr)^2 } \, .
\ee
The 95\% C.L. exclusion reach on $\overline{\sigma}_n$ for a light (effectively massless) and heavy mediator are shown in Figs.~\ref{fig:ultralight_sigman} and \ref{fig:massive_sigman} respectively, assuming $f_p=f_n$, an exposure of one kg-yr, and zero background events.  In the rest of this section we explain in detail the features in these plots.

\subsection{Single Phonon Excitations}
\label{sec:hadro_single_phonon}

We first consider DM creating a single phonon via the nucleon coupling. As shown in Ref.~\cite{Trickle2019a},
\be
\vect{Y}_j = \vect{q}\,\biggl(\frac{f_j}{f_n}\biggr)\, F_{N_j}(q)\,,
\label{eq:nucleon_coupling}
\ee
where $f_j = f_p Z_j+f_n(A_j-Z_j)$ for the nucleus at site $j$ in a primitive cell, and $F_{N_j}(q)$ is the nuclear form factor given by Eq.~\eqref{eq:helm}. As before, the total rate is calculated from Eq.~\eqref{eq:single_phonon_rate}. However, a major difference compared to the dark photon mediator model is that, if $f_p$ and $f_n$ have the same sign, the rate is dominated by acoustic and not optical phonons, assuming the energy threshold is low enough to access the acoustic phonons. This is because $\vect{Y}_j$ points in the same direction for all $j$, resulting in stronger in-phase oscillations as discussed in Sec.~\ref{sec:acoustic_v_optical}.

We first discuss Fig.~\ref{fig:ultralight_sigman}, for the light mediator case, when the energy threshold, $\omega_\text{min}$, is 1 meV.  While such a low threshold is experimentally challenging, the curves are easier to understand conceptually compared to the higher $\omega_\text{min}$ curves. In fact, over most of the mass range, for most materials, the rate is dominated by single longitudinal acoustic (LA) phonon production. At the high mass end, the reach is material-independent, understood analytically as follows.  The mediator form factor $\mathcal{F}_\text{med} \propto 1/q^2$, and therefore the rate is dominated by the lowest detectable momentum transfer.  In this case, we can set $\vect{G}=\vect{0}$ (or equivalently, $\vect{q} = \vect{k}$ in Eq.~\eqref{eq:single_phonon_rate}), $W_j\simeq 0$, $\omega_\text{LA} = c_s^\text{LA} q$, $F_{N_j}\simeq 1$, and $g(q, \omega) \propto q^{-1}$. Lastly, in this limit $\vect{q} \cdot \vect{\epsilon}_{\text{LA}, j, \vect{k}} \simeq q \sqrt{m_j/m_\text{cell}}$. Thus the rate
\be
R \propto  \frac{m_\chi^3}{m_\text{cell}^2} \frac{\overline{\sigma}_n}{\mu_{\chi n}^2} \biggl( \sum_j \frac{f_{j}}{f_n} \biggr)^2 \frac{1}{\omega_\text{min}} \, .
\label{eq:phonon_hadro_massless_rate}
\ee
For $f_p=f_n$, we have $f_j \propto A_j \propto m_\text{cell}/m_n$, and the dependence on the target properties drops out. The reference cross section $\overline{\sigma}_n$ corresponding to a given event rate $R$ scales with mass as $\mu_{\chi n}^2/m_\chi^3$, as we see in Fig.~\ref{fig:ultralight_sigman}. Note that as we go to higher $m_\chi$ the reach on the couplings $f_n^2 y_\chi^2$ gets worse as $\mu_{\chi n}^2 m_\chi$; the apparent better reach at higher mass in Fig.~\ref{fig:ultralight_sigman} is due to the definition of $\overline{\sigma}_n\propto m_\chi^{-4}$.

For DM masses below $\sim 0.1$ MeV in Fig.~\ref{fig:ultralight_sigman}, kinematics causes the reach to diminish: the maximum momentum transfer, $2 m_\chi v_\text{max}$, must be large enough to reach the minimum momentum transfer set by the detector threshold, $\omega_\text{min}/c_s^\text{LA}$. This sets the minimum reachable DM mass
\be
m_{\chi,\text{min}}
\sim 20\,\text{keV}\, \biggl(\frac{\omega_\text{min}}{\text{meV}}\biggr)\biggl(\frac{10^{-5}}{c_s^\text{LA}}\biggr) \, .
\label{eq:mmin_speed_of_sound}
\ee
To reach the lightest dark matter particle at low thresholds, an ideal material is then diamond, as it has the highest speed of sound. AlN and \ce{SiO2} are the next best candidates from our search.

As we move on to the curves with higher energy thresholds, $\omega_\text{min} = 20$~meV and 100~meV, the materials with lower sound speed lose reach altogether. (The $\omega_\text{min} = 500$ meV curves are derived from nuclear recoil; this is discussed in the next subsection.)  The reason is that acoustic phonons are accessible only when $\omega_\text{min} \lesssim c_s^{\text{LA}}/a$, where $a$ is the lattice spacing. For materials with lower sound speed, the energy threshold may simply never be low enough to have any reach with an acoustic phonon.  In addition, one can see where optical phonons start to play a role, as the slope of the reach curve changes at lower masses, {\em e.g.}\ Si with an energy threshold of 20 meV. This feature will be present for all materials if the lowest kinematically reachable DM mass from optical phonon excitations, given in Eq.~\eqref{eq:mxminopt}, is smaller than the lowest kinematically reachable DM mass from acoustic phonon excitations, given in Eq.~\eqref{eq:mmin_speed_of_sound}.

Next we turn our attention to Fig.~\ref{fig:massive_sigman}, for the same hadrophilic scalar mediator benchmark, but with a heavy mediator. Again, we first focus on the case of a 1~meV threshold, as here the acoustic phonon contributions dominate and analytic simplifications can be made since the integrals are dominated by the high momentum behavior. There are four distinct regions in mass and we now discuss the mass and material parameters dependence of each of them. 

In the lowest mass regime, $m_\chi \lesssim 10^{-1}$ MeV, the reach ends when the acoustic modes are no longer kinematically available, just as in the massless mediator case, with minimum reachable mass again set by Eq.~\eqref{eq:mmin_speed_of_sound}. Between $10^{-1}$ and $1$ MeV, the reach curves flatten and the order of the curves reverses: materials with a higher speed of sound have worse reach, which can be understood analytically starting with Eq.~\eqref{eq:single_phonon_rate}. For $m_\chi \lesssim 1$ MeV the momentum transfer is within the 1BZ, so we can take $\vect{q} = \vect{k}$, $W_j \simeq 0$, $\omega = c_s q$ and $g(q, \omega) \propto 1/q$ as in the light mediator case. For simplicity we ignore angular dependence, assume the ions are the same, $A_j \equiv A$, $m_j \equiv m$, set $f_n = f_p$, and consider only the longitudinal mode so that $\vect{q} \cdot \vect{\varepsilon} \propto q$.  Then we have
\begin{align}
R & \propto \frac{\overline{\sigma}_n}{m_\text{cell} m_\chi^3 c_s} \int^{2m_\chi v} d^3k \frac{1}{k^2} \left(\frac{k A}{\sqrt{m}} \right)^2 \propto \frac{\overline{\sigma}_n}{c_s}, 
\label{eq:massive_rate_scaling_1}
\end{align}
where the upper cutoff is due to kinematics and manifests in the $g$ function, which goes to zero as $k$ reaches the maximum allowed momentum transfer. 

A similar derivation applies to the mass dependence in the next two regimes. For $1 \text{ MeV}\lesssim m_X \lesssim 10 \text{ MeV}$, the dominant momentum transfer is outside of the 1BZ, which means that $\omega$ can no longer be approximated by $c_s q$. In fact, since $\omega$ is only a function of the phonon momentum in the 1BZ, it will vary rapidly as $q$ increases. We therefore exchange $\omega$ with a $q$ independent quantity, roughly thought of as the average of $\omega$ over the whole 1BZ, $\langle \omega \rangle$. The rate becomes
\begin{align}
R & \propto \frac{\overline{\sigma}_n}{m_\text{cell} m_\chi^3 \langle \omega \rangle} \int^{2m_\chi v} d^3k \frac{1}{k} \left(\frac{k A}{\sqrt{m}} \right)^2 \propto \frac{\overline{\sigma}_n m_\chi}{\langle \omega \rangle } .
\label{eq:massive_rate_scaling_2}
\end{align}
Since the rate scales inversely with $\langle \omega \rangle$, materials with lower energy phonon modes are preferred. As $\langle\omega\rangle$ is usually correlated with $c_s$, the ordering of the curves is the same as in the previous regime. We have neglected the Debye-Waller factor in the analytic estimates above, because the momentum transfer is on the order of $m_\chi v$, and is less than the Debye-Waller cut-off around $\sqrt{m_N \langle \omega \rangle}$. However, for the last mass regime, above $\sim$\,10 MeV, this is no longer the case, and the momentum integral is cut-off by the Debye-Waller factor,
\begin{align}
R & \propto \frac{\overline{\sigma}_n}{m_\text{cell} m_\chi \mu_{\chi n}^2 \langle \omega \rangle} \int^{\sqrt{m_N \langle \omega \rangle}}_0 d^3k \frac{1}{k} \left(\frac{k A}{\sqrt{m}} \right)^2  \nonumber \\
& \propto \frac{\overline{\sigma}_n A^2 \langle \omega \rangle^2 }{m_\chi \mu_{\chi n}^2}.
\label{eq:massive_rate_scaling_3}
\end{align}
Therefore, materials with heavier elements and higher phonon energies are preferred. In our search, \ce{CaWO4} has the highest factor of $A \langle \omega \rangle$, with \ce{PbTe} following, which is the reason we choose to highlight \ce{PbTe} in Fig.~\ref{fig:massive_sigman}.

For higher thresholds, the optical phonon modes contribute to a greater degree, so the scaling arguments given above for the first two mass regimes no longer hold, but for the last two they do, which is why the curves are almost parallel.

\subsection{Nuclear Recoils}

For DM heavier than $\OO(100\,\text{MeV})$, nuclear recoils offer a complementary detection channel to phonon excitations. The low mass behavior of the reach curves is understood in the same way as in Sec.~\ref{sec:darkphoton-nr} (see Eq.~\eqref{eq:mXmin_nuclear}), and lighter elements are advantageous. At higher masses, the $\overline{\sigma}_n$ reach depends on the mediator mass. To show this analytically we again consider a single nucleus species, $A_N = A$, and $f_n = f_p$. In the case of a light mediator the differential rate in Eq.~\eqref{eq:diff_rate_NR} becomes
\begin{align}
\frac{dR}{d\omega} \propto \frac{\overline{\sigma}_n m_\chi^3}{\mu_{\chi n}^2}\,\frac{1}{\omega^2} \eta \left( v_\text{min} \right) \,.
\label{eq:n-nr-l}
\end{align}
For DM heavier than a few hundred MeV, the $m_N$ dependence via $\eta(v_\text{min})$ is weak, as in the dark photon mediator case. The rate is then
\begin{align}
R \propto \frac{\overline{\sigma}_n m_\chi^3}{\mu_{\chi n}^2}\,\frac{1}{\omega_\text{min}} \,,
\label{eq:R-nr-l}
\end{align}
which is material independent. This is why all the reach curves coincide for large DM masses. We also see that as in the case of acoustic phonons, achieving lower energy thresholds is crucial for improving the reach. 

If the mediator is heavy, we have
\beq
\frac{dR}{d\omega} &\propto& \frac{\overline{\sigma}_n A^2}{m_\chi \mu_{\chi n}^2} \eta \left( v_\text{min} \right) ,\\
R &\propto& \frac{\overline{\sigma}_n A^2}{m_\chi \mu_{\chi n}^2} \omega_\text{max} \propto \frac{\overline{\sigma}_n A \mu_{\chi N}^2 }{m_\chi \mu_{\chi n}^2}  \,  ,
\label{eq:R-nr-h}
\eeq
where for simplicity we take the $\eta$ function to decrease sharply at the kinematic bound. We reach the conclusion that heavier nuclei are preferred, similar to the case of single phonon excitations with a heavy mediator. Note also that there is no threshold dependence for larger masses. Therefore a lower threshold only helps to reach lower DM masses, as opposed to the case of the light mediator.

\section{Conclusions}
\label{sec:conclusions}

\renewcommand\arraystretch{1.55}
\begin{table*}[t]
\centering
\begin{tabular}{|c|c|c|c|}
\hline
\multicolumn{4}{|c|}{Light dark photon mediator (Sec.~\ref{sec:darkphoton}, Fig.~\ref{fig:ultralight_sigmae})} \\
\hline
\multirow{2}{*}{Detection channel} & \multicolumn{2}{c|}{Quantity to maximize to reach ...} & \multirow{2}{*}{Best materials} \\
\cline{2-3}
& ... lower $m_\chi$ & ... lower $\overline\sigma_e$ & \\
\hline
(Optical) phonons & $\omega_O^{-1}$ (Eq.~\eqref{eq:mxminopt}) & quality factor $Q$ defined in Eq.~\eqref{eq:quality_factor} & SiO$_2$, Al$_2$O$_3$, CaWO$_4$ \\
\hline
Electron transitions & $E_g^{-1}$ (Eq.~\eqref{eq:mxmin_electron}) & depends on details of electron wavefunctions & InSb, Si \\
\hline
Nuclear recoils & $\left(A \omega_\text{min} \right)^{-1}$ (Eq.~\eqref{eq:mXmin_nuclear}) & $(Z/A)^2\,\omega_\text{min}^{-1}$ (Eq.~\eqref{eq:R-dp-nr}) & diamond, LiF \\
\hline
\hline
\multicolumn{4}{|c|}{Hadrophilic scalar mediator (Sec.~\ref{sec:hadrophilic}, Figs.~\ref{fig:ultralight_sigman}, \ref{fig:massive_sigman})} \\
\hline
\multirow{2}{*}{Detection channel} & \multicolumn{2}{c|}{Quantity to maximize to reach ...} & \multirow{2}{*}{Best materials} \\
\cline{2-3}
& ... lower $m_\chi$ & ... lower $\overline\sigma_n$ & \\
\hline
\multirow{3}{*}{(Acoustic) phonons} & \multirow{3}{*}{$c_s/\omega_\text{min}$ (Eq.~\eqref{eq:mmin_speed_of_sound})} & Light mediator: $\omega_\text{min}^{-1}$ (Eq.~\eqref{eq:phonon_hadro_massless_rate}) & diamond, \ce{SiO2} \\
\cline{3-4}
& & Heavy mediator: $c_s^{-1}$ or $\omega_\text{ph}^{-1}$ or $A\omega_\text{ph}$ & \multirow{2}{*}{all complementary} \\
& & depending on $m_\chi$ (Eqs.~\eqref{eq:massive_rate_scaling_1}, \eqref{eq:massive_rate_scaling_2}, \eqref{eq:massive_rate_scaling_3}) &  \\
\hline
\multirow{2}{*}{Nuclear recoils} & \multirow{2}{*}{$\left(A \omega_\text{min} \right)^{-1}$ (Eq.~\eqref{eq:mXmin_nuclear})} & Light mediator: $\omega_\text{min}^{-1}$ (Eq.~\eqref{eq:n-nr-l}) & diamond, LiF \\
\cline{3-4}
& & Heavy mediator: $A$ (Eq.~\eqref{eq:R-nr-h}) & CsI, Pb compounds \\
\hline
\end{tabular}
\caption{\label{tab:sum}
Summary of our results. The material properties relevant for the optimization of target are: atomic mass number $A$, proton number $Z$, electronic band gap $E_g$, speed of sound $c_s$, optical phonon energy $\omega_O$, average phonon energy $\omega_\text{ph}$, as well as Born effective charges and the high-frequency dielectric constant that enter the quality factor $Q$. Achieving lower detector energy thresholds $\omega_\text{min}$ is also crucial in several cases.
}
\end{table*}

We considered spin independent DM direct detection through three channels -- single phonon excitations, electron transitions, and nuclear recoils -- in a wide variety of crystal target materials, and two well motivated DM models. Many of these materials are already being discussed for DM detection, but we have presented some new targets for consideration.

For each type of interaction, we specified the target material parameters which should be optimized in order to maximize the reach, and we found complementarity between targets depending on {\em (i)}  the experimental threshold, {\em (ii)} the mass range, and {\em (iii)} the model.  The experimental threshold dictates which modes are available: at higher recoil energies, only electron transitions and nuclear recoils are possible; as the threshold drops, optical and acoustic phonons become accessible.  The phonon modes in materials with high sound speed become kinematically available at higher thresholds than in materials with lower sound speeds. Also, for a given threshold, materials with higher sound speeds have reach to lighter dark matter. Regarding the mass range, the smallest detectable masses are always set by a kinematic constraint, and the dependence on material parameters, and detection threshold, can be found in Eqs.~\eqref{eq:mxminopt},~\eqref{eq:mxmin_electron},~\eqref{eq:mmin_speed_of_sound},~\eqref{eq:mXmin_nuclear} for optical phonon, electron, acoustic phonon excitations, and nuclear recoils respectively. As for the model, we defined a quality factor (in Eq.~\eqref{eq:quality_factor}) for single optical phonon excitations from dark photon mediated scattering to indicate which targets will have the best sensitivity. On the other hand, for a hadrophilic mediator, target optimization for acoustic phonon excitations depends on the mediator and DM masses. We summarize our results in Table~\ref{tab:sum}.

An attractive feature of phonon and electron excitations is the possible daily modulation of event rates, as the dynamic structure factors in Eqs.~\eqref{eq:structure_electron} and~\eqref{eq:structure_phonon} are generically anisotropic. In the context of phonon excitations, \ce{Al2O3} has been considered in Ref.~\cite{Griffin:2018bjn}, and in our companion paper \cite{Trickle2019a} we have discussed hexagonal boron nitride as an example of an $\OO(\text{eV})$-gap target which exhibits daily modulation in electron transitions. We plan on identifying other promising targets for daily modulation in the future.

\vspace{10pt}
{\em Acknowledgments.}
We thank Kyle Bystrom, Rouven Essig, Matt Pyle, Adrian Soto and Tien-Tien Yu for useful discussions. T.T.\ and K.Z.\ are supported by the Quantum Information Science Enabled Discovery (QuantISED) for High Energy Physics (KA2401032) at LBNL. Z.Z.\ is supported by the NSF Grant PHY-1638509 and DoE Contract DE-AC02-05CH11231. Development of the materials calculations in this paper (S.G.\ and K.I.)\ was supported by the Laboratory Directed Research and Development Program of LBNL under the DoE Contract No.\ DE-AC02-05CH11231. Computational resources were provided by the National Energy Research Scientific Computing Center and the Molecular Foundry, DoE Office of Science User Facilities supported by the Office of Science of the U.S.\ Department of Energy under Contract No.\ DE-AC02-05CH11231. The work performed at the Molecular Foundry was supported by the Office of Science, Office of Basic Energy Sciences, of the U.S.\ Department of Energy under the same contract. 

\newpage
\appendix

\section{Calculations of Target Properties}
\label{app:dft}

We obtain the materials-specific responses using first principles calculations based on density functional theory (DFT)~\cite{Martin2004a}. DFT is a standard method for obtaining solutions to the many-electron interaction problem, and can accurately predict materials properties \textit{ab initio} ranging from electronic and magnetic  to mechanical and vibrational properties. For this work, we used DFT to calculate the full electronic and phonon spectra for a range of materials, with the calculation details given below. However, since DFT is a ground-state method, it suffers from the famous `band gap' problem where excited-state properties, including band gaps, are not accurately treated using standard DFT methods. We correct for this in two ways: (i) we performed beyond-DFT calculations (hybrid functional calculations) for several of the compounds where standard DFT gave a zero band gap, and (ii) we adjusted the band gaps to experimentally-reported values for all compounds. We note that the convergence parameters used for the electronic and phonon calculations are different owing to the different physical properties being calculated.

\begin{figure*}[t]
\centering
\includegraphics[width=0.9\textwidth]{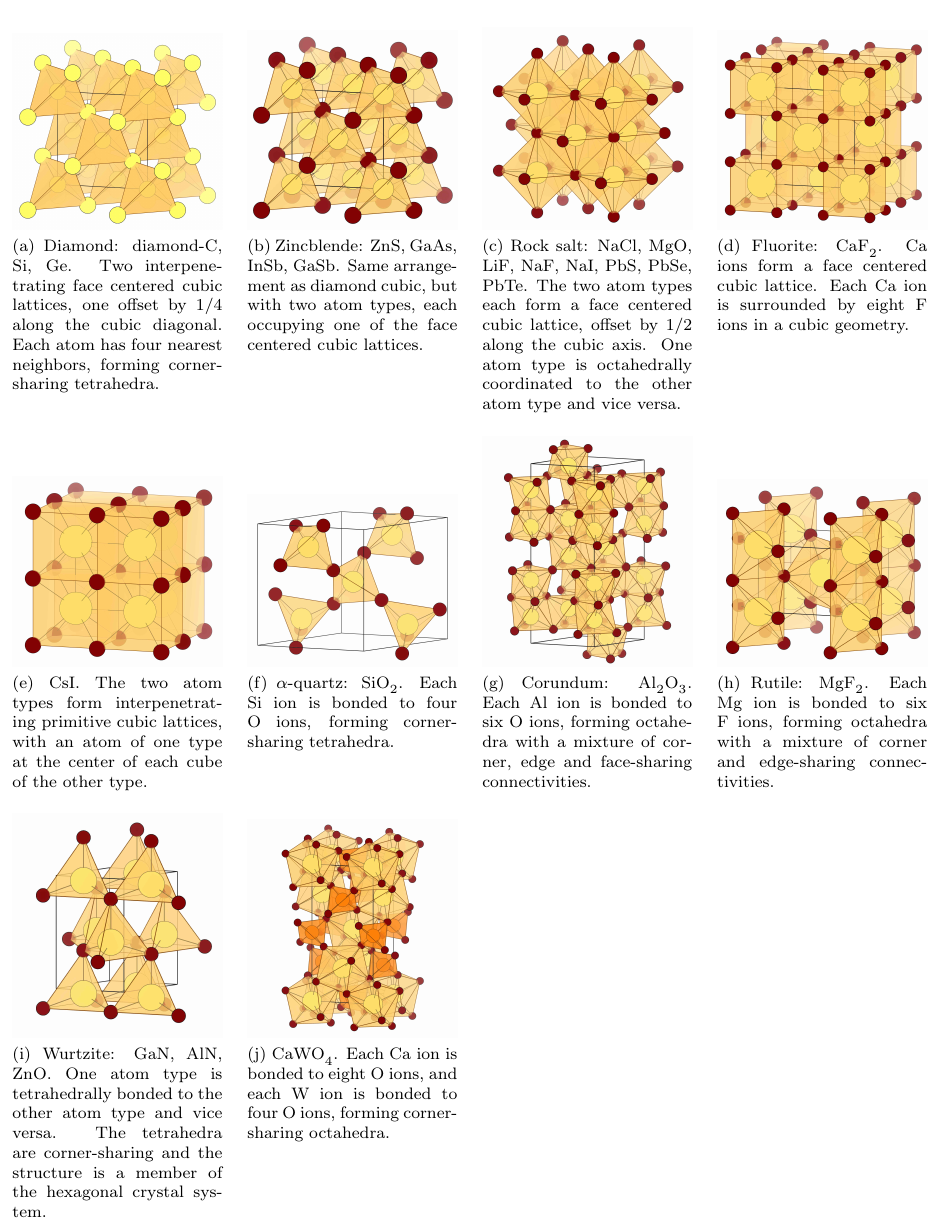}
\caption{Crystal structures of targets in Table~\ref{table:one}.}
\label{Crystalstructures}
\end{figure*}

\begin{figure*}[t!]
\centering
\includegraphics[width=\textwidth]{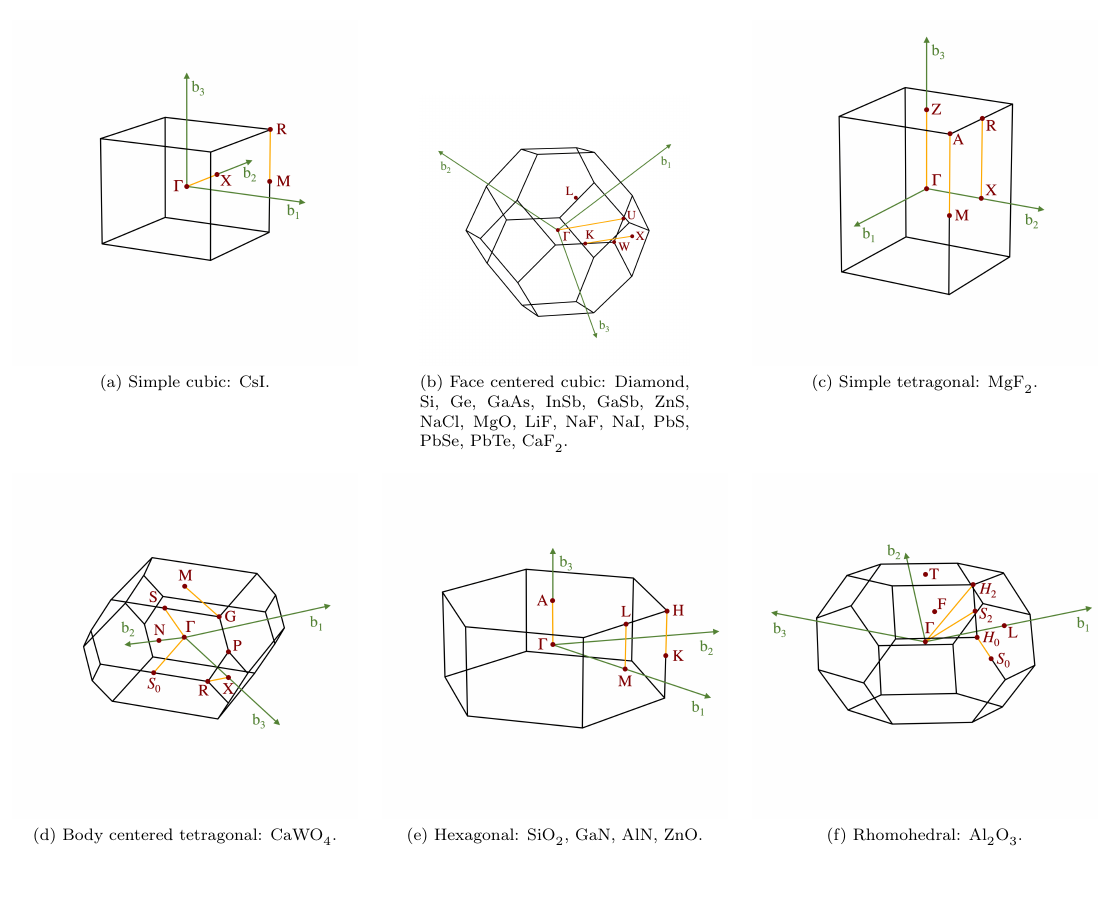}
\caption{First Brillouin zones of targets in Table~\ref{table:one}, with high symmetry points labeled.}
\label{BZs}
\end{figure*}

The list of materials calculated with their corresponding space groups and space group numbers is given in Table~\ref{tab:mat_calc_details}, with the crystal structures depicted in Fig.~\ref{Crystalstructures}. For compounds where several structural isomorphs exist, we considered the reported low-temperature ground state structure. The Brillouin zones for the crystal structures considered in this work are depicted in Fig.~\ref{BZs} with the high-symmetry points labelled. Both the electronic and phonon band structure plots take paths through these high-symmetry points.

All DFT calculations were performed using the Vienna \textit{Ab initio} Simulation Package (VASP)~\cite{Kresse1993a,Kresse1994a,Kresse1996a,Kresse1996a} with projector augmented wave (PAW) pseudopotentials~\cite{Bloechl1994a,Kresse1999a} using the Perdew-Becke-Ernzerhof (PBE) exchange-correlation functional~\cite{Perdew1996a}.  In the PAW scheme, we treated \textit{s} and \textit{p} electrons as valence for Li, C, N, O, F, Na, Al, Si, S, Cl, Ca, I, Cs and W, \textit{p} electrons as valence for Mg and \textit{d} electrons as valence for Zn, Ga, Ge, As, In and Sb. Below we summarize the convergence criteria used for the {\it (i)} electronic structure and wavefunctions, and {\it (ii)} phonon calculations. 

\renewcommand\arraystretch{1.15}
\begin{table*}[t]
\begin{tabular}{lcccccc}
\toprule
 & \multicolumn{2}{c}{Lattice Parameters (\AA)} & Space group  &  & Structure & $E_g$ \\
Material & a (el., ph., exp.) & c (el., ph. exp.) & (number) & Calc. $E_g$ (eV) & Ref. & Ref. \\
\hline
\ce{Al2O3}  & 4.808, 4.805, 4.759 & 13.121, 13.116, 12.991 & R$\overline{3}c$ (167) & 5.84 & ~\cite{Wang1994a} & ~\cite{French1990a} \\
AlN  & 3.130, 3.128, 3.111 & 5.020, 5.016, 4.978 & P$6_{3}mc$ (186) & 4.02 & ~\cite{Zemann1965a} & ~\cite{Yim1973a} \\
\ce{CaF2}  & 5.507, 5.499, 5.463 & -- & F$m\overline{3}m$ (225) & 11.81 & ~\cite{Speziale2002a} & ~\cite{Rubloff1972a} \\
\ce{CaWO4}  & 5.317, 5.320, 5.243 & 11.534, 11.444, 11.376 & I$4_{1}/a$  (88) & 4.04 & ~\cite{Senyshyn2004a} & ~\cite{Mikhailik2004a} \\
CsI & 4.671, 4.669, 4.567 & -- & P$m\overline{3}m$ (221) & 3.67 & ~\cite{Zemann1965a} & ~\cite{Lipp2006a} \\
Diamond & 3.572, 3.572, 3.567 & -- & F$d\overline{3}m$ (227) & 4.12 & ~\cite{Zemann1965a} & ~\cite{Clark1964a} \\ 
GaAs & 5.751, 5.756, 5.653 & -- & F$\overline{4}3m$ (216) & 0.141 & ~\cite{Zemann1965a} & ~\cite{Levinshtein1996a} \\
GaN & 3.129, 3.247, 3.189 & 5.246, 5.280, 5.186 & P$6_{3}mc$ (186) & 1.71 & ~\cite{Paszkowicz2004a} & ~\cite{Teisseyre1994a} \\
GaSb  & 6.217, 6.223, 6.118 & -- & F$\overline{4}3m$ (216) & 0.47 & ~\cite{Zemann1965a} & ~\cite{Adachi1989a} \\
Ge & 5.763, 5.782, 5.657 & -- & F$d\overline{3}m$ (227) & 0.37 & ~\cite{Zemann1965a} & ~\cite{Sze1968a} \\
InSb  & 6.635, 6.634, 6.478 & -- & F$\overline{4}3m$ (216) & 0.06 & ~\cite{Zemann1965a} & ~\cite{Littler1985a} \\
LiF  & 4.063, 4.065, 4.020 & -- & F$m\overline{3}m$ (225) & 8.85 & ~\cite{Ullrich1992a} & ~\cite{Singh1984a} \\
\ce{MgF2}  & 4.702, 4.684, 4.623 & 3.097, 3.081, 3.052 & P$4_{2}/mnm$ (136) & 6.79 & ~\cite{Zemann1965a} & ~\cite{Thomas1973a} \\
MgO  & 4.258, 4.250, 4.211 & -- & F$m\overline{3}m$ (225) & 4.43 & ~\cite{MgO} & ~\cite{Whited1973a} \\
NaCl  & 5.670,  5.696, 5.641 & -- & F$m\overline{3}m$ (225) & 5.05 & ~\cite{Zemann1965a} & ~\cite{Brown1970a} \\
NaF  & 4.682, 4.619, 4.634 & -- & F$m\overline{3}m$ (225) & 6.14 & ~\cite{Deshpande1961a} & ~\cite{Roy1985a} \\
NaI  & 6.498, 6.530, 6.473 & -- & F$m\overline{3}m$ (225) & 3.61 &  ~\cite{Zemann1965a} & ~\cite{Rodnyi1997a} \\
PbS  & --, 5.994, 5.936 & -- & F$m\overline{3}m$ (225) & -- \tnote{*} & ~\cite{Zemann1965a} & ~\cite{Strehlow1973a} \\
PbSe  & --, 6.206, 6.124 & -- & F$m\overline{3}m$ (225) & -- \tnote{*} & ~\cite{Zemann1965a} & ~\cite{Strehlow1973a} \\
PbTe  & --, 6.561, 6.454 & -- & F$m\overline{3}m$ (225) & -- \tnote{*} & ~\cite{Zemann1965a} & ~\cite{Strehlow1973a} \\
Si & 5.469, 5.469, 5.431 & -- & F$d\overline{3}m$ (227) & 0.75 & ~\cite{Zemann1965a} & ~\cite{Ahrenkiel2007a} \\ 
\ce{SiO2}  & 5.038, 5.016, 4.913 & 5.526, 5.507, 5.405 & P$3_{2}21$ (154) & 5.66 & ~\cite{Page1976a} & ~\cite{Sato2015a} \\
ZnO  & 3.288, 3.287, 3.250 & 5.308, 5.304, 5.207 & P$6_{3}mc$ (186) & 0.72 & ~\cite{Zemann1965a} & ~\cite{Srikant1998a} \\
ZnS & 5.449, 5.443, 5.420 & -- & F$\overline{4}3m$ (216) & 2.01 & ~\cite{Jumpertz1955} & ~\cite{Khamala2016a} \\ \hline
\end{tabular}
\caption{List of material properties used in DFT calculations. The calculated lattice parameters ($a$ and $c$) are listed for both those used in the electronic (el.)\ and phonon (ph.)\ excitation calculations, along with reported experimental values (exp.). The space group and corresponding space group number are included for the crystal strucrures considered. The PBE-level calculated band gaps are also listed, with details explained in the text.}
\label{tab:mat_calc_details}
\end{table*}

\subsection{Calculation details for electronic band structures and wavefunctions}
\label{app:dft-electron}

For structural optimizations, a plane wave cutoff-energy of 950 eV was used with a $12\times12\times12$ $\Gamma$-centered k-point grid. The energy and force convergence criteria were $\SI{1E-8}{\electronvolt}$ and $\SI{1}{\milli\electronvolt\per\angstrom}$ respectively.

All-electron wavefunction coefficients were extracted from PAW calculations using a modification of the pawpyseed code~\cite{Bystrom2019a}. This enabled recovery of the full wavefunctions as normalized single-particle Kohn-Sham states from the pseudo-wavefunctions obtained by the PAW-method. Initial PAW wavefunctions were calculated with a plane wave energy-cutoff of 1000 eV, from which the all-electron wavefunctions were constructed with a minimum energy-cutoff of 450 eV. Calculations were performed using $\Gamma$-centered Monkhorst-Pack grids, with a k-point density of at least $\SI{0.27}{\per\angstrom}$.  Energy bands were included up to 60 eV above and below the valence band maximum. However, since there is no pseudopotential containing the low-lying 4\textit{d}-states  for indium in VASP, these bands are neglected from the calculations. In NaI and CsI the I 4\textit{d} states are positioned at approximately 43 eV and 42 eV below the valence band maxima respectively. A scissor operator was applied to match the experimental band gaps given in Table~\ref{table:one}. For Ge, InSb and GaSb, the PBE functional gave partially occupied bands due to underestimation of the band gap. In these cases the HSE06 hybrid functional~\cite{Krukau2006a} was applied in a static calculation to introduce a band gap before applying the scissor correction. Electronic band structures were computed on a discrete k-mesh along the high-symmetry directions. 

PbS, PbSe and PbTe were excluded from the electron calculations because spin-orbit interactions are required to capture important features of the band structures and spin-orbit coupling is not yet implemented within the pawpyseed code.

Multiple k-point densities, energy-cutoffs, and energy bands included were tested for all materials to ensure convergence of the scattering rates to less than $2 \%$ at $m_\chi = 10$ GeV, less than $3 \%$ at $m_\chi = 10$ MeV, and less than $28 \%, 18 \%, 18 \%$ and $10 \%$ for GaAs, GaSb, Ge, and Si at $m_\chi = 1$ MeV respectively. InSb was tested with a $12\times12\times12$ and $14\times14\times14$ k-point grid, plotted as dotted and solid curves in Fig.~\ref{fig:ultralight_sigmae} respectively. At $m_\chi = 1$ MeV the rate convergence is $5 \%$, and decreases for larger masses. However at smaller masses the $\vect{G} = \vect{0}$ contribution, from momentum transfers within the 1BZ, and energy depositions below $\sim 1$ eV, dominate the rate. The slow convergence here is due to the fact that InSb has rapidly changing band structure near the $\Gamma$ point, and more k-points are needed for better convergence. These uncertainties are plotted as shaded bands in Fig.~\ref{fig:ultralight_sigmae}, and accompanying figures in Appendix~\ref{app:additional_plots}, although most are invisible due to the plots being log-log.

\subsection{Calculation details for phonon spectra}
\label{app:dft-phonon}

We obtained the phonon dispersions from Phonopy~\cite{phonopy} using the `frozen-phonon' method by diagonalizing the force matrix using VASP as the force calculator. For the VASP calculations, the electronic wavefunctions were expanded in a plane-wave basis with a kinetic-energy cutoff of 600 eV. The Brillouin zone sampling was no less than $\SI{0.8}{\per\angstrom}$ in each direction of the unit cell with Monkhorst-Pack grids, and was correspondingly scaled for phonon supercell calculations. Born effective charges were calculated for polar materials using density-functional perturbation theory as implemented in VASP.

\subsection{Parameters in Table~\ref{table:one}}
\label{app:table_params_detail}

The experimental electronic band gaps, $E_g$, are taken from references cited in Table~\ref{tab:mat_calc_details}. The speed of sound, $\overline{c}_s^{\text{LA}}$, was calculated by averaging $\omega^{\text{LA}}/q$ over a uniform $20\times20$ grid on the surface of a sphere in reciprocal space with radius $q \approx 10$ eV centered at the $\Gamma$ point. The same averaging procedure was used in calculating the range of optical modes, $\overline{\omega}_O$, when a range exists. The average Born effective charge, $\overline{Z}^*$, is defined as $\text{Tr}\left[\mathbf{Z_+^*}\right]/3$, where $\mathbf{Z_+^*}$ is the Born effective charge of the positive ion(s) (the other charges can be found by requiring that the primitive cell is neutral). The average high frequency dielectric constant, $\overline{\varepsilon}_\infty$, is defined as $\text{Tr}\left[ \boldsymbol{\varepsilon}_\infty \right]/3$.

\section{Additional Target Comparison Plots}
\label{app:additional_plots}

In this Appendix, we provide plots for the remainder of the materials in Table~\ref{table:one} not presented in the main text. For concreteness, in all figures we take the local DM density to be $\rho_\chi = 0.4$ GeV$/$cm$^3$, and assume a Maxwell-Boltzmann distribution with velocity dispersion $v_0 = 230$ km$/$s, truncated at the escape velocity $v_\text{esc} = 600$ km$/$s, and boosted to the target rest frame by the Earth velocity in the galactic rest frame $v_E = 240$ km$/$s. We take the direction of the Earth's velocity to be in the $\hat{z}$ direction with respect to the crystal coordinates when computing the reach (for most of the target materials we consider here we expect modulation effects from the Earth's motion to be small). The constraints on $\overline{\sigma}$ correspond to a $95 \%$ confidence level (C.L.) assuming Poisson distributed counts and no events are seen (equivalently, the constraint corresponds to the cross section needed to obtain three events). We chose the 95\% C.L. for easier comparison with previous literature. Because it is also standard to compute the 90\% C.L. exclusion reach, we note that one simply has to mulitply the 95\% C.L. exclusion reach by $2.3/3$ (as the 90\% C.L. constraint corresponds to the cross section needed to obtain 2.3 events).  We also assume an exposure of one kg-yr.

\begin{figure*}[t!]
\centering
\includegraphics[width=\textwidth]{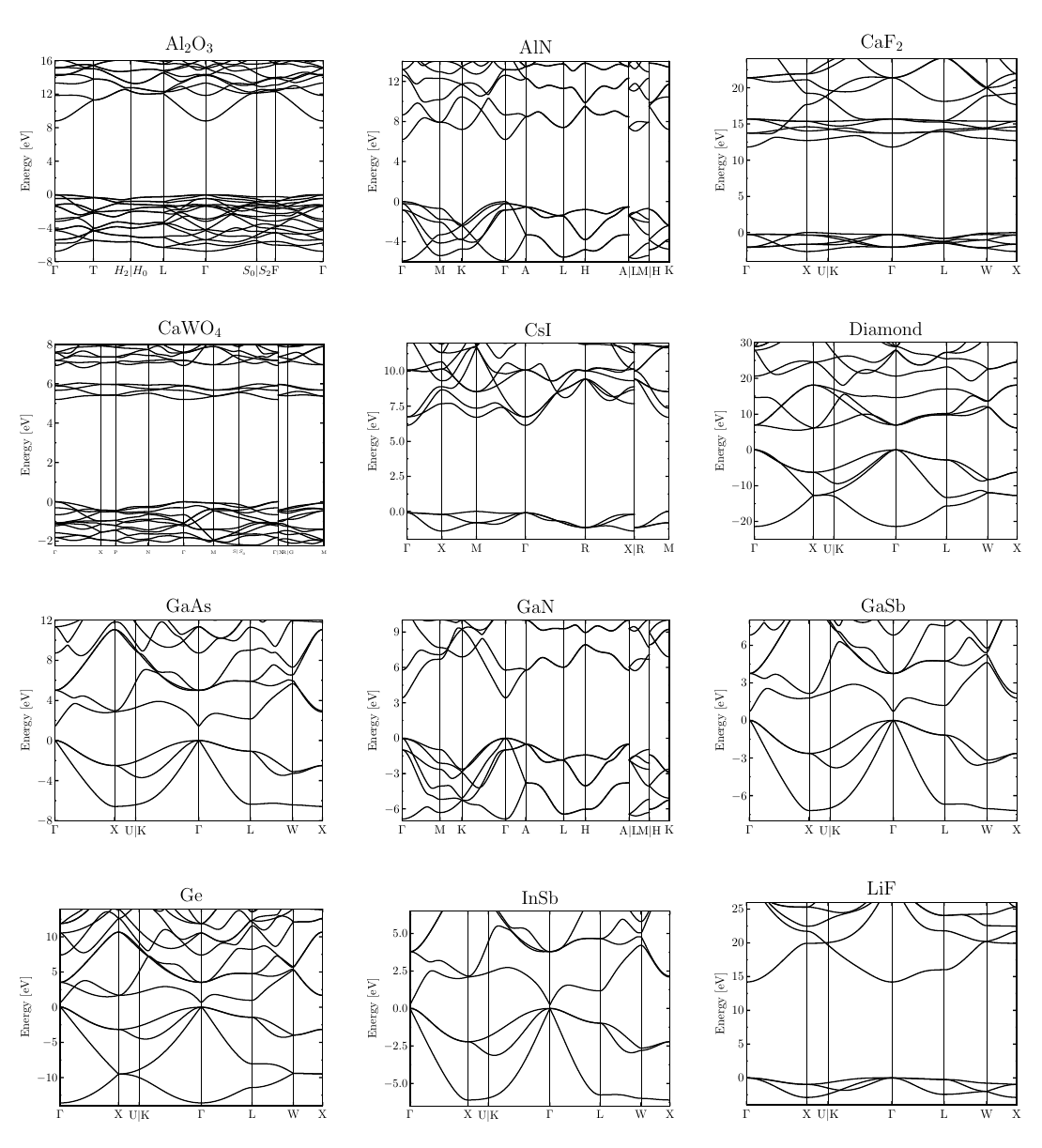}
\caption{Calculated electronic band structures of targets in Table~\ref{table:one}.}
\label{fig:electron_bands_1}
\end{figure*}

\begin{figure*}[t!] 
\centering
\includegraphics[width=\textwidth]{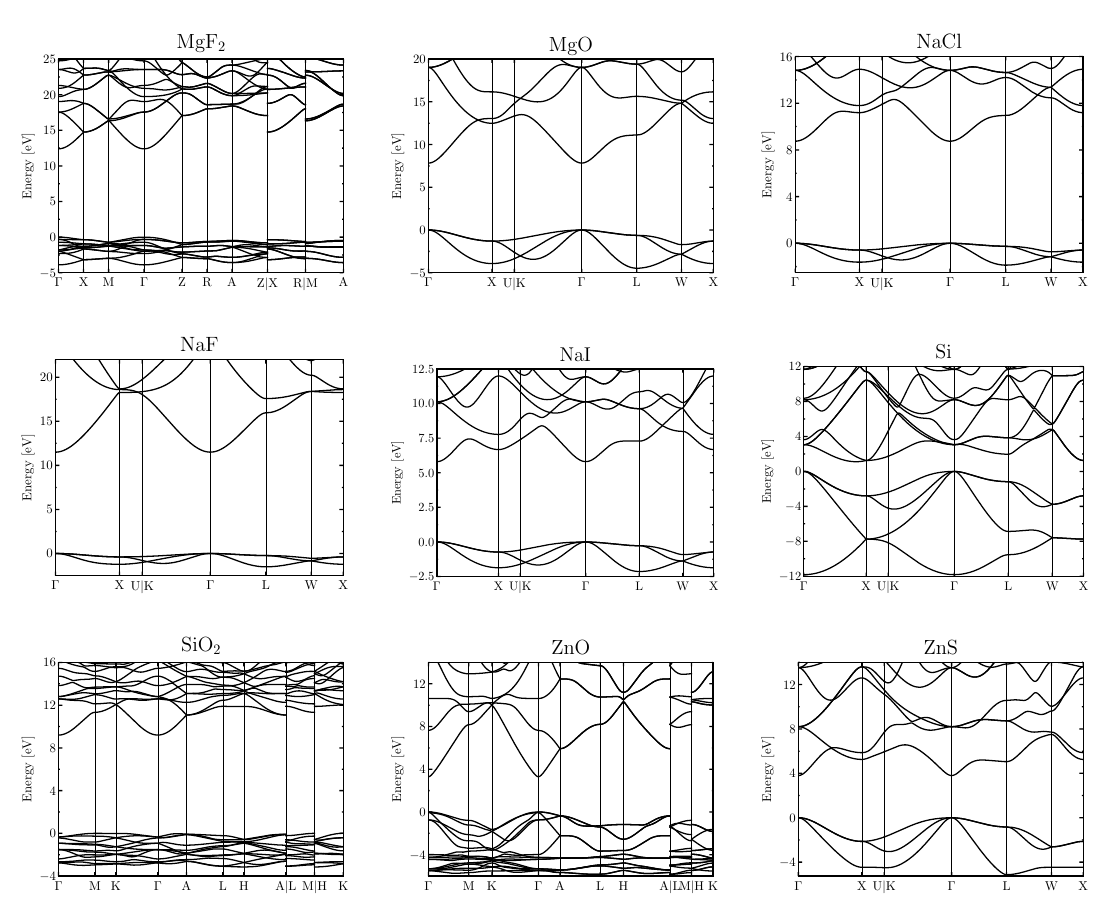}
\caption{Calculated electronic band structures of targets in Table~\ref{table:one}.}
\label{fig:electron_bands_2}
\end{figure*}

\begin{figure*}[t!] 
\centering
\includegraphics[width=\textwidth]{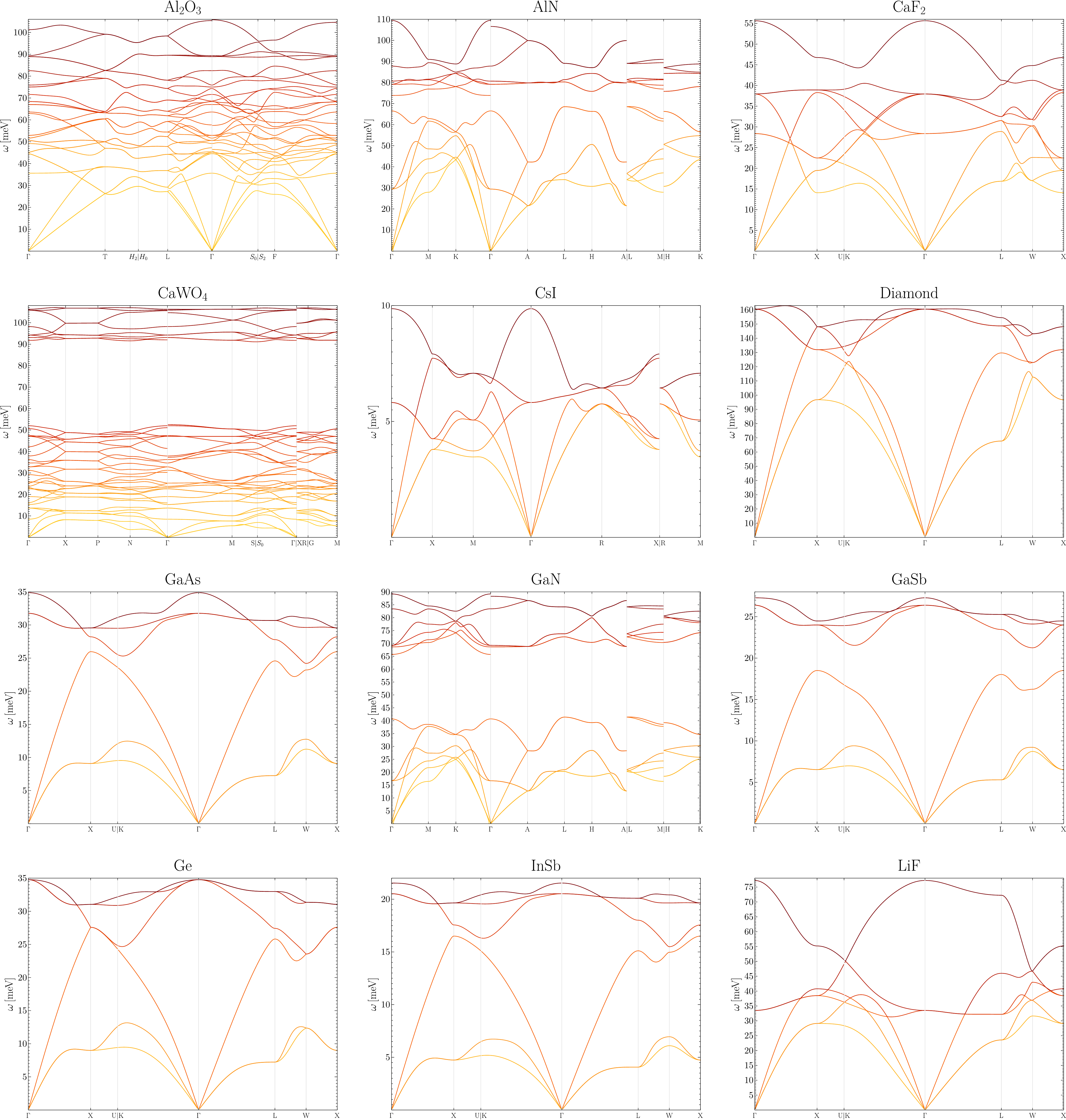}
\caption{Phonon dispersions calculated with VASP and phonopy~\cite{phonopy} including non-analytic corrections. The path through the high symmetry points is found using SeeK-path \cite{Hinuma:aa}.}
\label{fig:phonon_dispersion_1}
\end{figure*}

\begin{figure*}[t!] 
\centering
\includegraphics[width=\textwidth]{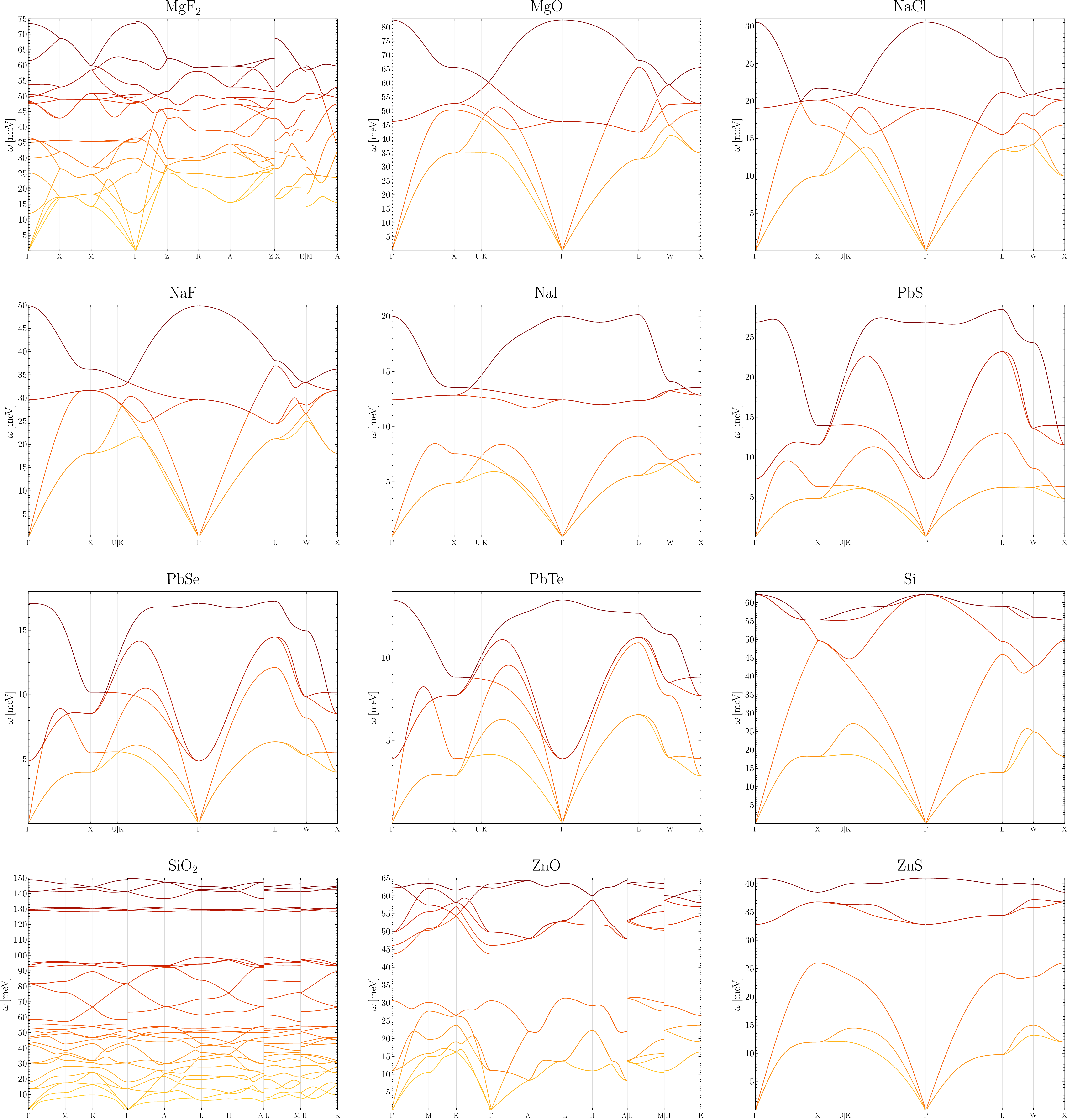}
\caption{Phonon dispersions calculated with VASP and phonopy~\cite{phonopy} including non-analytic corrections. The path through the high symmetry points is found using SeeK-path \cite{Hinuma:aa}.}
\label{fig:phonon_dispersion_2}
\end{figure*}

\begin{figure*}[t!]
\centering
\includegraphics[width=0.95\textwidth]{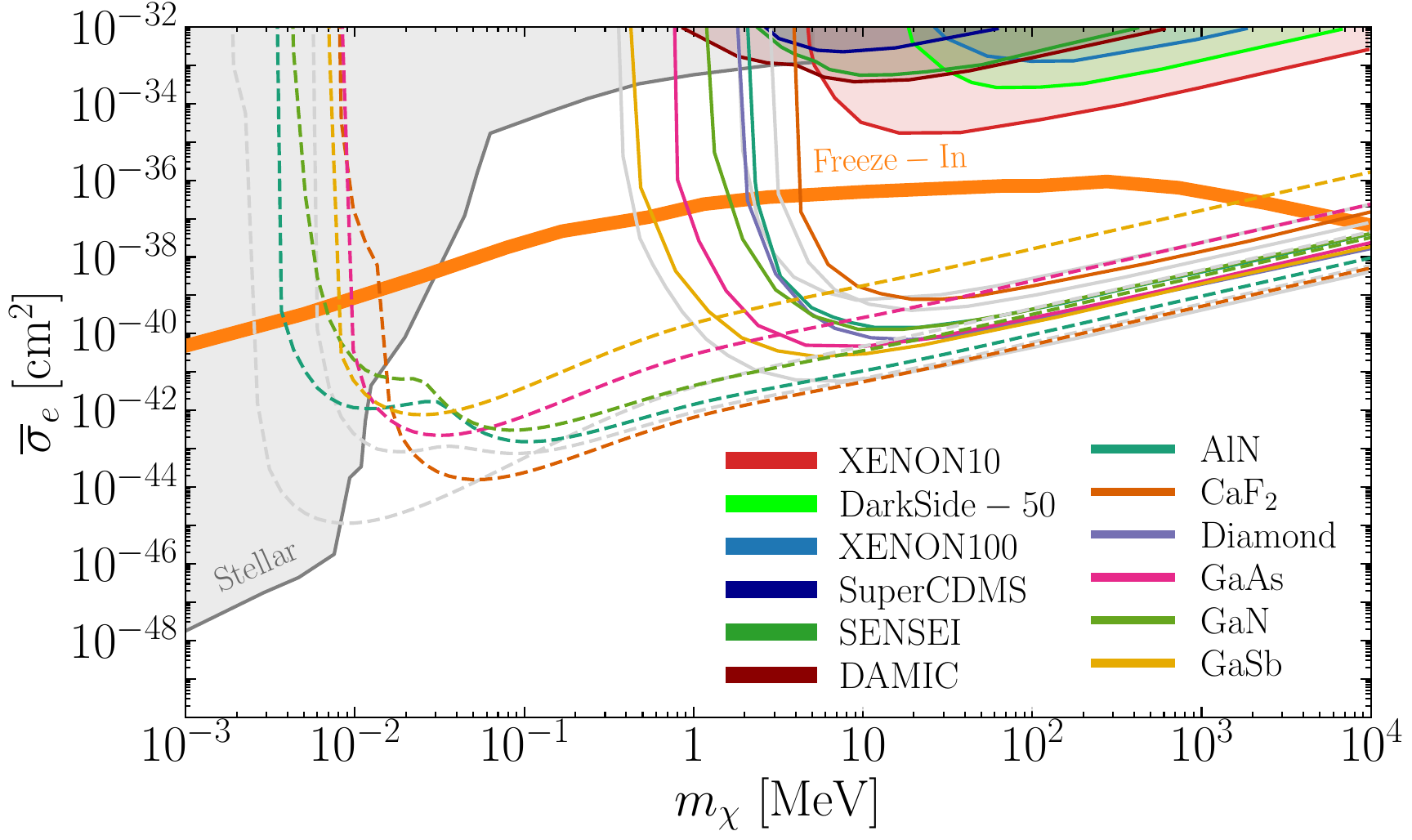}
\caption{Same as Fig.~\ref{fig:ultralight_sigmae}, but with different materials. For reference, gray lines are \ce{CsI}, \ce{Si}, and \ce{Al2O3} taken from Fig.~\ref{fig:ultralight_sigmae}.}
\label{fig:fig10}
\end{figure*}
\begin{figure*}[h!]
\centering
\includegraphics[width=0.95\textwidth]{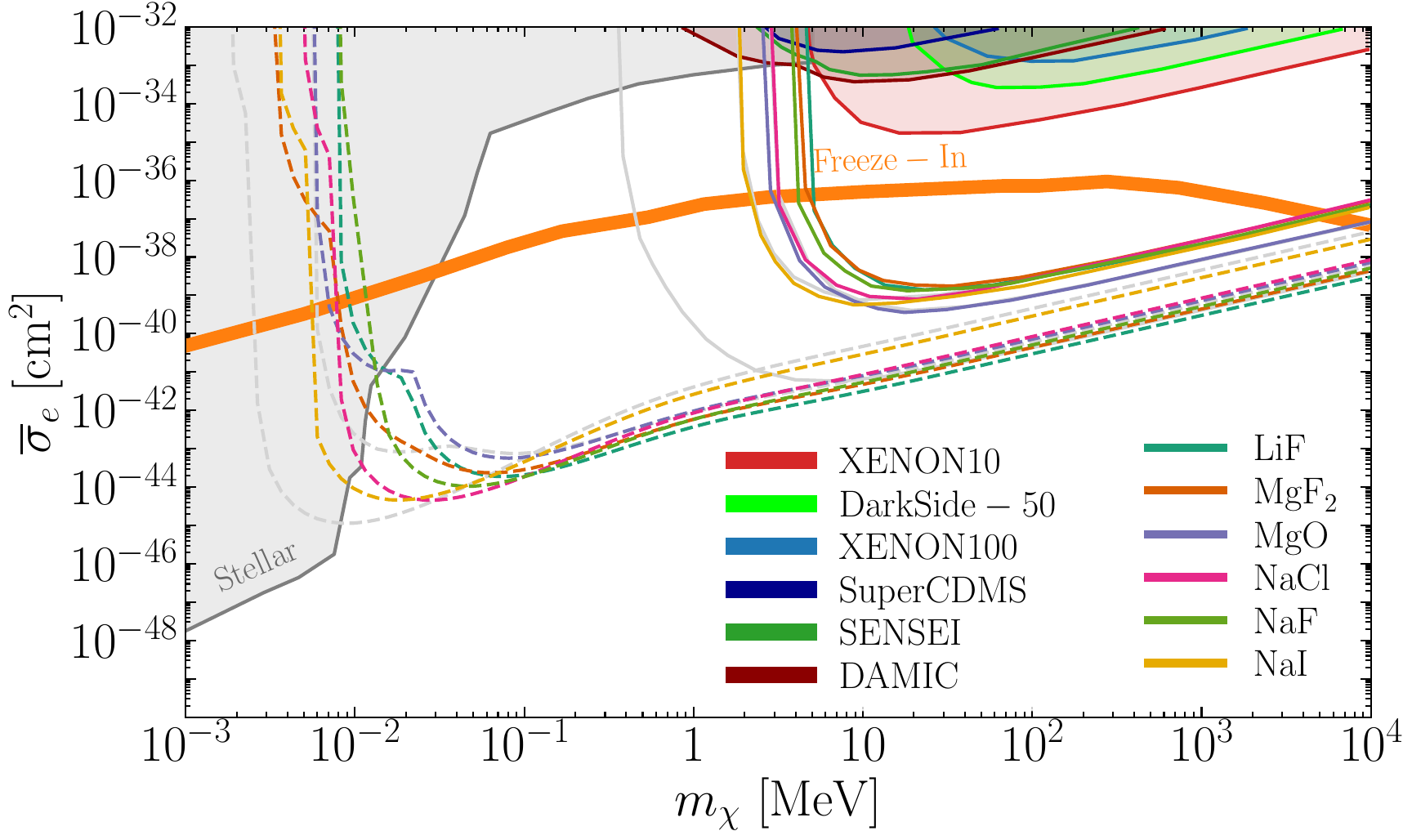}
\caption{Same as Fig.~\ref{fig:ultralight_sigmae}, but with different materials. For reference, gray lines are \ce{CsI}, \ce{Si}, and \ce{Al2O3} taken from Fig.~\ref{fig:ultralight_sigmae}.}
\label{fig:fig11}
\end{figure*}

\begin{figure*}[t!]
\centering
\includegraphics[width=0.95\textwidth]{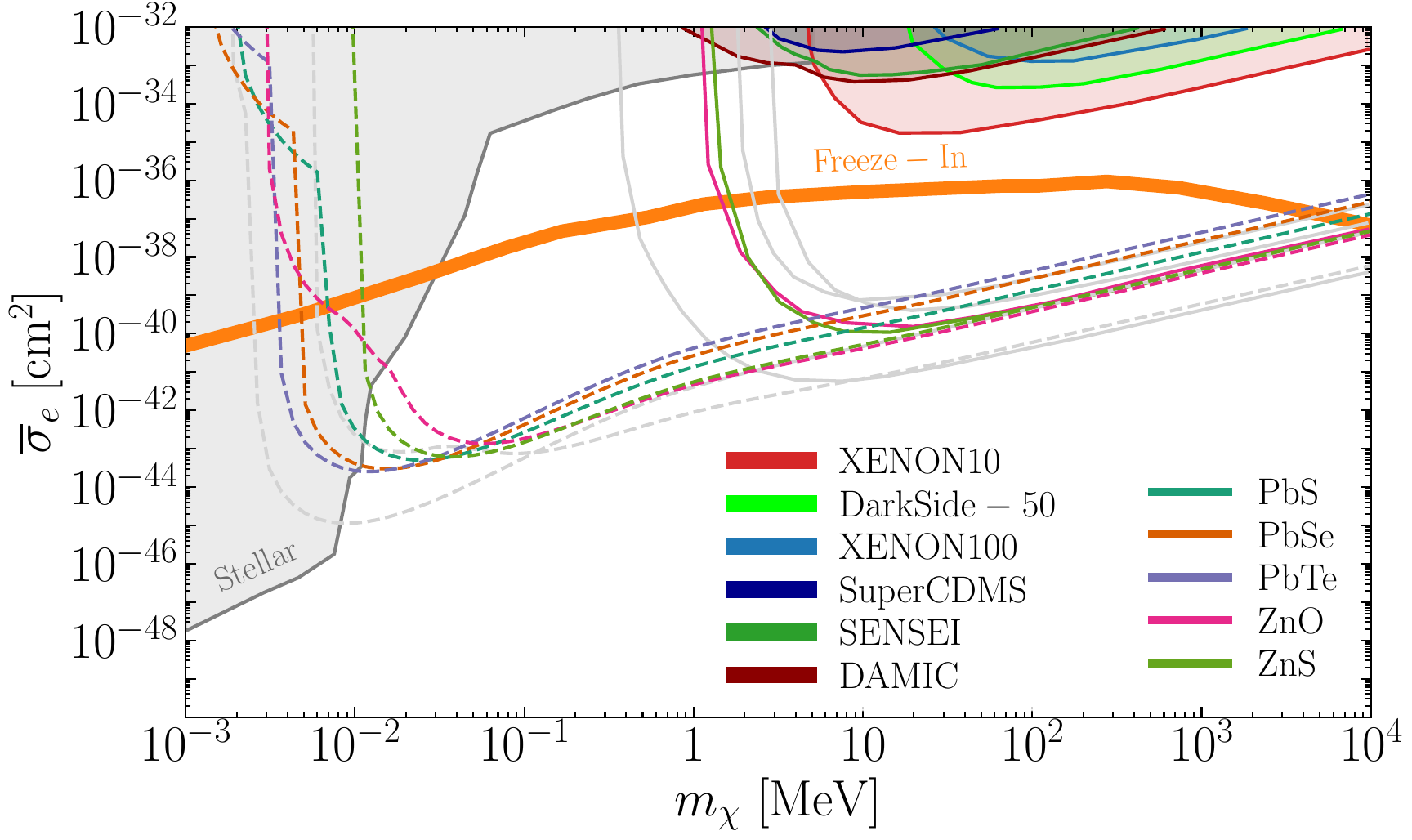}
\caption{Same as Fig.~\ref{fig:ultralight_sigmae}, but with different materials. For reference, gray lines are \ce{CsI}, \ce{Si}, and \ce{Al2O3} taken from Fig.~\ref{fig:ultralight_sigmae}.}
\label{fig:fig12}
\end{figure*}
\begin{figure*}[h!]
\centering
\includegraphics[width=0.95\textwidth]{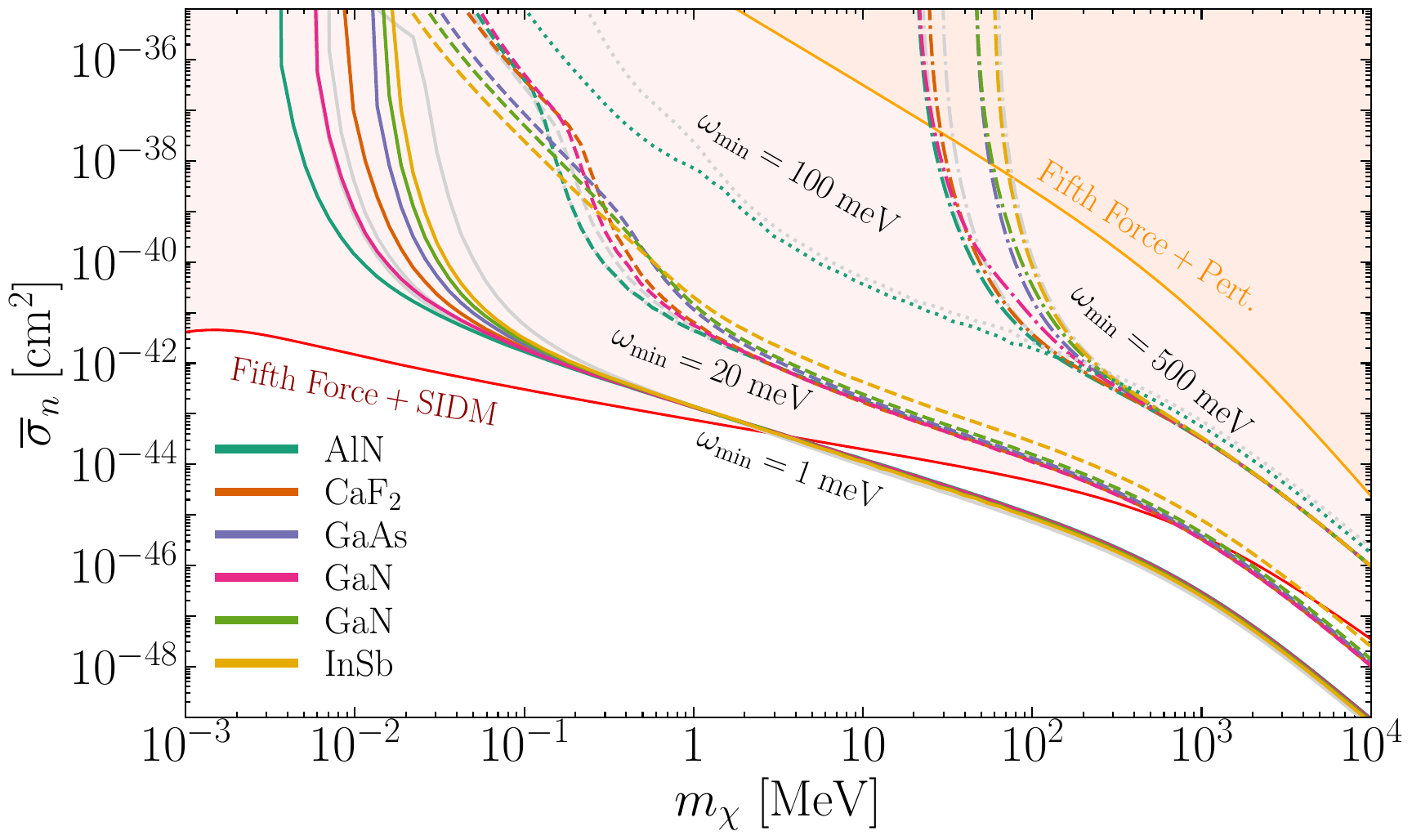}
\caption{Same as Fig.~\ref{fig:ultralight_sigman}, but with different materials. For reference, gray lines are \ce{CsI}, \ce{Si}, and \ce{Al2O3} taken from Fig.~\ref{fig:ultralight_sigman}.}
\label{fig:fig13}
\end{figure*}

\begin{figure*}[t!]
\centering
\includegraphics[width=0.95\textwidth]{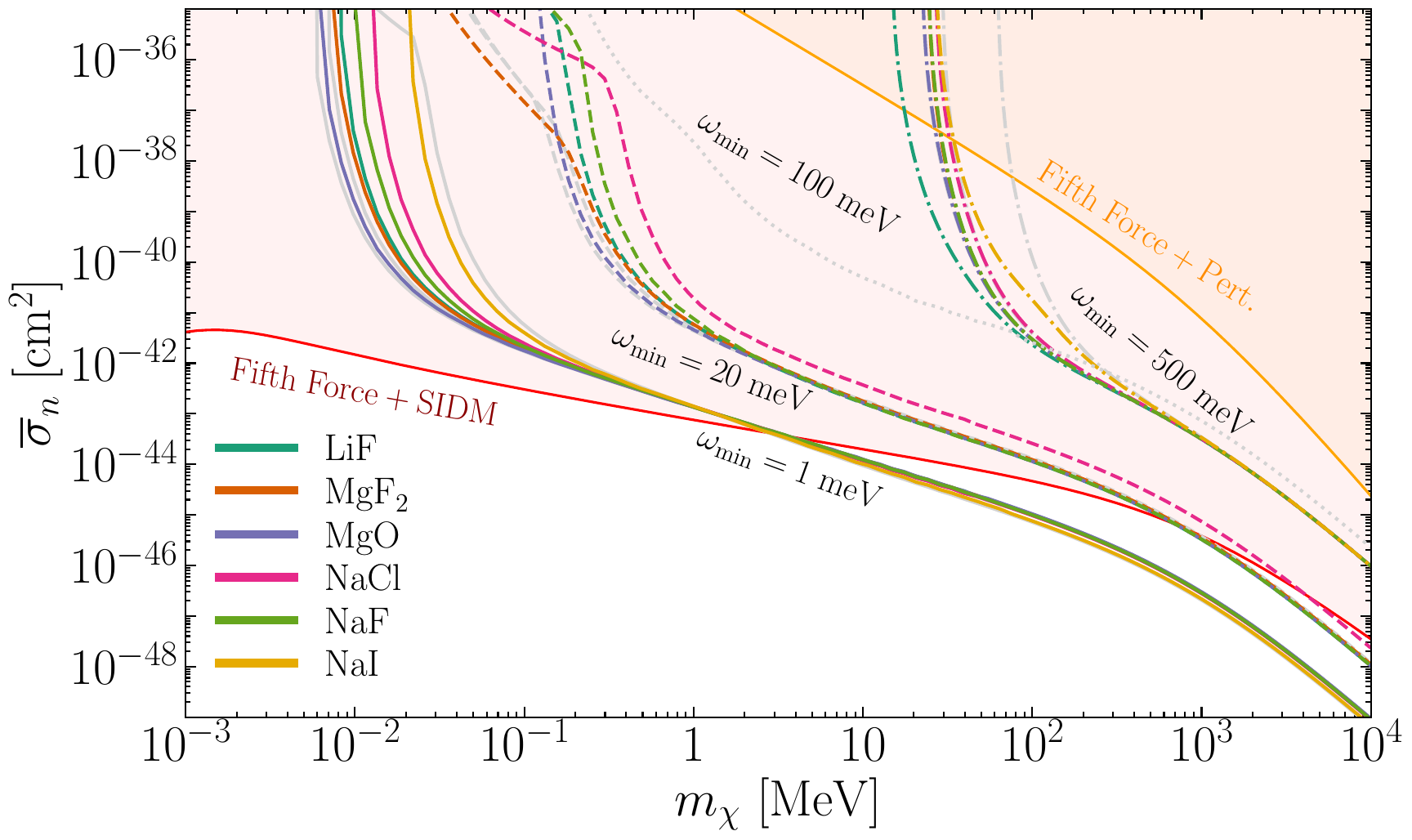}
\caption{Same as Fig.~\ref{fig:ultralight_sigman}, but with different materials. For reference, gray lines are \ce{CsI}, \ce{Si}, and \ce{Al2O3} taken from Fig.~\ref{fig:ultralight_sigman}.}
\label{fig:fig14}
\end{figure*}
\begin{figure*}[h!]
\centering
\includegraphics[width=0.95\textwidth]{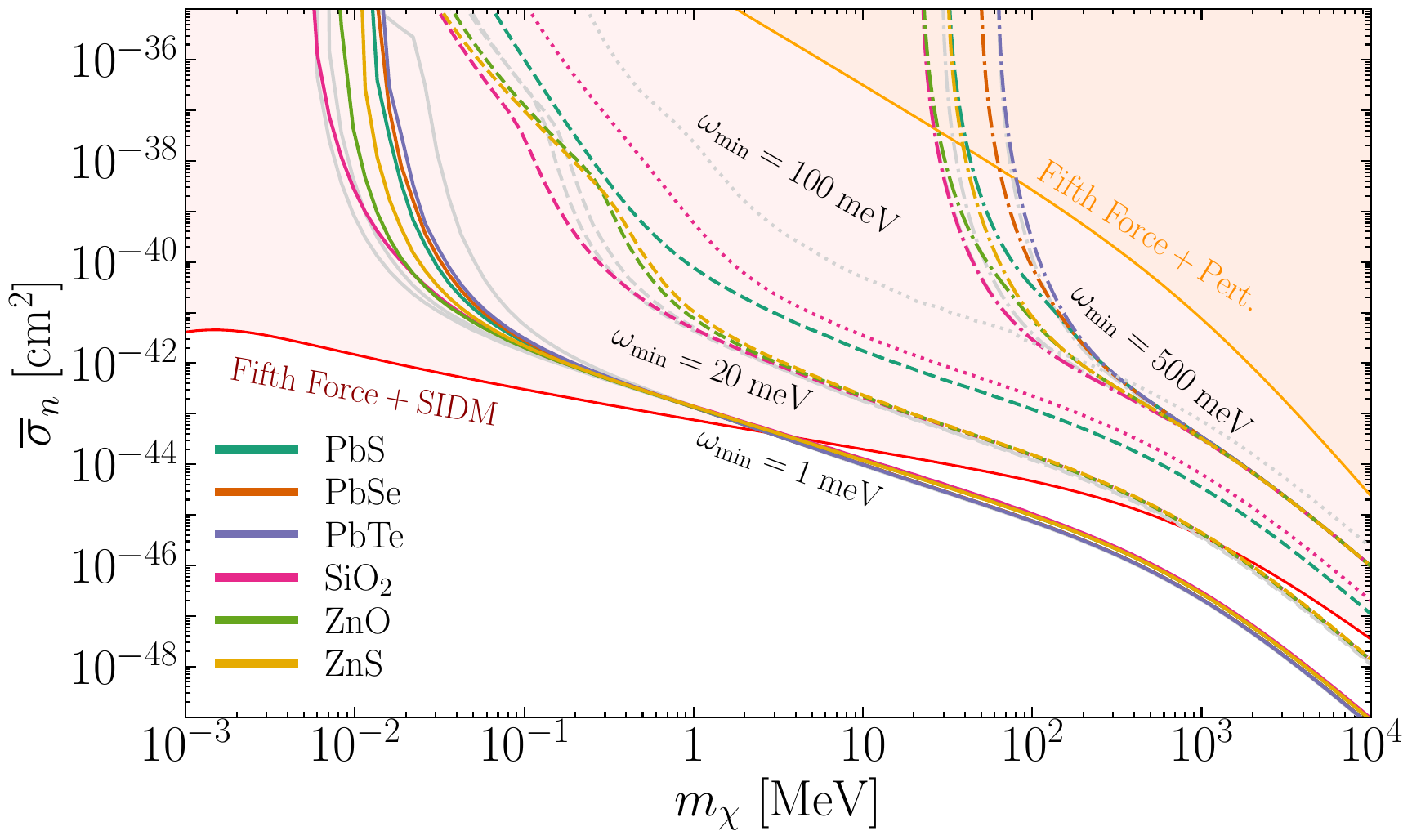}
\caption{Same as Fig.~\ref{fig:ultralight_sigman}, but with different materials. For reference, gray lines are \ce{CsI}, \ce{Si}, and \ce{Al2O3} taken from Fig.~\ref{fig:ultralight_sigman}.}
\label{fig:fig15}
\end{figure*}

\begin{figure*}[t!]
\centering
\includegraphics[width=0.95\textwidth]{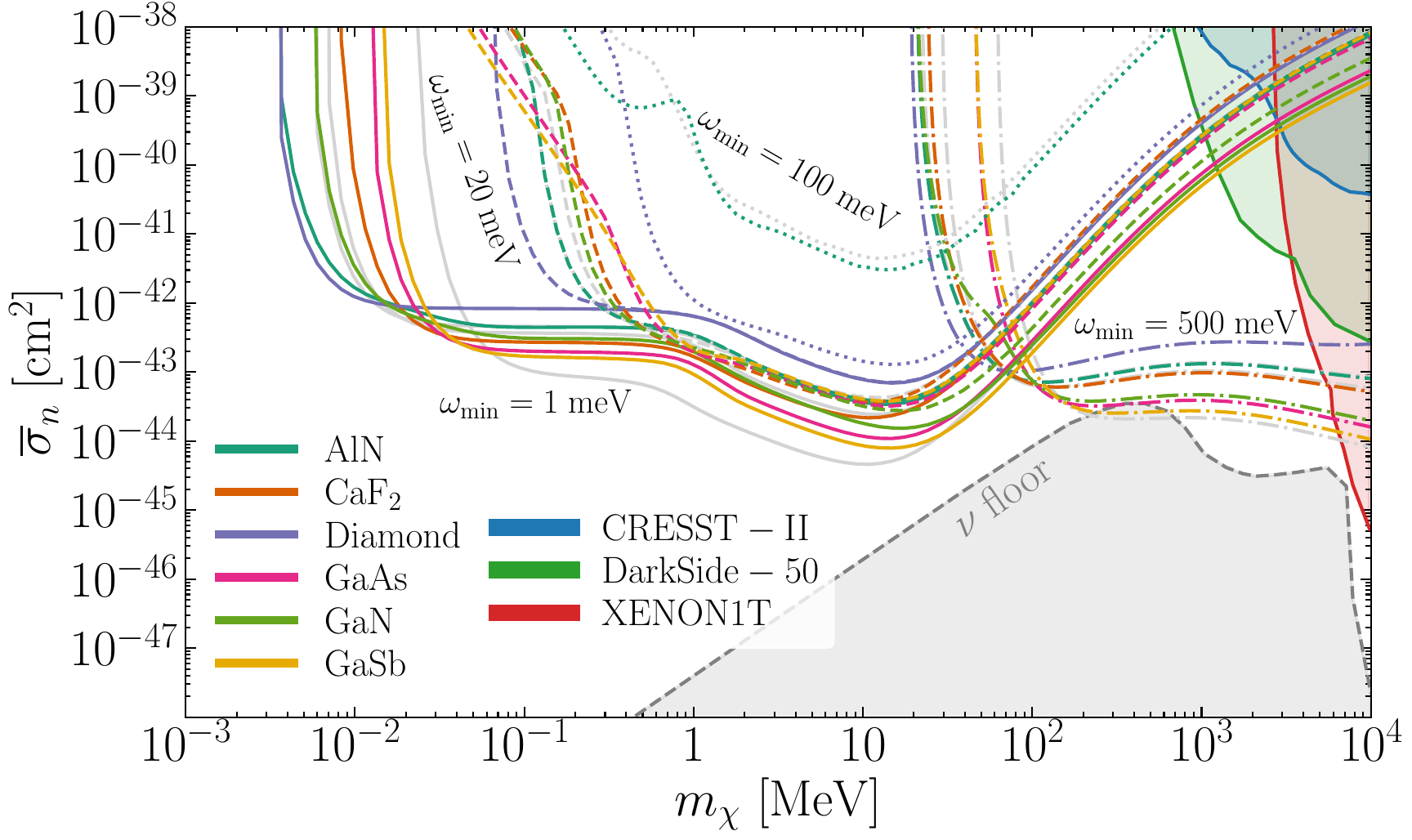}
\caption{Same as Fig.~\ref{fig:massive_sigman}, but with different materials. For reference, gray lines are \ce{CsI}, \ce{Si}, and \ce{Al2O3} taken from Fig.~\ref{fig:massive_sigman}.}
\label{fig:fig16}
\end{figure*}
\begin{figure*}[h!]
\centering
\includegraphics[width=0.95\textwidth]{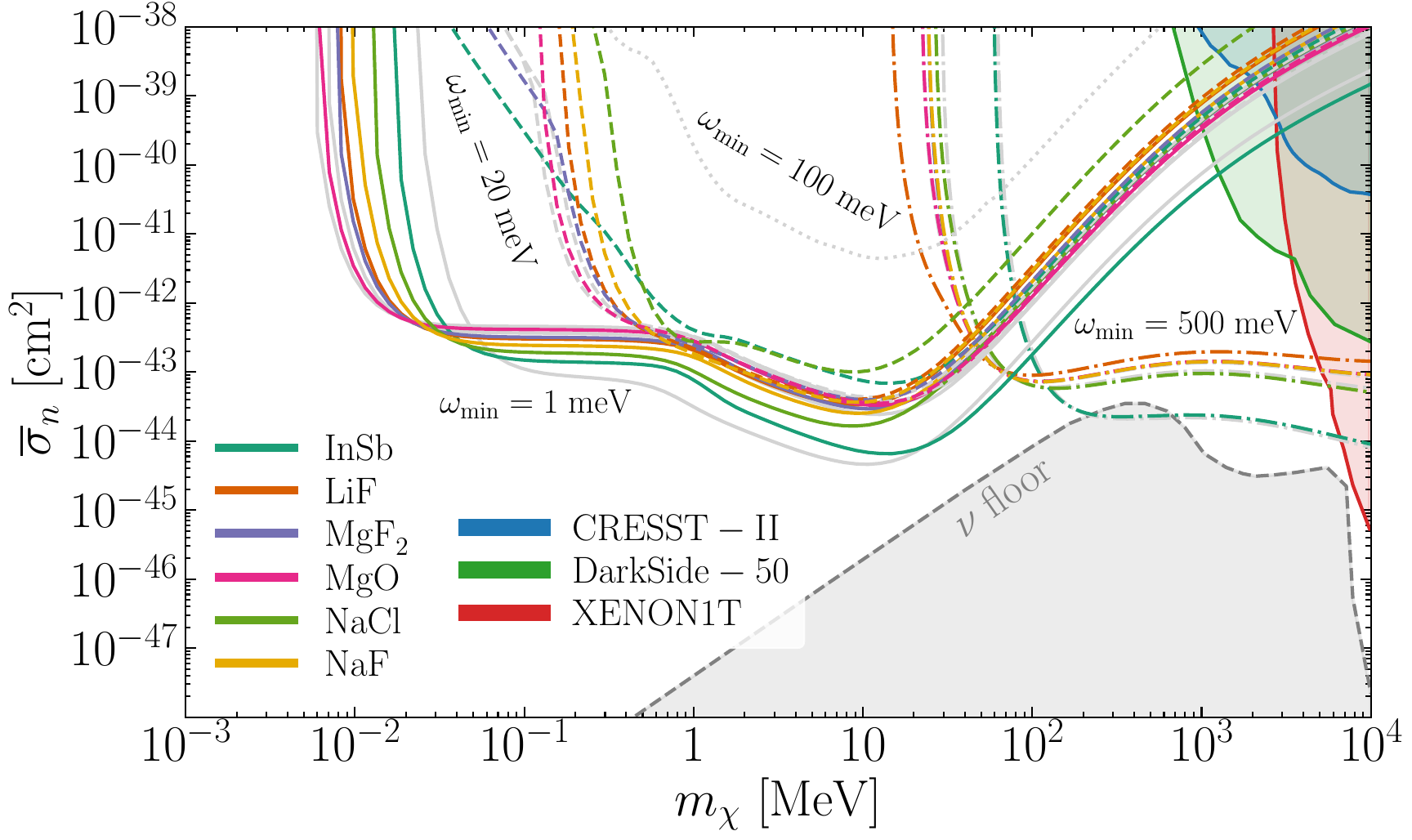}
\caption{Same as Fig.~\ref{fig:massive_sigman}, but with different materials. For reference, gray lines are \ce{CsI}, \ce{Si}, and \ce{Al2O3} taken from Fig.~\ref{fig:massive_sigman}.}
\label{fig:fig17}
\end{figure*}

\begin{figure*}[t!]
\centering
\includegraphics[width=0.95\textwidth]{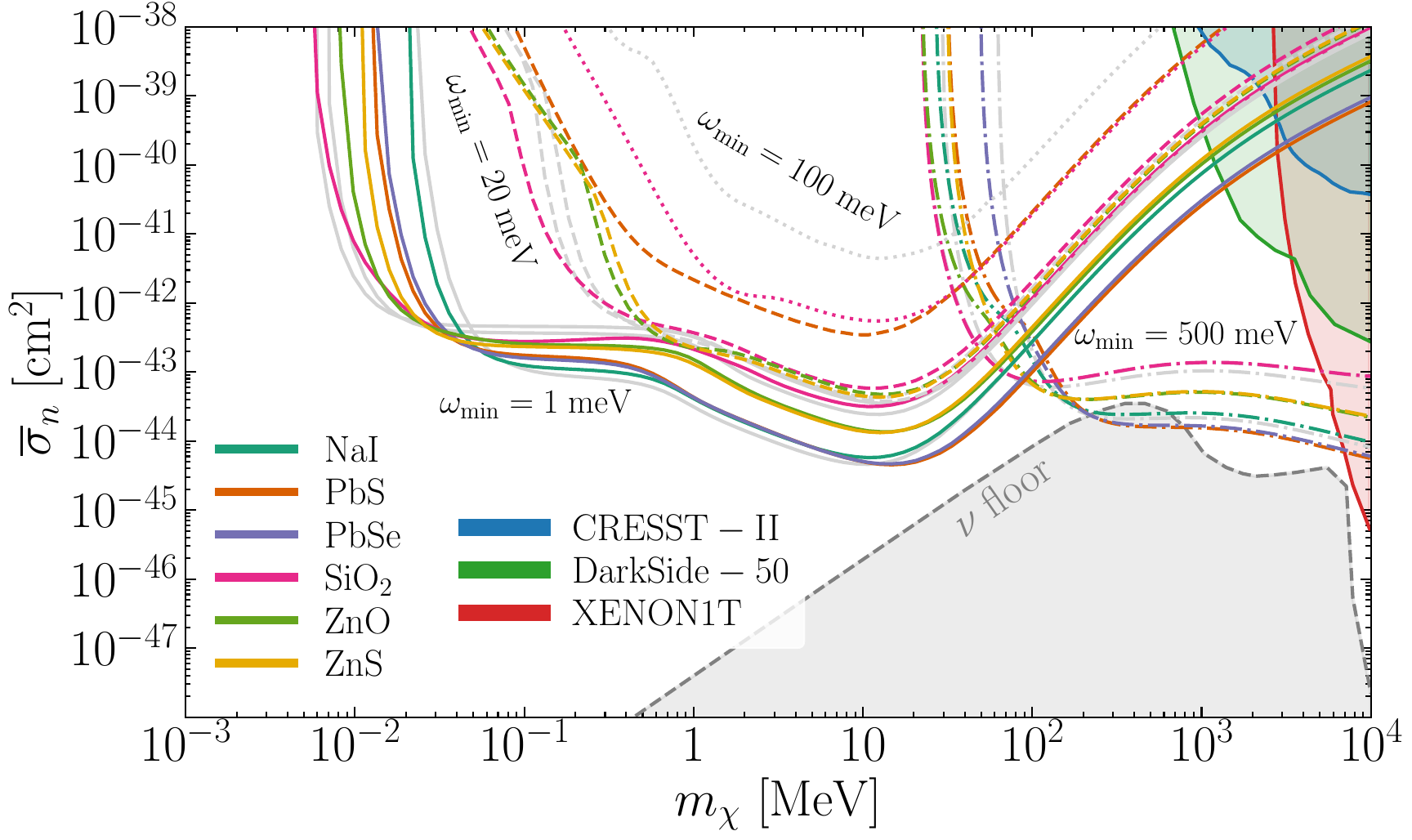}
\caption{Fig.~\ref{fig:massive_sigman} with different materials. Gray lines are \ce{CsI}, \ce{Si}, and \ce{Al2O3} taken from Fig.~\ref{fig:massive_sigman}.}
\label{fig:fig18}
\end{figure*}

\bibliographystyle{apsrev4-1}
\bibliography{refs_target_comparison}

\begin{thebibliography}{136}%
\makeatletter
\providecommand \@ifxundefined [1]{%
 \@ifx{#1\undefined}
}%
\providecommand \@ifnum [1]{%
 \ifnum #1\expandafter \@firstoftwo
 \else \expandafter \@secondoftwo
 \fi
}%
\providecommand \@ifx [1]{%
 \ifx #1\expandafter \@firstoftwo
 \else \expandafter \@secondoftwo
 \fi
}%
\providecommand \natexlab [1]{#1}%
\providecommand \enquote  [1]{``#1''}%
\providecommand \bibnamefont  [1]{#1}%
\providecommand \bibfnamefont [1]{#1}%
\providecommand \citenamefont [1]{#1}%
\providecommand \href@noop [0]{\@secondoftwo}%
\providecommand \href [0]{\begingroup \@sanitize@url \@href}%
\providecommand \@href[1]{\@@startlink{#1}\@@href}%
\providecommand \@@href[1]{\endgroup#1\@@endlink}%
\providecommand \@sanitize@url [0]{\catcode `\\12\catcode `\$12\catcode
  `\&12\catcode `\#12\catcode `\^12\catcode `\_12\catcode `\%12\relax}%
\providecommand \@@startlink[1]{}%
\providecommand \@@endlink[0]{}%
\providecommand \url  [0]{\begingroup\@sanitize@url \@url }%
\providecommand \@url [1]{\endgroup\@href {#1}{\urlprefix }}%
\providecommand \urlprefix  [0]{URL }%
\providecommand \Eprint [0]{\href }%
\providecommand \doibase [0]{http://dx.doi.org/}%
\providecommand \selectlanguage [0]{\@gobble}%
\providecommand \bibinfo  [0]{\@secondoftwo}%
\providecommand \bibfield  [0]{\@secondoftwo}%
\providecommand \translation [1]{[#1]}%
\providecommand \BibitemOpen [0]{}%
\providecommand \bibitemStop [0]{}%
\providecommand \bibitemNoStop [0]{.\EOS\space}%
\providecommand \EOS [0]{\spacefactor3000\relax}%
\providecommand \BibitemShut  [1]{\csname bibitem#1\endcsname}%
\let\auto@bib@innerbib\@empty
\bibitem [{\citenamefont {Boehm}\ and\ \citenamefont
  {Fayet}(2004)}]{Boehm2004a}%
  \BibitemOpen
  \bibfield  {author} {\bibinfo {author} {\bibfnamefont {C.}~\bibnamefont
  {Boehm}}\ and\ \bibinfo {author} {\bibfnamefont {P.}~\bibnamefont {Fayet}},\
  }\href {\doibase 10.1016/j.nuclphysb.2004.01.015} {\bibfield  {journal}
  {\bibinfo  {journal} {Nucl. Phys.}\ }\textbf {\bibinfo {volume} {B683}},\
  \bibinfo {pages} {219} (\bibinfo {year} {2004})},\ \Eprint
  {http://arxiv.org/abs/hep-ph/0305261} {arXiv:hep-ph/0305261 [hep-ph]}
  \BibitemShut {NoStop}%
\bibitem [{\citenamefont {Pospelov}\ \emph {et~al.}(2008)\citenamefont
  {Pospelov}, \citenamefont {Ritz},\ and\ \citenamefont
  {Voloshin}}]{Pospelov:2007mp}%
  \BibitemOpen
  \bibfield  {author} {\bibinfo {author} {\bibfnamefont {M.}~\bibnamefont
  {Pospelov}}, \bibinfo {author} {\bibfnamefont {A.}~\bibnamefont {Ritz}}, \
  and\ \bibinfo {author} {\bibfnamefont {M.~B.}\ \bibnamefont {Voloshin}},\
  }\href {\doibase 10.1016/j.physletb.2008.02.052} {\bibfield  {journal}
  {\bibinfo  {journal} {Phys. Lett.}\ }\textbf {\bibinfo {volume} {B662}},\
  \bibinfo {pages} {53} (\bibinfo {year} {2008})},\ \Eprint
  {http://arxiv.org/abs/0711.4866} {arXiv:0711.4866 [hep-ph]} \BibitemShut
  {NoStop}%
\bibitem [{\citenamefont {Hooper}\ and\ \citenamefont
  {Zurek}(2008)}]{Hooper:2008im}%
  \BibitemOpen
  \bibfield  {author} {\bibinfo {author} {\bibfnamefont {D.}~\bibnamefont
  {Hooper}}\ and\ \bibinfo {author} {\bibfnamefont {K.~M.}\ \bibnamefont
  {Zurek}},\ }\href {\doibase 10.1103/PhysRevD.77.087302} {\bibfield  {journal}
  {\bibinfo  {journal} {Phys. Rev.}\ }\textbf {\bibinfo {volume} {D77}},\
  \bibinfo {pages} {087302} (\bibinfo {year} {2008})},\ \Eprint
  {http://arxiv.org/abs/0801.3686} {arXiv:0801.3686 [hep-ph]} \BibitemShut
  {NoStop}%
\bibitem [{\citenamefont {Kumar}\ and\ \citenamefont
  {Feng}(2010)}]{Kumar:2009bw}%
  \BibitemOpen
  \bibfield  {author} {\bibinfo {author} {\bibfnamefont {J.}~\bibnamefont
  {Kumar}}\ and\ \bibinfo {author} {\bibfnamefont {J.~L.}\ \bibnamefont
  {Feng}},\ }\bibfield  {booktitle} {\emph {\bibinfo {booktitle} {{Proceedings,
  7th International Conference on Supersymmetry and the Unification of
  Fundamental Interactions (SUSY09): Boston, USA, June 5-10, 2009}}},\ }\href
  {\doibase 10.1063/1.3327538} {\bibfield  {journal} {\bibinfo  {journal} {AIP
  Conf. Proc.}\ }\textbf {\bibinfo {volume} {1200}},\ \bibinfo {pages} {1059}
  (\bibinfo {year} {2010})},\ \Eprint {http://arxiv.org/abs/0909.2877}
  {arXiv:0909.2877 [hep-ph]} \BibitemShut {NoStop}%
\bibitem [{\citenamefont {Arkani-Hamed}\ and\ \citenamefont
  {Weiner}(2008)}]{ArkaniHamed:2008qp}%
  \BibitemOpen
  \bibfield  {author} {\bibinfo {author} {\bibfnamefont {N.}~\bibnamefont
  {Arkani-Hamed}}\ and\ \bibinfo {author} {\bibfnamefont {N.}~\bibnamefont
  {Weiner}},\ }\href {\doibase 10.1088/1126-6708/2008/12/104} {\bibfield
  {journal} {\bibinfo  {journal} {JHEP}\ }\textbf {\bibinfo {volume} {12}},\
  \bibinfo {pages} {104} (\bibinfo {year} {2008})},\ \Eprint
  {http://arxiv.org/abs/0810.0714} {arXiv:0810.0714 [hep-ph]} \BibitemShut
  {NoStop}%
\bibitem [{\citenamefont {Cheung}\ \emph {et~al.}(2009)\citenamefont {Cheung},
  \citenamefont {Ruderman}, \citenamefont {Wang},\ and\ \citenamefont
  {Yavin}}]{Cheung:2009qd}%
  \BibitemOpen
  \bibfield  {author} {\bibinfo {author} {\bibfnamefont {C.}~\bibnamefont
  {Cheung}}, \bibinfo {author} {\bibfnamefont {J.~T.}\ \bibnamefont
  {Ruderman}}, \bibinfo {author} {\bibfnamefont {L.-T.}\ \bibnamefont {Wang}},
  \ and\ \bibinfo {author} {\bibfnamefont {I.}~\bibnamefont {Yavin}},\ }\href
  {\doibase 10.1103/PhysRevD.80.035008} {\bibfield  {journal} {\bibinfo
  {journal} {Phys. Rev.}\ }\textbf {\bibinfo {volume} {D80}},\ \bibinfo {pages}
  {035008} (\bibinfo {year} {2009})},\ \Eprint {http://arxiv.org/abs/0902.3246}
  {arXiv:0902.3246 [hep-ph]} \BibitemShut {NoStop}%
\bibitem [{\citenamefont {Morrissey}\ \emph {et~al.}(2009)\citenamefont
  {Morrissey}, \citenamefont {Poland},\ and\ \citenamefont
  {Zurek}}]{Morrissey:2009ur}%
  \BibitemOpen
  \bibfield  {author} {\bibinfo {author} {\bibfnamefont {D.~E.}\ \bibnamefont
  {Morrissey}}, \bibinfo {author} {\bibfnamefont {D.}~\bibnamefont {Poland}}, \
  and\ \bibinfo {author} {\bibfnamefont {K.~M.}\ \bibnamefont {Zurek}},\ }\href
  {\doibase 10.1088/1126-6708/2009/07/050} {\bibfield  {journal} {\bibinfo
  {journal} {JHEP}\ }\textbf {\bibinfo {volume} {07}},\ \bibinfo {pages} {050}
  (\bibinfo {year} {2009})},\ \Eprint {http://arxiv.org/abs/0904.2567}
  {arXiv:0904.2567 [hep-ph]} \BibitemShut {NoStop}%
\bibitem [{\citenamefont {Kaplan}\ \emph {et~al.}(2009)\citenamefont {Kaplan},
  \citenamefont {Luty},\ and\ \citenamefont {Zurek}}]{Kaplan:2009ag}%
  \BibitemOpen
  \bibfield  {author} {\bibinfo {author} {\bibfnamefont {D.~E.}\ \bibnamefont
  {Kaplan}}, \bibinfo {author} {\bibfnamefont {M.~A.}\ \bibnamefont {Luty}}, \
  and\ \bibinfo {author} {\bibfnamefont {K.~M.}\ \bibnamefont {Zurek}},\ }\href
  {\doibase 10.1103/PhysRevD.79.115016} {\bibfield  {journal} {\bibinfo
  {journal} {Phys. Rev.}\ }\textbf {\bibinfo {volume} {D79}},\ \bibinfo {pages}
  {115016} (\bibinfo {year} {2009})},\ \Eprint {http://arxiv.org/abs/0901.4117}
  {arXiv:0901.4117 [hep-ph]} \BibitemShut {NoStop}%
\bibitem [{\citenamefont {Cohen}\ \emph {et~al.}(2010)\citenamefont {Cohen},
  \citenamefont {Phalen}, \citenamefont {Pierce},\ and\ \citenamefont
  {Zurek}}]{Cohen:2010kn}%
  \BibitemOpen
  \bibfield  {author} {\bibinfo {author} {\bibfnamefont {T.}~\bibnamefont
  {Cohen}}, \bibinfo {author} {\bibfnamefont {D.~J.}\ \bibnamefont {Phalen}},
  \bibinfo {author} {\bibfnamefont {A.}~\bibnamefont {Pierce}}, \ and\ \bibinfo
  {author} {\bibfnamefont {K.~M.}\ \bibnamefont {Zurek}},\ }\href {\doibase
  10.1103/PhysRevD.82.056001} {\bibfield  {journal} {\bibinfo  {journal} {Phys.
  Rev.}\ }\textbf {\bibinfo {volume} {D82}},\ \bibinfo {pages} {056001}
  (\bibinfo {year} {2010})},\ \Eprint {http://arxiv.org/abs/1005.1655}
  {arXiv:1005.1655 [hep-ph]} \BibitemShut {NoStop}%
\bibitem [{\citenamefont {Hall}\ \emph {et~al.}(2010)\citenamefont {Hall},
  \citenamefont {Jedamzik}, \citenamefont {March-Russell},\ and\ \citenamefont
  {West}}]{Hall:2009bx}%
  \BibitemOpen
  \bibfield  {author} {\bibinfo {author} {\bibfnamefont {L.~J.}\ \bibnamefont
  {Hall}}, \bibinfo {author} {\bibfnamefont {K.}~\bibnamefont {Jedamzik}},
  \bibinfo {author} {\bibfnamefont {J.}~\bibnamefont {March-Russell}}, \ and\
  \bibinfo {author} {\bibfnamefont {S.~M.}\ \bibnamefont {West}},\ }\href
  {\doibase 10.1007/JHEP03(2010)080} {\bibfield  {journal} {\bibinfo  {journal}
  {JHEP}\ }\textbf {\bibinfo {volume} {03}},\ \bibinfo {pages} {080} (\bibinfo
  {year} {2010})},\ \Eprint {http://arxiv.org/abs/0911.1120} {arXiv:0911.1120
  [hep-ph]} \BibitemShut {NoStop}%
\bibitem [{\citenamefont {Hochberg}\ \emph {et~al.}(2014)\citenamefont
  {Hochberg}, \citenamefont {Kuflik}, \citenamefont {Volansky},\ and\
  \citenamefont {Wacker}}]{Hochberg:2014dra}%
  \BibitemOpen
  \bibfield  {author} {\bibinfo {author} {\bibfnamefont {Y.}~\bibnamefont
  {Hochberg}}, \bibinfo {author} {\bibfnamefont {E.}~\bibnamefont {Kuflik}},
  \bibinfo {author} {\bibfnamefont {T.}~\bibnamefont {Volansky}}, \ and\
  \bibinfo {author} {\bibfnamefont {J.~G.}\ \bibnamefont {Wacker}},\ }\href
  {\doibase 10.1103/PhysRevLett.113.171301} {\bibfield  {journal} {\bibinfo
  {journal} {Phys. Rev. Lett.}\ }\textbf {\bibinfo {volume} {113}},\ \bibinfo
  {pages} {171301} (\bibinfo {year} {2014})},\ \Eprint
  {http://arxiv.org/abs/1402.5143} {arXiv:1402.5143 [hep-ph]} \BibitemShut
  {NoStop}%
\bibitem [{\citenamefont {Essig}\ \emph
  {et~al.}(2012{\natexlab{a}})\citenamefont {Essig}, \citenamefont {Mardon},\
  and\ \citenamefont {Volansky}}]{Essig:2011nj}%
  \BibitemOpen
  \bibfield  {author} {\bibinfo {author} {\bibfnamefont {R.}~\bibnamefont
  {Essig}}, \bibinfo {author} {\bibfnamefont {J.}~\bibnamefont {Mardon}}, \
  and\ \bibinfo {author} {\bibfnamefont {T.}~\bibnamefont {Volansky}},\ }\href
  {\doibase 10.1103/PhysRevD.85.076007} {\bibfield  {journal} {\bibinfo
  {journal} {Phys. Rev.}\ }\textbf {\bibinfo {volume} {D85}},\ \bibinfo {pages}
  {076007} (\bibinfo {year} {2012}{\natexlab{a}})},\ \Eprint
  {http://arxiv.org/abs/1108.5383} {arXiv:1108.5383 [hep-ph]} \BibitemShut
  {NoStop}%
\bibitem [{\citenamefont {Graham}\ \emph {et~al.}(2012)\citenamefont {Graham},
  \citenamefont {Kaplan}, \citenamefont {Rajendran},\ and\ \citenamefont
  {Walters}}]{Graham:2012su}%
  \BibitemOpen
  \bibfield  {author} {\bibinfo {author} {\bibfnamefont {P.~W.}\ \bibnamefont
  {Graham}}, \bibinfo {author} {\bibfnamefont {D.~E.}\ \bibnamefont {Kaplan}},
  \bibinfo {author} {\bibfnamefont {S.}~\bibnamefont {Rajendran}}, \ and\
  \bibinfo {author} {\bibfnamefont {M.~T.}\ \bibnamefont {Walters}},\ }\href
  {\doibase 10.1016/j.dark.2012.09.001} {\bibfield  {journal} {\bibinfo
  {journal} {Phys. Dark Univ.}\ }\textbf {\bibinfo {volume} {1}},\ \bibinfo
  {pages} {32} (\bibinfo {year} {2012})},\ \Eprint
  {http://arxiv.org/abs/1203.2531} {arXiv:1203.2531 [hep-ph]} \BibitemShut
  {NoStop}%
\bibitem [{\citenamefont {Essig}\ \emph
  {et~al.}(2012{\natexlab{b}})\citenamefont {Essig}, \citenamefont
  {Manalaysay}, \citenamefont {Mardon}, \citenamefont {Sorensen},\ and\
  \citenamefont {Volansky}}]{Essig:2012yx}%
  \BibitemOpen
  \bibfield  {author} {\bibinfo {author} {\bibfnamefont {R.}~\bibnamefont
  {Essig}}, \bibinfo {author} {\bibfnamefont {A.}~\bibnamefont {Manalaysay}},
  \bibinfo {author} {\bibfnamefont {J.}~\bibnamefont {Mardon}}, \bibinfo
  {author} {\bibfnamefont {P.}~\bibnamefont {Sorensen}}, \ and\ \bibinfo
  {author} {\bibfnamefont {T.}~\bibnamefont {Volansky}},\ }\href {\doibase
  10.1103/PhysRevLett.109.021301} {\bibfield  {journal} {\bibinfo  {journal}
  {Phys. Rev. Lett.}\ }\textbf {\bibinfo {volume} {109}},\ \bibinfo {pages}
  {021301} (\bibinfo {year} {2012}{\natexlab{b}})},\ \Eprint
  {http://arxiv.org/abs/1206.2644} {arXiv:1206.2644 [astro-ph.CO]} \BibitemShut
  {NoStop}%
\bibitem [{\citenamefont {Lee}\ \emph {et~al.}(2015)\citenamefont {Lee},
  \citenamefont {Lisanti}, \citenamefont {Mishra-Sharma},\ and\ \citenamefont
  {Safdi}}]{Lee:2015qva}%
  \BibitemOpen
  \bibfield  {author} {\bibinfo {author} {\bibfnamefont {S.~K.}\ \bibnamefont
  {Lee}}, \bibinfo {author} {\bibfnamefont {M.}~\bibnamefont {Lisanti}},
  \bibinfo {author} {\bibfnamefont {S.}~\bibnamefont {Mishra-Sharma}}, \ and\
  \bibinfo {author} {\bibfnamefont {B.~R.}\ \bibnamefont {Safdi}},\ }\href
  {\doibase 10.1103/PhysRevD.92.083517} {\bibfield  {journal} {\bibinfo
  {journal} {Phys. Rev.}\ }\textbf {\bibinfo {volume} {D92}},\ \bibinfo {pages}
  {083517} (\bibinfo {year} {2015})},\ \Eprint
  {http://arxiv.org/abs/1508.07361} {arXiv:1508.07361 [hep-ph]} \BibitemShut
  {NoStop}%
\bibitem [{\citenamefont {Essig}\ \emph {et~al.}(2016)\citenamefont {Essig},
  \citenamefont {Fernandez-Serra}, \citenamefont {Mardon}, \citenamefont
  {Soto}, \citenamefont {Volansky},\ and\ \citenamefont {Yu}}]{Essig:2015cda}%
  \BibitemOpen
  \bibfield  {author} {\bibinfo {author} {\bibfnamefont {R.}~\bibnamefont
  {Essig}}, \bibinfo {author} {\bibfnamefont {M.}~\bibnamefont
  {Fernandez-Serra}}, \bibinfo {author} {\bibfnamefont {J.}~\bibnamefont
  {Mardon}}, \bibinfo {author} {\bibfnamefont {A.}~\bibnamefont {Soto}},
  \bibinfo {author} {\bibfnamefont {T.}~\bibnamefont {Volansky}}, \ and\
  \bibinfo {author} {\bibfnamefont {T.-T.}\ \bibnamefont {Yu}},\ }\href
  {\doibase 10.1007/JHEP05(2016)046} {\bibfield  {journal} {\bibinfo  {journal}
  {JHEP}\ }\textbf {\bibinfo {volume} {05}},\ \bibinfo {pages} {046} (\bibinfo
  {year} {2016})},\ \Eprint {http://arxiv.org/abs/1509.01598} {arXiv:1509.01598
  [hep-ph]} \BibitemShut {NoStop}%
\bibitem [{\citenamefont {Hochberg}\ \emph
  {et~al.}(2017{\natexlab{a}})\citenamefont {Hochberg}, \citenamefont {Kahn},
  \citenamefont {Lisanti}, \citenamefont {Tully},\ and\ \citenamefont
  {Zurek}}]{Hochberg:2016ntt}%
  \BibitemOpen
  \bibfield  {author} {\bibinfo {author} {\bibfnamefont {Y.}~\bibnamefont
  {Hochberg}}, \bibinfo {author} {\bibfnamefont {Y.}~\bibnamefont {Kahn}},
  \bibinfo {author} {\bibfnamefont {M.}~\bibnamefont {Lisanti}}, \bibinfo
  {author} {\bibfnamefont {C.~G.}\ \bibnamefont {Tully}}, \ and\ \bibinfo
  {author} {\bibfnamefont {K.~M.}\ \bibnamefont {Zurek}},\ }\href {\doibase
  10.1016/j.physletb.2017.06.051} {\bibfield  {journal} {\bibinfo  {journal}
  {Phys. Lett.}\ }\textbf {\bibinfo {volume} {B772}},\ \bibinfo {pages} {239}
  (\bibinfo {year} {2017}{\natexlab{a}})},\ \Eprint
  {http://arxiv.org/abs/1606.08849} {arXiv:1606.08849 [hep-ph]} \BibitemShut
  {NoStop}%
\bibitem [{\citenamefont {Derenzo}\ \emph {et~al.}(2017)\citenamefont
  {Derenzo}, \citenamefont {Essig}, \citenamefont {Massari}, \citenamefont
  {Soto},\ and\ \citenamefont {Yu}}]{Derenzo:2016fse}%
  \BibitemOpen
  \bibfield  {author} {\bibinfo {author} {\bibfnamefont {S.}~\bibnamefont
  {Derenzo}}, \bibinfo {author} {\bibfnamefont {R.}~\bibnamefont {Essig}},
  \bibinfo {author} {\bibfnamefont {A.}~\bibnamefont {Massari}}, \bibinfo
  {author} {\bibfnamefont {A.}~\bibnamefont {Soto}}, \ and\ \bibinfo {author}
  {\bibfnamefont {T.-T.}\ \bibnamefont {Yu}},\ }\href {\doibase
  10.1103/PhysRevD.96.016026} {\bibfield  {journal} {\bibinfo  {journal} {Phys.
  Rev.}\ }\textbf {\bibinfo {volume} {D96}},\ \bibinfo {pages} {016026}
  (\bibinfo {year} {2017})},\ \Eprint {http://arxiv.org/abs/1607.01009}
  {arXiv:1607.01009 [hep-ph]} \BibitemShut {NoStop}%
\bibitem [{\citenamefont {Hochberg}\ \emph
  {et~al.}(2017{\natexlab{b}})\citenamefont {Hochberg}, \citenamefont {Lin},\
  and\ \citenamefont {Zurek}}]{Hochberg:2016sqx}%
  \BibitemOpen
  \bibfield  {author} {\bibinfo {author} {\bibfnamefont {Y.}~\bibnamefont
  {Hochberg}}, \bibinfo {author} {\bibfnamefont {T.}~\bibnamefont {Lin}}, \
  and\ \bibinfo {author} {\bibfnamefont {K.~M.}\ \bibnamefont {Zurek}},\ }\href
  {\doibase 10.1103/PhysRevD.95.023013} {\bibfield  {journal} {\bibinfo
  {journal} {Phys. Rev.}\ }\textbf {\bibinfo {volume} {D95}},\ \bibinfo {pages}
  {023013} (\bibinfo {year} {2017}{\natexlab{b}})},\ \Eprint
  {http://arxiv.org/abs/1608.01994} {arXiv:1608.01994 [hep-ph]} \BibitemShut
  {NoStop}%
\bibitem [{\citenamefont {Bloch}\ \emph {et~al.}(2017)\citenamefont {Bloch},
  \citenamefont {Essig}, \citenamefont {Tobioka}, \citenamefont {Volansky},\
  and\ \citenamefont {Yu}}]{Bloch:2016sjj}%
  \BibitemOpen
  \bibfield  {author} {\bibinfo {author} {\bibfnamefont {I.~M.}\ \bibnamefont
  {Bloch}}, \bibinfo {author} {\bibfnamefont {R.}~\bibnamefont {Essig}},
  \bibinfo {author} {\bibfnamefont {K.}~\bibnamefont {Tobioka}}, \bibinfo
  {author} {\bibfnamefont {T.}~\bibnamefont {Volansky}}, \ and\ \bibinfo
  {author} {\bibfnamefont {T.-T.}\ \bibnamefont {Yu}},\ }\href {\doibase
  10.1007/JHEP06(2017)087} {\bibfield  {journal} {\bibinfo  {journal} {JHEP}\
  }\textbf {\bibinfo {volume} {06}},\ \bibinfo {pages} {087} (\bibinfo {year}
  {2017})},\ \Eprint {http://arxiv.org/abs/1608.02123} {arXiv:1608.02123
  [hep-ph]} \BibitemShut {NoStop}%
\bibitem [{\citenamefont {Essig}\ \emph
  {et~al.}(2017{\natexlab{a}})\citenamefont {Essig}, \citenamefont {Volansky},\
  and\ \citenamefont {Yu}}]{Essig:2017kqs}%
  \BibitemOpen
  \bibfield  {author} {\bibinfo {author} {\bibfnamefont {R.}~\bibnamefont
  {Essig}}, \bibinfo {author} {\bibfnamefont {T.}~\bibnamefont {Volansky}}, \
  and\ \bibinfo {author} {\bibfnamefont {T.-T.}\ \bibnamefont {Yu}},\ }\href
  {\doibase 10.1103/PhysRevD.96.043017} {\bibfield  {journal} {\bibinfo
  {journal} {Phys. Rev.}\ }\textbf {\bibinfo {volume} {D96}},\ \bibinfo {pages}
  {043017} (\bibinfo {year} {2017}{\natexlab{a}})},\ \Eprint
  {http://arxiv.org/abs/1703.00910} {arXiv:1703.00910 [hep-ph]} \BibitemShut
  {NoStop}%
\bibitem [{\citenamefont {Kadribasic}\ \emph {et~al.}(2018)\citenamefont
  {Kadribasic}, \citenamefont {Mirabolfathi}, \citenamefont {Nordlund},
  \citenamefont {Sand}, \citenamefont {Holmstr{\"o}m},\ and\ \citenamefont
  {Djurabekova}}]{Kadribasic:2017obi}%
  \BibitemOpen
  \bibfield  {author} {\bibinfo {author} {\bibfnamefont {F.}~\bibnamefont
  {Kadribasic}}, \bibinfo {author} {\bibfnamefont {N.}~\bibnamefont
  {Mirabolfathi}}, \bibinfo {author} {\bibfnamefont {K.}~\bibnamefont
  {Nordlund}}, \bibinfo {author} {\bibfnamefont {A.~E.}\ \bibnamefont {Sand}},
  \bibinfo {author} {\bibfnamefont {E.}~\bibnamefont {Holmstr{\"o}m}}, \ and\
  \bibinfo {author} {\bibfnamefont {F.}~\bibnamefont {Djurabekova}},\ }\href
  {\doibase 10.1103/PhysRevLett.120.111301} {\bibfield  {journal} {\bibinfo
  {journal} {Phys. Rev. Lett.}\ }\textbf {\bibinfo {volume} {120}},\ \bibinfo
  {pages} {111301} (\bibinfo {year} {2018})},\ \Eprint
  {http://arxiv.org/abs/1703.05371} {arXiv:1703.05371 [physics.ins-det]}
  \BibitemShut {NoStop}%
\bibitem [{\citenamefont {Kurinsky}\ \emph {et~al.}(2019)\citenamefont
  {Kurinsky}, \citenamefont {Yu}, \citenamefont {Hochberg},\ and\ \citenamefont
  {Cabrera}}]{Kurinsky:2019pgb}%
  \BibitemOpen
  \bibfield  {author} {\bibinfo {author} {\bibfnamefont {N.~A.}\ \bibnamefont
  {Kurinsky}}, \bibinfo {author} {\bibfnamefont {T.~C.}\ \bibnamefont {Yu}},
  \bibinfo {author} {\bibfnamefont {Y.}~\bibnamefont {Hochberg}}, \ and\
  \bibinfo {author} {\bibfnamefont {B.}~\bibnamefont {Cabrera}},\ }\href
  {\doibase 10.1103/PhysRevD.99.123005} {\bibfield  {journal} {\bibinfo
  {journal} {Phys. Rev.}\ }\textbf {\bibinfo {volume} {D99}},\ \bibinfo {pages}
  {123005} (\bibinfo {year} {2019})},\ \Eprint
  {http://arxiv.org/abs/1901.07569} {arXiv:1901.07569 [hep-ex]} \BibitemShut
  {NoStop}%
\bibitem [{\citenamefont {Heikinheimo}\ \emph {et~al.}(2019)\citenamefont
  {Heikinheimo}, \citenamefont {Nordlund}, \citenamefont {Tuominen},\ and\
  \citenamefont {Mirabolfathi}}]{Heikinheimo:2019lwg}%
  \BibitemOpen
  \bibfield  {author} {\bibinfo {author} {\bibfnamefont {M.}~\bibnamefont
  {Heikinheimo}}, \bibinfo {author} {\bibfnamefont {K.}~\bibnamefont
  {Nordlund}}, \bibinfo {author} {\bibfnamefont {K.}~\bibnamefont {Tuominen}},
  \ and\ \bibinfo {author} {\bibfnamefont {N.}~\bibnamefont {Mirabolfathi}},\
  }\href {\doibase 10.1103/PhysRevD.99.103018} {\bibfield  {journal} {\bibinfo
  {journal} {Phys. Rev.}\ }\textbf {\bibinfo {volume} {D99}},\ \bibinfo {pages}
  {103018} (\bibinfo {year} {2019})},\ \Eprint
  {http://arxiv.org/abs/1903.08654} {arXiv:1903.08654 [hep-ph]} \BibitemShut
  {NoStop}%
\bibitem [{\citenamefont {Emken}\ \emph {et~al.}(2019)\citenamefont {Emken},
  \citenamefont {Essig}, \citenamefont {Kouvaris},\ and\ \citenamefont
  {Sholapurkar}}]{Emken:2019tni}%
  \BibitemOpen
  \bibfield  {author} {\bibinfo {author} {\bibfnamefont {T.}~\bibnamefont
  {Emken}}, \bibinfo {author} {\bibfnamefont {R.}~\bibnamefont {Essig}},
  \bibinfo {author} {\bibfnamefont {C.}~\bibnamefont {Kouvaris}}, \ and\
  \bibinfo {author} {\bibfnamefont {M.}~\bibnamefont {Sholapurkar}},\ }\href
  {\doibase 10.1088/1475-7516/2019/09/070} {\bibfield  {journal} {\bibinfo
  {journal} {JCAP}\ }\textbf {\bibinfo {volume} {1909}},\ \bibinfo {pages}
  {070} (\bibinfo {year} {2019})},\ \Eprint {http://arxiv.org/abs/1905.06348}
  {arXiv:1905.06348 [hep-ph]} \BibitemShut {NoStop}%
\bibitem [{\citenamefont {Hochberg}\ \emph
  {et~al.}(2016{\natexlab{a}})\citenamefont {Hochberg}, \citenamefont {Zhao},\
  and\ \citenamefont {Zurek}}]{Hochberg:2015pha}%
  \BibitemOpen
  \bibfield  {author} {\bibinfo {author} {\bibfnamefont {Y.}~\bibnamefont
  {Hochberg}}, \bibinfo {author} {\bibfnamefont {Y.}~\bibnamefont {Zhao}}, \
  and\ \bibinfo {author} {\bibfnamefont {K.~M.}\ \bibnamefont {Zurek}},\ }\href
  {\doibase 10.1103/PhysRevLett.116.011301} {\bibfield  {journal} {\bibinfo
  {journal} {Phys. Rev. Lett.}\ }\textbf {\bibinfo {volume} {116}},\ \bibinfo
  {pages} {011301} (\bibinfo {year} {2016}{\natexlab{a}})},\ \Eprint
  {http://arxiv.org/abs/1504.07237} {arXiv:1504.07237 [hep-ph]} \BibitemShut
  {NoStop}%
\bibitem [{\citenamefont {Hochberg}\ \emph
  {et~al.}(2016{\natexlab{b}})\citenamefont {Hochberg}, \citenamefont {Pyle},
  \citenamefont {Zhao},\ and\ \citenamefont {Zurek}}]{Hochberg:2015fth}%
  \BibitemOpen
  \bibfield  {author} {\bibinfo {author} {\bibfnamefont {Y.}~\bibnamefont
  {Hochberg}}, \bibinfo {author} {\bibfnamefont {M.}~\bibnamefont {Pyle}},
  \bibinfo {author} {\bibfnamefont {Y.}~\bibnamefont {Zhao}}, \ and\ \bibinfo
  {author} {\bibfnamefont {K.~M.}\ \bibnamefont {Zurek}},\ }\href {\doibase
  10.1007/JHEP08(2016)057} {\bibfield  {journal} {\bibinfo  {journal} {JHEP}\
  }\textbf {\bibinfo {volume} {08}},\ \bibinfo {pages} {057} (\bibinfo {year}
  {2016}{\natexlab{b}})},\ \Eprint {http://arxiv.org/abs/1512.04533}
  {arXiv:1512.04533 [hep-ph]} \BibitemShut {NoStop}%
\bibitem [{\citenamefont {Hochberg}\ \emph
  {et~al.}(2016{\natexlab{c}})\citenamefont {Hochberg}, \citenamefont {Lin},\
  and\ \citenamefont {Zurek}}]{Hochberg:2016ajh}%
  \BibitemOpen
  \bibfield  {author} {\bibinfo {author} {\bibfnamefont {Y.}~\bibnamefont
  {Hochberg}}, \bibinfo {author} {\bibfnamefont {T.}~\bibnamefont {Lin}}, \
  and\ \bibinfo {author} {\bibfnamefont {K.~M.}\ \bibnamefont {Zurek}},\ }\href
  {\doibase 10.1103/PhysRevD.94.015019} {\bibfield  {journal} {\bibinfo
  {journal} {Phys. Rev.}\ }\textbf {\bibinfo {volume} {D94}},\ \bibinfo {pages}
  {015019} (\bibinfo {year} {2016}{\natexlab{c}})},\ \Eprint
  {http://arxiv.org/abs/1604.06800} {arXiv:1604.06800 [hep-ph]} \BibitemShut
  {NoStop}%
\bibitem [{\citenamefont {Hochberg}\ \emph {et~al.}(2018)\citenamefont
  {Hochberg}, \citenamefont {Kahn}, \citenamefont {Lisanti}, \citenamefont
  {Zurek}, \citenamefont {Grushin}, \citenamefont {Ilan}, \citenamefont
  {Griffin}, \citenamefont {Liu}, \citenamefont {Weber},\ and\ \citenamefont
  {Neaton}}]{Hochberg:2017wce}%
  \BibitemOpen
  \bibfield  {author} {\bibinfo {author} {\bibfnamefont {Y.}~\bibnamefont
  {Hochberg}}, \bibinfo {author} {\bibfnamefont {Y.}~\bibnamefont {Kahn}},
  \bibinfo {author} {\bibfnamefont {M.}~\bibnamefont {Lisanti}}, \bibinfo
  {author} {\bibfnamefont {K.~M.}\ \bibnamefont {Zurek}}, \bibinfo {author}
  {\bibfnamefont {A.~G.}\ \bibnamefont {Grushin}}, \bibinfo {author}
  {\bibfnamefont {R.}~\bibnamefont {Ilan}}, \bibinfo {author} {\bibfnamefont
  {S.~M.}\ \bibnamefont {Griffin}}, \bibinfo {author} {\bibfnamefont {Z.-F.}\
  \bibnamefont {Liu}}, \bibinfo {author} {\bibfnamefont {S.~F.}\ \bibnamefont
  {Weber}}, \ and\ \bibinfo {author} {\bibfnamefont {J.~B.}\ \bibnamefont
  {Neaton}},\ }\href {\doibase 10.1103/PhysRevD.97.015004} {\bibfield
  {journal} {\bibinfo  {journal} {Phys. Rev.}\ }\textbf {\bibinfo {volume}
  {D97}},\ \bibinfo {pages} {015004} (\bibinfo {year} {2018})},\ \Eprint
  {http://arxiv.org/abs/1708.08929} {arXiv:1708.08929 [hep-ph]} \BibitemShut
  {NoStop}%
\bibitem [{\citenamefont {Coskuner}\ \emph
  {et~al.}(2019{\natexlab{a}})\citenamefont {Coskuner}, \citenamefont
  {Mitridate}, \citenamefont {Olivares},\ and\ \citenamefont
  {Zurek}}]{Coskuner:2019odd}%
  \BibitemOpen
  \bibfield  {author} {\bibinfo {author} {\bibfnamefont {A.}~\bibnamefont
  {Coskuner}}, \bibinfo {author} {\bibfnamefont {A.}~\bibnamefont {Mitridate}},
  \bibinfo {author} {\bibfnamefont {A.}~\bibnamefont {Olivares}}, \ and\
  \bibinfo {author} {\bibfnamefont {K.~M.}\ \bibnamefont {Zurek}},\ }\href@noop
  {} {\  (\bibinfo {year} {2019}{\natexlab{a}})},\ \Eprint
  {http://arxiv.org/abs/1909.09170} {arXiv:1909.09170 [hep-ph]} \BibitemShut
  {NoStop}%
\bibitem [{\citenamefont {Geilhufe}\ \emph {et~al.}(2019)\citenamefont
  {Geilhufe}, \citenamefont {Kahlhoefer},\ and\ \citenamefont
  {Winkler}}]{Geilhufe:2019ndy}%
  \BibitemOpen
  \bibfield  {author} {\bibinfo {author} {\bibfnamefont {R.~M.}\ \bibnamefont
  {Geilhufe}}, \bibinfo {author} {\bibfnamefont {F.}~\bibnamefont
  {Kahlhoefer}}, \ and\ \bibinfo {author} {\bibfnamefont {M.~W.}\ \bibnamefont
  {Winkler}},\ }\href@noop {} {\  (\bibinfo {year} {2019})},\ \Eprint
  {http://arxiv.org/abs/1910.02091} {arXiv:1910.02091 [hep-ph]} \BibitemShut
  {NoStop}%
\bibitem [{\citenamefont {Ibe}\ \emph {et~al.}(2018)\citenamefont {Ibe},
  \citenamefont {Nakano}, \citenamefont {Shoji},\ and\ \citenamefont
  {Suzuki}}]{Ibe:2017yqa}%
  \BibitemOpen
  \bibfield  {author} {\bibinfo {author} {\bibfnamefont {M.}~\bibnamefont
  {Ibe}}, \bibinfo {author} {\bibfnamefont {W.}~\bibnamefont {Nakano}},
  \bibinfo {author} {\bibfnamefont {Y.}~\bibnamefont {Shoji}}, \ and\ \bibinfo
  {author} {\bibfnamefont {K.}~\bibnamefont {Suzuki}},\ }\href {\doibase
  10.1007/JHEP03(2018)194} {\bibfield  {journal} {\bibinfo  {journal} {JHEP}\
  }\textbf {\bibinfo {volume} {03}},\ \bibinfo {pages} {194} (\bibinfo {year}
  {2018})},\ \Eprint {http://arxiv.org/abs/1707.07258} {arXiv:1707.07258
  [hep-ph]} \BibitemShut {NoStop}%
\bibitem [{\citenamefont {Dolan}\ \emph {et~al.}(2018)\citenamefont {Dolan},
  \citenamefont {Kahlhoefer},\ and\ \citenamefont {McCabe}}]{Dolan:2017xbu}%
  \BibitemOpen
  \bibfield  {author} {\bibinfo {author} {\bibfnamefont {M.~J.}\ \bibnamefont
  {Dolan}}, \bibinfo {author} {\bibfnamefont {F.}~\bibnamefont {Kahlhoefer}}, \
  and\ \bibinfo {author} {\bibfnamefont {C.}~\bibnamefont {McCabe}},\ }\href
  {\doibase 10.1103/PhysRevLett.121.101801} {\bibfield  {journal} {\bibinfo
  {journal} {Phys. Rev. Lett.}\ }\textbf {\bibinfo {volume} {121}},\ \bibinfo
  {pages} {101801} (\bibinfo {year} {2018})},\ \Eprint
  {http://arxiv.org/abs/1711.09906} {arXiv:1711.09906 [hep-ph]} \BibitemShut
  {NoStop}%
\bibitem [{\citenamefont {Bell}\ \emph {et~al.}(2019)\citenamefont {Bell},
  \citenamefont {Dent}, \citenamefont {Newstead}, \citenamefont {Sabharwale},\
  and\ \citenamefont {Weiler}}]{Bell:2019egg}%
  \BibitemOpen
  \bibfield  {author} {\bibinfo {author} {\bibfnamefont {N.~F.}\ \bibnamefont
  {Bell}}, \bibinfo {author} {\bibfnamefont {J.~B.}\ \bibnamefont {Dent}},
  \bibinfo {author} {\bibfnamefont {J.~L.}\ \bibnamefont {Newstead}}, \bibinfo
  {author} {\bibfnamefont {S.}~\bibnamefont {Sabharwale}}, \ and\ \bibinfo
  {author} {\bibfnamefont {T.~J.}\ \bibnamefont {Weiler}},\ }\href@noop {} {\
  (\bibinfo {year} {2019})},\ \Eprint {http://arxiv.org/abs/1905.00046}
  {arXiv:1905.00046 [hep-ph]} \BibitemShut {NoStop}%
\bibitem [{\citenamefont {Baxter}\ \emph {et~al.}(2019)\citenamefont {Baxter},
  \citenamefont {Kahn},\ and\ \citenamefont {Krnjaic}}]{Baxter:2019pnz}%
  \BibitemOpen
  \bibfield  {author} {\bibinfo {author} {\bibfnamefont {D.}~\bibnamefont
  {Baxter}}, \bibinfo {author} {\bibfnamefont {Y.}~\bibnamefont {Kahn}}, \ and\
  \bibinfo {author} {\bibfnamefont {G.}~\bibnamefont {Krnjaic}},\ }\href@noop
  {} {\  (\bibinfo {year} {2019})},\ \Eprint {http://arxiv.org/abs/1908.00012}
  {arXiv:1908.00012 [hep-ph]} \BibitemShut {NoStop}%
\bibitem [{\citenamefont {Essig}\ \emph
  {et~al.}(2019{\natexlab{a}})\citenamefont {Essig}, \citenamefont {Pradler},
  \citenamefont {Sholapurkar},\ and\ \citenamefont {Yu}}]{Essig:2019xkx}%
  \BibitemOpen
  \bibfield  {author} {\bibinfo {author} {\bibfnamefont {R.}~\bibnamefont
  {Essig}}, \bibinfo {author} {\bibfnamefont {J.}~\bibnamefont {Pradler}},
  \bibinfo {author} {\bibfnamefont {M.}~\bibnamefont {Sholapurkar}}, \ and\
  \bibinfo {author} {\bibfnamefont {T.-T.}\ \bibnamefont {Yu}},\ }\href@noop {}
  {\  (\bibinfo {year} {2019}{\natexlab{a}})},\ \Eprint
  {http://arxiv.org/abs/1908.10881} {arXiv:1908.10881 [hep-ph]} \BibitemShut
  {NoStop}%
\bibitem [{\citenamefont {Essig}\ \emph
  {et~al.}(2017{\natexlab{b}})\citenamefont {Essig}, \citenamefont {Mardon},
  \citenamefont {Slone},\ and\ \citenamefont {Volansky}}]{Essig:2016crl}%
  \BibitemOpen
  \bibfield  {author} {\bibinfo {author} {\bibfnamefont {R.}~\bibnamefont
  {Essig}}, \bibinfo {author} {\bibfnamefont {J.}~\bibnamefont {Mardon}},
  \bibinfo {author} {\bibfnamefont {O.}~\bibnamefont {Slone}}, \ and\ \bibinfo
  {author} {\bibfnamefont {T.}~\bibnamefont {Volansky}},\ }\href {\doibase
  10.1103/PhysRevD.95.056011} {\bibfield  {journal} {\bibinfo  {journal} {Phys.
  Rev.}\ }\textbf {\bibinfo {volume} {D95}},\ \bibinfo {pages} {056011}
  (\bibinfo {year} {2017}{\natexlab{b}})},\ \Eprint
  {http://arxiv.org/abs/1608.02940} {arXiv:1608.02940 [hep-ph]} \BibitemShut
  {NoStop}%
\bibitem [{\citenamefont {Arvanitaki}\ \emph {et~al.}(2018)\citenamefont
  {Arvanitaki}, \citenamefont {Dimopoulos},\ and\ \citenamefont
  {Van~Tilburg}}]{Arvanitaki:2017nhi}%
  \BibitemOpen
  \bibfield  {author} {\bibinfo {author} {\bibfnamefont {A.}~\bibnamefont
  {Arvanitaki}}, \bibinfo {author} {\bibfnamefont {S.}~\bibnamefont
  {Dimopoulos}}, \ and\ \bibinfo {author} {\bibfnamefont {K.}~\bibnamefont
  {Van~Tilburg}},\ }\href {\doibase 10.1103/PhysRevX.8.041001} {\bibfield
  {journal} {\bibinfo  {journal} {Phys. Rev.}\ }\textbf {\bibinfo {volume}
  {X8}},\ \bibinfo {pages} {041001} (\bibinfo {year} {2018})},\ \Eprint
  {http://arxiv.org/abs/1709.05354} {arXiv:1709.05354 [hep-ph]} \BibitemShut
  {NoStop}%
\bibitem [{\citenamefont {Essig}\ \emph
  {et~al.}(2019{\natexlab{b}})\citenamefont {Essig}, \citenamefont
  {P{\'e}rez-R{\'\i}os}, \citenamefont {Ramani},\ and\ \citenamefont
  {Slone}}]{Essig:2019kfe}%
  \BibitemOpen
  \bibfield  {author} {\bibinfo {author} {\bibfnamefont {R.}~\bibnamefont
  {Essig}}, \bibinfo {author} {\bibfnamefont {J.}~\bibnamefont
  {P{\'e}rez-R{\'\i}os}}, \bibinfo {author} {\bibfnamefont {H.}~\bibnamefont
  {Ramani}}, \ and\ \bibinfo {author} {\bibfnamefont {O.}~\bibnamefont
  {Slone}},\ }\href@noop {} {\  (\bibinfo {year} {2019}{\natexlab{b}})},\
  \Eprint {http://arxiv.org/abs/1907.07682} {arXiv:1907.07682 [hep-ph]}
  \BibitemShut {NoStop}%
\bibitem [{\citenamefont {Schutz}\ and\ \citenamefont
  {Zurek}(2016)}]{Schutz:2016tid}%
  \BibitemOpen
  \bibfield  {author} {\bibinfo {author} {\bibfnamefont {K.}~\bibnamefont
  {Schutz}}\ and\ \bibinfo {author} {\bibfnamefont {K.~M.}\ \bibnamefont
  {Zurek}},\ }\href {\doibase 10.1103/PhysRevLett.117.121302} {\bibfield
  {journal} {\bibinfo  {journal} {Phys. Rev. Lett.}\ }\textbf {\bibinfo
  {volume} {117}},\ \bibinfo {pages} {121302} (\bibinfo {year} {2016})},\
  \Eprint {http://arxiv.org/abs/1604.08206} {arXiv:1604.08206 [hep-ph]}
  \BibitemShut {NoStop}%
\bibitem [{\citenamefont {Knapen}\ \emph
  {et~al.}(2017{\natexlab{a}})\citenamefont {Knapen}, \citenamefont {Lin},\
  and\ \citenamefont {Zurek}}]{Knapen:2016cue}%
  \BibitemOpen
  \bibfield  {author} {\bibinfo {author} {\bibfnamefont {S.}~\bibnamefont
  {Knapen}}, \bibinfo {author} {\bibfnamefont {T.}~\bibnamefont {Lin}}, \ and\
  \bibinfo {author} {\bibfnamefont {K.~M.}\ \bibnamefont {Zurek}},\ }\href
  {\doibase 10.1103/PhysRevD.95.056019} {\bibfield  {journal} {\bibinfo
  {journal} {Phys. Rev.}\ }\textbf {\bibinfo {volume} {D95}},\ \bibinfo {pages}
  {056019} (\bibinfo {year} {2017}{\natexlab{a}})},\ \Eprint
  {http://arxiv.org/abs/1611.06228} {arXiv:1611.06228 [hep-ph]} \BibitemShut
  {NoStop}%
\bibitem [{\citenamefont {Acanfora}\ \emph {et~al.}(2019)\citenamefont
  {Acanfora}, \citenamefont {Esposito},\ and\ \citenamefont
  {Polosa}}]{Acanfora:2019con}%
  \BibitemOpen
  \bibfield  {author} {\bibinfo {author} {\bibfnamefont {F.}~\bibnamefont
  {Acanfora}}, \bibinfo {author} {\bibfnamefont {A.}~\bibnamefont {Esposito}},
  \ and\ \bibinfo {author} {\bibfnamefont {A.~D.}\ \bibnamefont {Polosa}},\
  }\href {\doibase 10.1140/epjc/s10052-019-7057-0} {\bibfield  {journal}
  {\bibinfo  {journal} {Eur. Phys. J.}\ }\textbf {\bibinfo {volume} {C79}},\
  \bibinfo {pages} {549} (\bibinfo {year} {2019})},\ \Eprint
  {http://arxiv.org/abs/1902.02361} {arXiv:1902.02361 [hep-ph]} \BibitemShut
  {NoStop}%
\bibitem [{\citenamefont {Caputo}\ \emph {et~al.}(2019)\citenamefont {Caputo},
  \citenamefont {Esposito},\ and\ \citenamefont {Polosa}}]{Caputo:2019cyg}%
  \BibitemOpen
  \bibfield  {author} {\bibinfo {author} {\bibfnamefont {A.}~\bibnamefont
  {Caputo}}, \bibinfo {author} {\bibfnamefont {A.}~\bibnamefont {Esposito}}, \
  and\ \bibinfo {author} {\bibfnamefont {A.~D.}\ \bibnamefont {Polosa}},\
  }\href@noop {} {\  (\bibinfo {year} {2019})},\ \Eprint
  {http://arxiv.org/abs/1907.10635} {arXiv:1907.10635 [hep-ph]} \BibitemShut
  {NoStop}%
\bibitem [{\citenamefont {Budnik}\ \emph {et~al.}(2018)\citenamefont {Budnik},
  \citenamefont {Chesnovsky}, \citenamefont {Slone},\ and\ \citenamefont
  {Volansky}}]{Budnik:2017sbu}%
  \BibitemOpen
  \bibfield  {author} {\bibinfo {author} {\bibfnamefont {R.}~\bibnamefont
  {Budnik}}, \bibinfo {author} {\bibfnamefont {O.}~\bibnamefont {Chesnovsky}},
  \bibinfo {author} {\bibfnamefont {O.}~\bibnamefont {Slone}}, \ and\ \bibinfo
  {author} {\bibfnamefont {T.}~\bibnamefont {Volansky}},\ }\href {\doibase
  10.1016/j.physletb.2018.04.063} {\bibfield  {journal} {\bibinfo  {journal}
  {Phys. Lett.}\ }\textbf {\bibinfo {volume} {B782}},\ \bibinfo {pages} {242}
  (\bibinfo {year} {2018})},\ \Eprint {http://arxiv.org/abs/1705.03016}
  {arXiv:1705.03016 [hep-ph]} \BibitemShut {NoStop}%
\bibitem [{\citenamefont {Rajendran}\ \emph {et~al.}(2017)\citenamefont
  {Rajendran}, \citenamefont {Zobrist}, \citenamefont {Sushkov}, \citenamefont
  {Walsworth},\ and\ \citenamefont {Lukin}}]{Rajendran:2017ynw}%
  \BibitemOpen
  \bibfield  {author} {\bibinfo {author} {\bibfnamefont {S.}~\bibnamefont
  {Rajendran}}, \bibinfo {author} {\bibfnamefont {N.}~\bibnamefont {Zobrist}},
  \bibinfo {author} {\bibfnamefont {A.~O.}\ \bibnamefont {Sushkov}}, \bibinfo
  {author} {\bibfnamefont {R.}~\bibnamefont {Walsworth}}, \ and\ \bibinfo
  {author} {\bibfnamefont {M.}~\bibnamefont {Lukin}},\ }\href {\doibase
  10.1103/PhysRevD.96.035009} {\bibfield  {journal} {\bibinfo  {journal} {Phys.
  Rev.}\ }\textbf {\bibinfo {volume} {D96}},\ \bibinfo {pages} {035009}
  (\bibinfo {year} {2017})},\ \Eprint {http://arxiv.org/abs/1705.09760}
  {arXiv:1705.09760 [hep-ph]} \BibitemShut {NoStop}%
\bibitem [{\citenamefont {Knapen}\ \emph {et~al.}(2018)\citenamefont {Knapen},
  \citenamefont {Lin}, \citenamefont {Pyle},\ and\ \citenamefont
  {Zurek}}]{Knapen:2017ekk}%
  \BibitemOpen
  \bibfield  {author} {\bibinfo {author} {\bibfnamefont {S.}~\bibnamefont
  {Knapen}}, \bibinfo {author} {\bibfnamefont {T.}~\bibnamefont {Lin}},
  \bibinfo {author} {\bibfnamefont {M.}~\bibnamefont {Pyle}}, \ and\ \bibinfo
  {author} {\bibfnamefont {K.~M.}\ \bibnamefont {Zurek}},\ }\href {\doibase
  10.1016/j.physletb.2018.08.064} {\bibfield  {journal} {\bibinfo  {journal}
  {Phys. Lett.}\ }\textbf {\bibinfo {volume} {B785}},\ \bibinfo {pages} {386}
  (\bibinfo {year} {2018})},\ \Eprint {http://arxiv.org/abs/1712.06598}
  {arXiv:1712.06598 [hep-ph]} \BibitemShut {NoStop}%
\bibitem [{\citenamefont {Griffin}\ \emph {et~al.}(2018)\citenamefont
  {Griffin}, \citenamefont {Knapen}, \citenamefont {Lin},\ and\ \citenamefont
  {Zurek}}]{Griffin:2018bjn}%
  \BibitemOpen
  \bibfield  {author} {\bibinfo {author} {\bibfnamefont {S.}~\bibnamefont
  {Griffin}}, \bibinfo {author} {\bibfnamefont {S.}~\bibnamefont {Knapen}},
  \bibinfo {author} {\bibfnamefont {T.}~\bibnamefont {Lin}}, \ and\ \bibinfo
  {author} {\bibfnamefont {K.~M.}\ \bibnamefont {Zurek}},\ }\href {\doibase
  10.1103/PhysRevD.98.115034} {\bibfield  {journal} {\bibinfo  {journal} {Phys.
  Rev.}\ }\textbf {\bibinfo {volume} {D98}},\ \bibinfo {pages} {115034}
  (\bibinfo {year} {2018})},\ \Eprint {http://arxiv.org/abs/1807.10291}
  {arXiv:1807.10291 [hep-ph]} \BibitemShut {NoStop}%
\bibitem [{\citenamefont {Trickle}\ \emph {et~al.}(2019)\citenamefont
  {Trickle}, \citenamefont {Zhang},\ and\ \citenamefont
  {Zurek}}]{Trickle:2019ovy}%
  \BibitemOpen
  \bibfield  {author} {\bibinfo {author} {\bibfnamefont {T.}~\bibnamefont
  {Trickle}}, \bibinfo {author} {\bibfnamefont {Z.}~\bibnamefont {Zhang}}, \
  and\ \bibinfo {author} {\bibfnamefont {K.~M.}\ \bibnamefont {Zurek}},\
  }\href@noop {} {\  (\bibinfo {year} {2019})},\ \Eprint
  {http://arxiv.org/abs/1905.13744} {arXiv:1905.13744 [hep-ph]} \BibitemShut
  {NoStop}%
\bibitem [{\citenamefont {Riedel}(2013)}]{Riedel:2012ur}%
  \BibitemOpen
  \bibfield  {author} {\bibinfo {author} {\bibfnamefont {C.~J.}\ \bibnamefont
  {Riedel}},\ }\href {\doibase 10.1103/PhysRevD.88.116005} {\bibfield
  {journal} {\bibinfo  {journal} {Phys. Rev.}\ }\textbf {\bibinfo {volume}
  {D88}},\ \bibinfo {pages} {116005} (\bibinfo {year} {2013})},\ \Eprint
  {http://arxiv.org/abs/1212.3061} {arXiv:1212.3061 [quant-ph]} \BibitemShut
  {NoStop}%
\bibitem [{\citenamefont {Kouvaris}\ and\ \citenamefont
  {Pradler}(2017)}]{Kouvaris:2016afs}%
  \BibitemOpen
  \bibfield  {author} {\bibinfo {author} {\bibfnamefont {C.}~\bibnamefont
  {Kouvaris}}\ and\ \bibinfo {author} {\bibfnamefont {J.}~\bibnamefont
  {Pradler}},\ }\href {\doibase 10.1103/PhysRevLett.118.031803} {\bibfield
  {journal} {\bibinfo  {journal} {Phys. Rev. Lett.}\ }\textbf {\bibinfo
  {volume} {118}},\ \bibinfo {pages} {031803} (\bibinfo {year} {2017})},\
  \Eprint {http://arxiv.org/abs/1607.01789} {arXiv:1607.01789 [hep-ph]}
  \BibitemShut {NoStop}%
\bibitem [{\citenamefont {Riedel}\ and\ \citenamefont
  {Yavin}(2017)}]{Riedel:2016acj}%
  \BibitemOpen
  \bibfield  {author} {\bibinfo {author} {\bibfnamefont {C.~J.}\ \bibnamefont
  {Riedel}}\ and\ \bibinfo {author} {\bibfnamefont {I.}~\bibnamefont {Yavin}},\
  }\href {\doibase 10.1103/PhysRevD.96.023007} {\bibfield  {journal} {\bibinfo
  {journal} {Phys. Rev.}\ }\textbf {\bibinfo {volume} {D96}},\ \bibinfo {pages}
  {023007} (\bibinfo {year} {2017})},\ \Eprint
  {http://arxiv.org/abs/1609.04145} {arXiv:1609.04145 [quant-ph]} \BibitemShut
  {NoStop}%
\bibitem [{\citenamefont {Cavoto}\ \emph {et~al.}(2018)\citenamefont {Cavoto},
  \citenamefont {Luchetta},\ and\ \citenamefont {Polosa}}]{Cavoto:2017otc}%
  \BibitemOpen
  \bibfield  {author} {\bibinfo {author} {\bibfnamefont {G.}~\bibnamefont
  {Cavoto}}, \bibinfo {author} {\bibfnamefont {F.}~\bibnamefont {Luchetta}}, \
  and\ \bibinfo {author} {\bibfnamefont {A.~D.}\ \bibnamefont {Polosa}},\
  }\href {\doibase 10.1016/j.physletb.2017.11.064} {\bibfield  {journal}
  {\bibinfo  {journal} {Phys. Lett.}\ }\textbf {\bibinfo {volume} {B776}},\
  \bibinfo {pages} {338} (\bibinfo {year} {2018})},\ \Eprint
  {http://arxiv.org/abs/1706.02487} {arXiv:1706.02487 [hep-ph]} \BibitemShut
  {NoStop}%
\bibitem [{\citenamefont {Marsh}\ \emph {et~al.}(2019)\citenamefont {Marsh},
  \citenamefont {Fong}, \citenamefont {Lentz}, \citenamefont {Smejkal},\ and\
  \citenamefont {Ali}}]{Marsh:2018dlj}%
  \BibitemOpen
  \bibfield  {author} {\bibinfo {author} {\bibfnamefont {D.~J.~E.}\
  \bibnamefont {Marsh}}, \bibinfo {author} {\bibfnamefont {K.-C.}\ \bibnamefont
  {Fong}}, \bibinfo {author} {\bibfnamefont {E.~W.}\ \bibnamefont {Lentz}},
  \bibinfo {author} {\bibfnamefont {L.}~\bibnamefont {Smejkal}}, \ and\
  \bibinfo {author} {\bibfnamefont {M.~N.}\ \bibnamefont {Ali}},\ }\href
  {\doibase 10.1103/PhysRevLett.123.121601} {\bibfield  {journal} {\bibinfo
  {journal} {Phys. Rev. Lett.}\ }\textbf {\bibinfo {volume} {123}},\ \bibinfo
  {pages} {121601} (\bibinfo {year} {2019})},\ \Eprint
  {http://arxiv.org/abs/1807.08810} {arXiv:1807.08810 [hep-ph]} \BibitemShut
  {NoStop}%
\bibitem [{\citenamefont {Alonso}\ \emph {et~al.}(2019)\citenamefont {Alonso},
  \citenamefont {Blas},\ and\ \citenamefont {Wolf}}]{Alonso:2018dxy}%
  \BibitemOpen
  \bibfield  {author} {\bibinfo {author} {\bibfnamefont {R.}~\bibnamefont
  {Alonso}}, \bibinfo {author} {\bibfnamefont {D.}~\bibnamefont {Blas}}, \ and\
  \bibinfo {author} {\bibfnamefont {P.}~\bibnamefont {Wolf}},\ }\href {\doibase
  10.1007/JHEP07(2019)069} {\bibfield  {journal} {\bibinfo  {journal} {JHEP}\
  }\textbf {\bibinfo {volume} {07}},\ \bibinfo {pages} {069} (\bibinfo {year}
  {2019})},\ \Eprint {http://arxiv.org/abs/1810.00889} {arXiv:1810.00889
  [hep-ph]} \BibitemShut {NoStop}%
\bibitem [{\citenamefont {Roberts}\ \emph {et~al.}(2019)\citenamefont {Roberts}
  \emph {et~al.}}]{Roberts:2019sfo}%
  \BibitemOpen
  \bibfield  {author} {\bibinfo {author} {\bibfnamefont {B.~M.}\ \bibnamefont
  {Roberts}} \emph {et~al.},\ }\href@noop {} {\  (\bibinfo {year} {2019})},\
  \Eprint {http://arxiv.org/abs/1907.02661} {arXiv:1907.02661 [astro-ph.CO]}
  \BibitemShut {NoStop}%
\bibitem [{\citenamefont {Zarei}\ \emph {et~al.}(2019)\citenamefont {Zarei},
  \citenamefont {Shakeri}, \citenamefont {Abdi}, \citenamefont {Marsh},\ and\
  \citenamefont {Matarrese}}]{Zarei:2019sva}%
  \BibitemOpen
  \bibfield  {author} {\bibinfo {author} {\bibfnamefont {M.}~\bibnamefont
  {Zarei}}, \bibinfo {author} {\bibfnamefont {S.}~\bibnamefont {Shakeri}},
  \bibinfo {author} {\bibfnamefont {M.}~\bibnamefont {Abdi}}, \bibinfo {author}
  {\bibfnamefont {D.~J.~E.}\ \bibnamefont {Marsh}}, \ and\ \bibinfo {author}
  {\bibfnamefont {S.}~\bibnamefont {Matarrese}},\ }\href@noop {} {\  (\bibinfo
  {year} {2019})},\ \Eprint {http://arxiv.org/abs/1910.09973} {arXiv:1910.09973
  [hep-ph]} \BibitemShut {NoStop}%
\bibitem [{\citenamefont {Goodman}\ and\ \citenamefont
  {Witten}(1985)}]{Goodman1985a}%
  \BibitemOpen
  \bibfield  {author} {\bibinfo {author} {\bibfnamefont {M.~W.}\ \bibnamefont
  {Goodman}}\ and\ \bibinfo {author} {\bibfnamefont {E.}~\bibnamefont
  {Witten}},\ }\href {\doibase 10.1103/physrevd.31.3059} {\bibfield  {journal}
  {\bibinfo  {journal} {Physical Review D}\ }\textbf {\bibinfo {volume} {31}},\
  \bibinfo {pages} {3059} (\bibinfo {year} {1985})}\BibitemShut {NoStop}%
\bibitem [{\citenamefont {Drukier}\ \emph {et~al.}(1986)\citenamefont
  {Drukier}, \citenamefont {Freese},\ and\ \citenamefont
  {Spergel}}]{Drukier1986a}%
  \BibitemOpen
  \bibfield  {author} {\bibinfo {author} {\bibfnamefont {A.~K.}\ \bibnamefont
  {Drukier}}, \bibinfo {author} {\bibfnamefont {K.}~\bibnamefont {Freese}}, \
  and\ \bibinfo {author} {\bibfnamefont {D.~N.}\ \bibnamefont {Spergel}},\
  }\href {\doibase 10.1103/physrevd.33.3495} {\bibfield  {journal} {\bibinfo
  {journal} {Physical Review D}\ }\textbf {\bibinfo {volume} {33}},\ \bibinfo
  {pages} {3495} (\bibinfo {year} {1986})}\BibitemShut {NoStop}%
\bibitem [{\citenamefont {Trickle}\ \emph {et~al.}()\citenamefont {Trickle},
  \citenamefont {Zhang}, \citenamefont {Zurek}, \citenamefont {Inzani},\ and\
  \citenamefont {Griffin}}]{Trickle2019a}%
  \BibitemOpen
  \bibfield  {author} {\bibinfo {author} {\bibfnamefont {T.}~\bibnamefont
  {Trickle}}, \bibinfo {author} {\bibfnamefont {Z.}~\bibnamefont {Zhang}},
  \bibinfo {author} {\bibfnamefont {K.~M.}\ \bibnamefont {Zurek}}, \bibinfo
  {author} {\bibfnamefont {K.}~\bibnamefont {Inzani}}, \ and\ \bibinfo {author}
  {\bibfnamefont {S.}~\bibnamefont {Griffin}},\ }\href@noop {} {\ }\Eprint
  {http://arxiv.org/abs/http://arxiv.org/abs/1910.08092v1}
  {http://arxiv.org/abs/1910.08092v1} \BibitemShut {NoStop}%
\bibitem [{\citenamefont {de~Mello~Neto}\ \emph {et~al.}(2016)\citenamefont
  {de~Mello~Neto} \emph {et~al.}}]{deMelloNeto:2015mca}%
  \BibitemOpen
  \bibfield  {author} {\bibinfo {author} {\bibfnamefont {J.~R.~T.}\
  \bibnamefont {de~Mello~Neto}} \emph {et~al.} (\bibinfo {collaboration}
  {DAMIC}),\ }\bibfield  {booktitle} {\emph {\bibinfo {booktitle}
  {{Proceedings, 34th International Cosmic Ray Conference (ICRC 2015): The
  Hague, The Netherlands, July 30-August 6, 2015}}},\ }\href {\doibase
  10.22323/1.236.1221} {\bibfield  {journal} {\bibinfo  {journal} {PoS}\
  }\textbf {\bibinfo {volume} {ICRC2015}},\ \bibinfo {pages} {1221} (\bibinfo
  {year} {2016})},\ \Eprint {http://arxiv.org/abs/1510.02126} {arXiv:1510.02126
  [physics.ins-det]} \BibitemShut {NoStop}%
\bibitem [{\citenamefont {Aguilar-Arevalo}\ \emph {et~al.}(2019)\citenamefont
  {Aguilar-Arevalo} \emph {et~al.}}]{AguilarArevalo2019a}%
  \BibitemOpen
  \bibfield  {author} {\bibinfo {author} {\bibfnamefont {A.}~\bibnamefont
  {Aguilar-Arevalo}} \emph {et~al.} (\bibinfo {collaboration} {DAMIC}),\ }\href
  {\doibase 10.1103/PhysRevLett.123.181802} {\bibfield  {journal} {\bibinfo
  {journal} {Phys. Rev. Lett.}\ }\textbf {\bibinfo {volume} {123}},\ \bibinfo
  {pages} {181802} (\bibinfo {year} {2019})},\ \Eprint
  {http://arxiv.org/abs/1907.12628} {arXiv:1907.12628 [astro-ph.CO]}
  \BibitemShut {NoStop}%
\bibitem [{\citenamefont {Abramoff}\ \emph {et~al.}(2019)\citenamefont
  {Abramoff} \emph {et~al.}}]{Abramoff2019a}%
  \BibitemOpen
  \bibfield  {author} {\bibinfo {author} {\bibfnamefont {O.}~\bibnamefont
  {Abramoff}} \emph {et~al.} (\bibinfo {collaboration} {SENSEI}),\ }\href
  {\doibase 10.1103/PhysRevLett.122.161801} {\bibfield  {journal} {\bibinfo
  {journal} {Phys. Rev. Lett.}\ }\textbf {\bibinfo {volume} {122}},\ \bibinfo
  {pages} {161801} (\bibinfo {year} {2019})},\ \Eprint
  {http://arxiv.org/abs/1901.10478} {arXiv:1901.10478 [hep-ex]} \BibitemShut
  {NoStop}%
\bibitem [{\citenamefont {Agnese}\ \emph {et~al.}(2014)\citenamefont {Agnese}
  \emph {et~al.}}]{Agnese:2014aze}%
  \BibitemOpen
  \bibfield  {author} {\bibinfo {author} {\bibfnamefont {R.}~\bibnamefont
  {Agnese}} \emph {et~al.} (\bibinfo {collaboration} {SuperCDMS}),\ }\href
  {\doibase 10.1103/PhysRevLett.112.241302} {\bibfield  {journal} {\bibinfo
  {journal} {Phys. Rev. Lett.}\ }\textbf {\bibinfo {volume} {112}},\ \bibinfo
  {pages} {241302} (\bibinfo {year} {2014})},\ \Eprint
  {http://arxiv.org/abs/1402.7137} {arXiv:1402.7137 [hep-ex]} \BibitemShut
  {NoStop}%
\bibitem [{\citenamefont {Agnese}\ \emph {et~al.}(2017)\citenamefont {Agnese}
  \emph {et~al.}}]{Agnese:2016cpb}%
  \BibitemOpen
  \bibfield  {author} {\bibinfo {author} {\bibfnamefont {R.}~\bibnamefont
  {Agnese}} \emph {et~al.} (\bibinfo {collaboration} {SuperCDMS}),\ }\href
  {\doibase 10.1103/PhysRevD.95.082002} {\bibfield  {journal} {\bibinfo
  {journal} {Phys. Rev.}\ }\textbf {\bibinfo {volume} {D95}},\ \bibinfo {pages}
  {082002} (\bibinfo {year} {2017})},\ \Eprint
  {http://arxiv.org/abs/1610.00006} {arXiv:1610.00006 [physics.ins-det]}
  \BibitemShut {NoStop}%
\bibitem [{\citenamefont {Agnese}\ \emph {et~al.}(2016)\citenamefont {Agnese}
  \emph {et~al.}}]{Agnese:2015nto}%
  \BibitemOpen
  \bibfield  {author} {\bibinfo {author} {\bibfnamefont {R.}~\bibnamefont
  {Agnese}} \emph {et~al.} (\bibinfo {collaboration} {SuperCDMS}),\ }\href
  {\doibase 10.1103/PhysRevLett.116.071301} {\bibfield  {journal} {\bibinfo
  {journal} {Phys. Rev. Lett.}\ }\textbf {\bibinfo {volume} {116}},\ \bibinfo
  {pages} {071301} (\bibinfo {year} {2016})},\ \Eprint
  {http://arxiv.org/abs/1509.02448} {arXiv:1509.02448 [astro-ph.CO]}
  \BibitemShut {NoStop}%
\bibitem [{\citenamefont {Agnese}\ \emph
  {et~al.}(2018{\natexlab{a}})\citenamefont {Agnese} \emph
  {et~al.}}]{Agnese2018a}%
  \BibitemOpen
  \bibfield  {author} {\bibinfo {author} {\bibfnamefont {R.}~\bibnamefont
  {Agnese}} \emph {et~al.} (\bibinfo {collaboration} {SuperCDMS}),\ }\href
  {\doibase 10.1103/PhysRevD.97.022002} {\bibfield  {journal} {\bibinfo
  {journal} {Phys. Rev.}\ }\textbf {\bibinfo {volume} {D97}},\ \bibinfo {pages}
  {022002} (\bibinfo {year} {2018}{\natexlab{a}})},\ \Eprint
  {http://arxiv.org/abs/1707.01632} {arXiv:1707.01632 [astro-ph.CO]}
  \BibitemShut {NoStop}%
\bibitem [{\citenamefont {Agnese}\ \emph {et~al.}(2019)\citenamefont {Agnese}
  \emph {et~al.}}]{Agnese2019a}%
  \BibitemOpen
  \bibfield  {author} {\bibinfo {author} {\bibfnamefont {R.}~\bibnamefont
  {Agnese}} \emph {et~al.} (\bibinfo {collaboration} {SuperCDMS}),\ }\href
  {\doibase 10.1103/PhysRevD.99.062001} {\bibfield  {journal} {\bibinfo
  {journal} {Phys. Rev.}\ }\textbf {\bibinfo {volume} {D99}},\ \bibinfo {pages}
  {062001} (\bibinfo {year} {2019})},\ \Eprint
  {http://arxiv.org/abs/1808.09098} {arXiv:1808.09098 [astro-ph.CO]}
  \BibitemShut {NoStop}%
\bibitem [{\citenamefont {Agnese}\ \emph
  {et~al.}(2018{\natexlab{b}})\citenamefont {Agnese} \emph
  {et~al.}}]{Agnese2018b}%
  \BibitemOpen
  \bibfield  {author} {\bibinfo {author} {\bibfnamefont {R.}~\bibnamefont
  {Agnese}} \emph {et~al.} (\bibinfo {collaboration} {SuperCDMS}),\ }\href
  {\doibase 10.1103/PhysRevLett.122.069901, 10.1103/PhysRevLett.121.051301}
  {\bibfield  {journal} {\bibinfo  {journal} {Phys. Rev. Lett.}\ }\textbf
  {\bibinfo {volume} {121}},\ \bibinfo {pages} {051301} (\bibinfo {year}
  {2018}{\natexlab{b}})},\ \bibinfo {note} {[erratum: Phys. Rev.
  Lett.122,no.6,069901(2019)]},\ \Eprint {http://arxiv.org/abs/1804.10697}
  {arXiv:1804.10697 [hep-ex]} \BibitemShut {NoStop}%
\bibitem [{\citenamefont {Baum}\ \emph {et~al.}(2019)\citenamefont {Baum},
  \citenamefont {Freese},\ and\ \citenamefont {Kelso}}]{Baum2019a}%
  \BibitemOpen
  \bibfield  {author} {\bibinfo {author} {\bibfnamefont {S.}~\bibnamefont
  {Baum}}, \bibinfo {author} {\bibfnamefont {K.}~\bibnamefont {Freese}}, \ and\
  \bibinfo {author} {\bibfnamefont {C.}~\bibnamefont {Kelso}},\ }\href
  {\doibase 10.1016/j.physletb.2018.12.036} {\bibfield  {journal} {\bibinfo
  {journal} {Physics Letters B}\ }\textbf {\bibinfo {volume} {789}},\ \bibinfo
  {pages} {262} (\bibinfo {year} {2019})}\BibitemShut {NoStop}%
\bibitem [{\citenamefont {Kim}(2015)}]{Kim:2015prm}%
  \BibitemOpen
  \bibfield  {author} {\bibinfo {author} {\bibfnamefont {K.}~\bibnamefont
  {Kim}} (\bibinfo {collaboration} {KIMS}),\ }in\ \href@noop {} {\emph
  {\bibinfo {booktitle} {{Proceedings, Meeting of the APS Division of Particles
  and Fields (DPF 2015): Ann Arbor, Michigan, USA, 4-8 Aug 2015}}}}\ (\bibinfo
  {year} {2015})\ \Eprint {http://arxiv.org/abs/1511.00023} {arXiv:1511.00023
  [physics.ins-det]} \BibitemShut {NoStop}%
\bibitem [{\citenamefont {Amar{\'e}}\ \emph {et~al.}(2019)\citenamefont
  {Amar{\'e}} \emph {et~al.}}]{Amare:2019jul}%
  \BibitemOpen
  \bibfield  {author} {\bibinfo {author} {\bibfnamefont {J.}~\bibnamefont
  {Amar{\'e}}} \emph {et~al.},\ }\href {\doibase
  10.1103/PhysRevLett.123.031301} {\bibfield  {journal} {\bibinfo  {journal}
  {Phys. Rev. Lett.}\ }\textbf {\bibinfo {volume} {123}},\ \bibinfo {pages}
  {031301} (\bibinfo {year} {2019})},\ \Eprint
  {http://arxiv.org/abs/1903.03973} {arXiv:1903.03973 [astro-ph.IM]}
  \BibitemShut {NoStop}%
\bibitem [{\citenamefont {Shields}\ \emph {et~al.}(2015)\citenamefont
  {Shields}, \citenamefont {Xu},\ and\ \citenamefont
  {Calaprice}}]{Shields2015a}%
  \BibitemOpen
  \bibfield  {author} {\bibinfo {author} {\bibfnamefont {E.}~\bibnamefont
  {Shields}}, \bibinfo {author} {\bibfnamefont {J.}~\bibnamefont {Xu}}, \ and\
  \bibinfo {author} {\bibfnamefont {F.}~\bibnamefont {Calaprice}},\ }\href
  {\doibase 10.1016/j.phpro.2014.12.028} {\bibfield  {journal} {\bibinfo
  {journal} {Physics Procedia}\ }\textbf {\bibinfo {volume} {61}},\ \bibinfo
  {pages} {169} (\bibinfo {year} {2015})}\BibitemShut {NoStop}%
\bibitem [{\citenamefont {Jo}(2017)}]{Jo:2016qql}%
  \BibitemOpen
  \bibfield  {author} {\bibinfo {author} {\bibfnamefont {J.~H.}\ \bibnamefont
  {Jo}} (\bibinfo {collaboration} {DM-Ice}),\ }\bibfield  {booktitle} {\emph
  {\bibinfo {booktitle} {{Proceedings, 38th International Conference on High
  Energy Physics (ICHEP 2016): Chicago, IL, USA, August 3-10, 2016}}},\ }\href
  {\doibase 10.22323/1.282.1223} {\bibfield  {journal} {\bibinfo  {journal}
  {PoS}\ }\textbf {\bibinfo {volume} {ICHEP2016}},\ \bibinfo {pages} {1223}
  (\bibinfo {year} {2017})},\ \Eprint {http://arxiv.org/abs/1612.07426}
  {arXiv:1612.07426 [physics.ins-det]} \BibitemShut {NoStop}%
\bibitem [{\citenamefont {Kim}\ and\ \citenamefont {the
  KIMS~Collaboration}(2008)}]{Kim2008a}%
  \BibitemOpen
  \bibfield  {author} {\bibinfo {author} {\bibfnamefont {S.~K.}\ \bibnamefont
  {Kim}}\ and\ \bibinfo {author} {\bibnamefont {the KIMS~Collaboration}},\
  }\href {\doibase 10.1088/1742-6596/120/4/042021} {\bibfield  {journal}
  {\bibinfo  {journal} {Journal of Physics: Conference Series}\ }\textbf
  {\bibinfo {volume} {120}},\ \bibinfo {pages} {042021} (\bibinfo {year}
  {2008})}\BibitemShut {NoStop}%
\bibitem [{\citenamefont {Cozzini}\ \emph {et~al.}(2002)\citenamefont {Cozzini}
  \emph {et~al.}}]{Cozzini2002}%
  \BibitemOpen
  \bibfield  {author} {\bibinfo {author} {\bibfnamefont {C.}~\bibnamefont
  {Cozzini}} \emph {et~al.},\ }\bibfield  {booktitle} {\emph {\bibinfo
  {booktitle} {{Low temperature detectors. Proceedings, 9th International
  Workshop, LTD-9, Madison, USA, July 22-27, 2001}}},\ }\href {\doibase
  10.1063/1.1457690} {\bibfield  {journal} {\bibinfo  {journal} {AIP Conf.
  Proc.}\ }\textbf {\bibinfo {volume} {605}},\ \bibinfo {pages} {481} (\bibinfo
  {year} {2002})}\BibitemShut {NoStop}%
\bibitem [{\citenamefont {Petricca}\ \emph {et~al.}(2017)\citenamefont
  {Petricca} \emph {et~al.}}]{Petricca:2017zdp}%
  \BibitemOpen
  \bibfield  {author} {\bibinfo {author} {\bibfnamefont {F.}~\bibnamefont
  {Petricca}} \emph {et~al.} (\bibinfo {collaboration} {CRESST}),\ }in\
  \href@noop {} {\emph {\bibinfo {booktitle} {{15th International Conference on
  Topics in Astroparticle and Underground Physics (TAUP 2017) Sudbury, Ontario,
  Canada, July 24-28, 2017}}}}\ (\bibinfo {year} {2017})\ \Eprint
  {http://arxiv.org/abs/1711.07692} {arXiv:1711.07692 [astro-ph.CO]}
  \BibitemShut {NoStop}%
\bibitem [{\citenamefont {Angloher}\ \emph {et~al.}(2016)\citenamefont
  {Angloher} \emph {et~al.}}]{Angloher2016a}%
  \BibitemOpen
  \bibfield  {author} {\bibinfo {author} {\bibfnamefont {G.}~\bibnamefont
  {Angloher}} \emph {et~al.} (\bibinfo {collaboration} {CRESST}),\ }\href
  {\doibase 10.1140/epjc/s10052-016-3877-3} {\bibfield  {journal} {\bibinfo
  {journal} {Eur. Phys. J.}\ }\textbf {\bibinfo {volume} {C76}},\ \bibinfo
  {pages} {25} (\bibinfo {year} {2016})},\ \Eprint
  {http://arxiv.org/abs/1509.01515} {arXiv:1509.01515 [astro-ph.CO]}
  \BibitemShut {NoStop}%
\bibitem [{\citenamefont {Coskuner}\ \emph
  {et~al.}(2019{\natexlab{b}})\citenamefont {Coskuner}, \citenamefont
  {Grabowska}, \citenamefont {Knapen},\ and\ \citenamefont
  {Zurek}}]{Coskuner:2018are}%
  \BibitemOpen
  \bibfield  {author} {\bibinfo {author} {\bibfnamefont {A.}~\bibnamefont
  {Coskuner}}, \bibinfo {author} {\bibfnamefont {D.~M.}\ \bibnamefont
  {Grabowska}}, \bibinfo {author} {\bibfnamefont {S.}~\bibnamefont {Knapen}}, \
  and\ \bibinfo {author} {\bibfnamefont {K.~M.}\ \bibnamefont {Zurek}},\ }\href
  {\doibase 10.1103/PhysRevD.100.035025} {\bibfield  {journal} {\bibinfo
  {journal} {Phys. Rev.}\ }\textbf {\bibinfo {volume} {D100}},\ \bibinfo
  {pages} {035025} (\bibinfo {year} {2019}{\natexlab{b}})},\ \Eprint
  {http://arxiv.org/abs/1812.07573} {arXiv:1812.07573 [hep-ph]} \BibitemShut
  {NoStop}%
\bibitem [{\citenamefont {Trickle}\ \emph {et~al.}(2021)\citenamefont
  {Trickle}, \citenamefont {Zhang},\ and\ \citenamefont {Zurek}}]{phonodark}%
  \BibitemOpen
  \bibfield  {author} {\bibinfo {author} {\bibfnamefont {T.}~\bibnamefont
  {Trickle}}, \bibinfo {author} {\bibfnamefont {Z.}~\bibnamefont {Zhang}}, \
  and\ \bibinfo {author} {\bibfnamefont {K.~M.}\ \bibnamefont {Zurek}},\ }\href
  {{https://phonodark.caltech.edu}} {\enquote {\bibinfo {title}
  {{https://phonodark.caltech.edu}},}\ } (\bibinfo {year} {2021})\BibitemShut
  {NoStop}%
\bibitem [{\citenamefont {Chu}\ \emph {et~al.}(2012)\citenamefont {Chu},
  \citenamefont {Hambye},\ and\ \citenamefont {Tytgat}}]{Chu:2011be}%
  \BibitemOpen
  \bibfield  {author} {\bibinfo {author} {\bibfnamefont {X.}~\bibnamefont
  {Chu}}, \bibinfo {author} {\bibfnamefont {T.}~\bibnamefont {Hambye}}, \ and\
  \bibinfo {author} {\bibfnamefont {M.~H.~G.}\ \bibnamefont {Tytgat}},\ }\href
  {\doibase 10.1088/1475-7516/2012/05/034} {\bibfield  {journal} {\bibinfo
  {journal} {JCAP}\ }\textbf {\bibinfo {volume} {1205}},\ \bibinfo {pages}
  {034} (\bibinfo {year} {2012})},\ \Eprint {http://arxiv.org/abs/1112.0493}
  {arXiv:1112.0493 [hep-ph]} \BibitemShut {NoStop}%
\bibitem [{\citenamefont {Dvorkin}\ \emph {et~al.}(2019)\citenamefont
  {Dvorkin}, \citenamefont {Lin},\ and\ \citenamefont
  {Schutz}}]{Dvorkin:2019zdi}%
  \BibitemOpen
  \bibfield  {author} {\bibinfo {author} {\bibfnamefont {C.}~\bibnamefont
  {Dvorkin}}, \bibinfo {author} {\bibfnamefont {T.}~\bibnamefont {Lin}}, \ and\
  \bibinfo {author} {\bibfnamefont {K.}~\bibnamefont {Schutz}},\ }\href
  {\doibase 10.1103/PhysRevD.99.115009} {\bibfield  {journal} {\bibinfo
  {journal} {Phys. Rev.}\ }\textbf {\bibinfo {volume} {D99}},\ \bibinfo {pages}
  {115009} (\bibinfo {year} {2019})},\ \Eprint
  {http://arxiv.org/abs/1902.08623} {arXiv:1902.08623 [hep-ph]} \BibitemShut
  {NoStop}%
\bibitem [{\citenamefont {Vogel}\ and\ \citenamefont
  {Redondo}(2014)}]{Vogel:2013raa}%
  \BibitemOpen
  \bibfield  {author} {\bibinfo {author} {\bibfnamefont {H.}~\bibnamefont
  {Vogel}}\ and\ \bibinfo {author} {\bibfnamefont {J.}~\bibnamefont
  {Redondo}},\ }\href {\doibase 10.1088/1475-7516/2014/02/029} {\bibfield
  {journal} {\bibinfo  {journal} {JCAP}\ }\textbf {\bibinfo {volume} {1402}},\
  \bibinfo {pages} {029} (\bibinfo {year} {2014})},\ \Eprint
  {http://arxiv.org/abs/1311.2600} {arXiv:1311.2600 [hep-ph]} \BibitemShut
  {NoStop}%
\bibitem [{\citenamefont {Agnes}\ \emph
  {et~al.}(2018{\natexlab{a}})\citenamefont {Agnes} \emph
  {et~al.}}]{Agnes2018a}%
  \BibitemOpen
  \bibfield  {author} {\bibinfo {author} {\bibfnamefont {P.}~\bibnamefont
  {Agnes}} \emph {et~al.} (\bibinfo {collaboration} {DarkSide}),\ }\href
  {\doibase 10.1103/PhysRevLett.121.111303} {\bibfield  {journal} {\bibinfo
  {journal} {Phys. Rev. Lett.}\ }\textbf {\bibinfo {volume} {121}},\ \bibinfo
  {pages} {111303} (\bibinfo {year} {2018}{\natexlab{a}})},\ \Eprint
  {http://arxiv.org/abs/1802.06998} {arXiv:1802.06998 [astro-ph.CO]}
  \BibitemShut {NoStop}%
\bibitem [{\citenamefont {Aprile}\ \emph {et~al.}(2014)\citenamefont {Aprile}
  \emph {et~al.}}]{Aprile2014a}%
  \BibitemOpen
  \bibfield  {author} {\bibinfo {author} {\bibfnamefont {E.}~\bibnamefont
  {Aprile}} \emph {et~al.} (\bibinfo {collaboration} {XENON100}),\ }\href
  {\doibase 10.1088/0954-3899/41/3/035201} {\bibfield  {journal} {\bibinfo
  {journal} {J. Phys.}\ }\textbf {\bibinfo {volume} {G41}},\ \bibinfo {pages}
  {035201} (\bibinfo {year} {2014})},\ \Eprint {http://arxiv.org/abs/1311.1088}
  {arXiv:1311.1088 [physics.ins-det]} \BibitemShut {NoStop}%
\bibitem [{\citenamefont {Miuchi}\ \emph {et~al.}(2003)\citenamefont {Miuchi},
  \citenamefont {Minowa}, \citenamefont {Takeda}, \citenamefont {Sekiya},
  \citenamefont {Shimizu}, \citenamefont {Inoue}, \citenamefont {Ootani},\ and\
  \citenamefont {Ootuka}}]{Miuchi:2002zp}%
  \BibitemOpen
  \bibfield  {author} {\bibinfo {author} {\bibfnamefont {K.}~\bibnamefont
  {Miuchi}}, \bibinfo {author} {\bibfnamefont {M.}~\bibnamefont {Minowa}},
  \bibinfo {author} {\bibfnamefont {A.}~\bibnamefont {Takeda}}, \bibinfo
  {author} {\bibfnamefont {H.}~\bibnamefont {Sekiya}}, \bibinfo {author}
  {\bibfnamefont {Y.}~\bibnamefont {Shimizu}}, \bibinfo {author} {\bibfnamefont
  {Y.}~\bibnamefont {Inoue}}, \bibinfo {author} {\bibfnamefont
  {W.}~\bibnamefont {Ootani}}, \ and\ \bibinfo {author} {\bibfnamefont
  {Y.}~\bibnamefont {Ootuka}},\ }\href {\doibase 10.1016/S0927-6505(02)00192-5}
  {\bibfield  {journal} {\bibinfo  {journal} {Astropart. Phys.}\ }\textbf
  {\bibinfo {volume} {19}},\ \bibinfo {pages} {135} (\bibinfo {year} {2003})},\
  \Eprint {http://arxiv.org/abs/astro-ph/0204411} {arXiv:astro-ph/0204411
  [astro-ph]} \BibitemShut {NoStop}%
\bibitem [{\citenamefont {Detraux}\ \emph {et~al.}(1997)\citenamefont
  {Detraux}, \citenamefont {Ghosez},\ and\ \citenamefont
  {Gonze}}]{Detraux_et_al:1997}%
  \BibitemOpen
  \bibfield  {author} {\bibinfo {author} {\bibfnamefont {F.}~\bibnamefont
  {Detraux}}, \bibinfo {author} {\bibfnamefont {P.}~\bibnamefont {Ghosez}}, \
  and\ \bibinfo {author} {\bibfnamefont {X.}~\bibnamefont {Gonze}},\ }\href
  {\doibase 10.1103/PhysRevB.56.983} {\bibfield  {journal} {\bibinfo  {journal}
  {Phys. Rev. B}\ }\textbf {\bibinfo {volume} {56}},\ \bibinfo {pages} {983}
  (\bibinfo {year} {1997})}\BibitemShut {NoStop}%
\bibitem [{\citenamefont {Peralta}\ \emph {et~al.}(2006)\citenamefont
  {Peralta}, \citenamefont {Heyd}, \citenamefont {Scuseria},\ and\
  \citenamefont {Martin}}]{Peralta2006a}%
  \BibitemOpen
  \bibfield  {author} {\bibinfo {author} {\bibfnamefont {J.~E.}\ \bibnamefont
  {Peralta}}, \bibinfo {author} {\bibfnamefont {J.}~\bibnamefont {Heyd}},
  \bibinfo {author} {\bibfnamefont {G.~E.}\ \bibnamefont {Scuseria}}, \ and\
  \bibinfo {author} {\bibfnamefont {R.~L.}\ \bibnamefont {Martin}},\ }\href
  {\doibase 10.1103/physrevb.74.073101} {\bibfield  {journal} {\bibinfo
  {journal} {Physical Review B}\ }\textbf {\bibinfo {volume} {74}} (\bibinfo
  {year} {2006}),\ 10.1103/physrevb.74.073101}\BibitemShut {NoStop}%
\bibitem [{\citenamefont {Knapen}\ \emph
  {et~al.}(2017{\natexlab{b}})\citenamefont {Knapen}, \citenamefont {Lin},\
  and\ \citenamefont {Zurek}}]{Knapen:2017xzo}%
  \BibitemOpen
  \bibfield  {author} {\bibinfo {author} {\bibfnamefont {S.}~\bibnamefont
  {Knapen}}, \bibinfo {author} {\bibfnamefont {T.}~\bibnamefont {Lin}}, \ and\
  \bibinfo {author} {\bibfnamefont {K.~M.}\ \bibnamefont {Zurek}},\ }\href
  {\doibase 10.1103/PhysRevD.96.115021} {\bibfield  {journal} {\bibinfo
  {journal} {Phys. Rev.}\ }\textbf {\bibinfo {volume} {D96}},\ \bibinfo {pages}
  {115021} (\bibinfo {year} {2017}{\natexlab{b}})},\ \Eprint
  {http://arxiv.org/abs/1709.07882} {arXiv:1709.07882 [hep-ph]} \BibitemShut
  {NoStop}%
\bibitem [{\citenamefont {Agnes}\ \emph
  {et~al.}(2018{\natexlab{b}})\citenamefont {Agnes} \emph
  {et~al.}}]{Agnes2018b}%
  \BibitemOpen
  \bibfield  {author} {\bibinfo {author} {\bibfnamefont {P.}~\bibnamefont
  {Agnes}} \emph {et~al.} (\bibinfo {collaboration} {DarkSide}),\ }\href
  {\doibase 10.1103/PhysRevLett.121.081307} {\bibfield  {journal} {\bibinfo
  {journal} {Phys. Rev. Lett.}\ }\textbf {\bibinfo {volume} {121}},\ \bibinfo
  {pages} {081307} (\bibinfo {year} {2018}{\natexlab{b}})},\ \Eprint
  {http://arxiv.org/abs/1802.06994} {arXiv:1802.06994 [astro-ph.HE]}
  \BibitemShut {NoStop}%
\bibitem [{\citenamefont {Aprile}\ \emph {et~al.}(2018)\citenamefont {Aprile}
  \emph {et~al.}}]{Aprile2018a}%
  \BibitemOpen
  \bibfield  {author} {\bibinfo {author} {\bibfnamefont {E.}~\bibnamefont
  {Aprile}} \emph {et~al.} (\bibinfo {collaboration} {XENON}),\ }\href
  {\doibase 10.1103/PhysRevLett.121.111302} {\bibfield  {journal} {\bibinfo
  {journal} {Phys. Rev. Lett.}\ }\textbf {\bibinfo {volume} {121}},\ \bibinfo
  {pages} {111302} (\bibinfo {year} {2018})},\ \Eprint
  {http://arxiv.org/abs/1805.12562} {arXiv:1805.12562 [astro-ph.CO]}
  \BibitemShut {NoStop}%
\bibitem [{\citenamefont {Aprile}\ \emph {et~al.}(2019)\citenamefont {Aprile}
  \emph {et~al.}}]{Aprile2019a}%
  \BibitemOpen
  \bibfield  {author} {\bibinfo {author} {\bibfnamefont {E.}~\bibnamefont
  {Aprile}} \emph {et~al.} (\bibinfo {collaboration} {XENON}),\ }\href
  {\doibase 10.1103/PhysRevLett.123.251801} {\bibfield  {journal} {\bibinfo
  {journal} {Phys. Rev. Lett.}\ }\textbf {\bibinfo {volume} {123}},\ \bibinfo
  {pages} {251801} (\bibinfo {year} {2019})},\ \Eprint
  {http://arxiv.org/abs/1907.11485} {arXiv:1907.11485 [hep-ex]} \BibitemShut
  {NoStop}%
\bibitem [{\citenamefont {Akerib}\ \emph {et~al.}(2019)\citenamefont {Akerib}
  \emph {et~al.}}]{Akerib2019b}%
  \BibitemOpen
  \bibfield  {author} {\bibinfo {author} {\bibfnamefont {D.~S.}\ \bibnamefont
  {Akerib}} \emph {et~al.} (\bibinfo {collaboration} {LUX}),\ }\href {\doibase
  10.1103/PhysRevLett.122.131301} {\bibfield  {journal} {\bibinfo  {journal}
  {Phys. Rev. Lett.}\ }\textbf {\bibinfo {volume} {122}},\ \bibinfo {pages}
  {131301} (\bibinfo {year} {2019})},\ \Eprint
  {http://arxiv.org/abs/1811.11241} {arXiv:1811.11241 [astro-ph.CO]}
  \BibitemShut {NoStop}%
\bibitem [{\citenamefont {Battaglieri}\ \emph {et~al.}(2017)\citenamefont
  {Battaglieri} \emph {et~al.}}]{Battaglieri:2017aum}%
  \BibitemOpen
  \bibfield  {author} {\bibinfo {author} {\bibfnamefont {M.}~\bibnamefont
  {Battaglieri}} \emph {et~al.},\ }in\ \href
  {http://lss.fnal.gov/archive/2017/conf/fermilab-conf-17-282-ae-ppd-t.pdf}
  {\emph {\bibinfo {booktitle} {{U.S. Cosmic Visions: New Ideas in Dark Matter
  College Park, MD, USA, March 23-25, 2017}}}}\ (\bibinfo {year} {2017})\
  \Eprint {http://arxiv.org/abs/1707.04591} {arXiv:1707.04591 [hep-ph]}
  \BibitemShut {NoStop}%
\bibitem [{\citenamefont {Martin}(2004)}]{Martin2004a}%
  \BibitemOpen
  \bibfield  {author} {\bibinfo {author} {\bibfnamefont {R.~M.}\ \bibnamefont
  {Martin}},\ }\href {\doibase 10.1017/cbo9780511805769} {\emph {\bibinfo
  {title} {Electronic Structure}}}\ (\bibinfo  {publisher} {Cambridge
  University Press},\ \bibinfo {year} {2004})\BibitemShut {NoStop}%
\bibitem [{\citenamefont {Kresse}\ and\ \citenamefont
  {Hafner}(1993)}]{Kresse1993a}%
  \BibitemOpen
  \bibfield  {author} {\bibinfo {author} {\bibfnamefont {G.}~\bibnamefont
  {Kresse}}\ and\ \bibinfo {author} {\bibfnamefont {J.}~\bibnamefont
  {Hafner}},\ }\href {\doibase 10.1103/physrevb.47.558} {\bibfield  {journal}
  {\bibinfo  {journal} {Physical Review B}\ }\textbf {\bibinfo {volume} {47}},\
  \bibinfo {pages} {558} (\bibinfo {year} {1993})}\BibitemShut {NoStop}%
\bibitem [{\citenamefont {Kresse}\ and\ \citenamefont
  {Hafner}(1994)}]{Kresse1994a}%
  \BibitemOpen
  \bibfield  {author} {\bibinfo {author} {\bibfnamefont {G.}~\bibnamefont
  {Kresse}}\ and\ \bibinfo {author} {\bibfnamefont {J.}~\bibnamefont
  {Hafner}},\ }\href {\doibase 10.1103/physrevb.49.14251} {\bibfield  {journal}
  {\bibinfo  {journal} {Physical Review B}\ }\textbf {\bibinfo {volume} {49}},\
  \bibinfo {pages} {14251} (\bibinfo {year} {1994})}\BibitemShut {NoStop}%
\bibitem [{\citenamefont {Kresse}\ and\ \citenamefont
  {Furthm{\"u}ller}(1996)}]{Kresse1996a}%
  \BibitemOpen
  \bibfield  {author} {\bibinfo {author} {\bibfnamefont {G.}~\bibnamefont
  {Kresse}}\ and\ \bibinfo {author} {\bibfnamefont {J.}~\bibnamefont
  {Furthm{\"u}ller}},\ }\href {\doibase 10.1016/0927-0256(96)00008-0}
  {\bibfield  {journal} {\bibinfo  {journal} {Computational Materials Science}\
  }\textbf {\bibinfo {volume} {6}},\ \bibinfo {pages} {15} (\bibinfo {year}
  {1996})}\BibitemShut {NoStop}%
\bibitem [{\citenamefont {Bl{\"o}chl}(1994)}]{Bloechl1994a}%
  \BibitemOpen
  \bibfield  {author} {\bibinfo {author} {\bibfnamefont {P.~E.}\ \bibnamefont
  {Bl{\"o}chl}},\ }\href {\doibase 10.1103/physrevb.50.17953} {\bibfield
  {journal} {\bibinfo  {journal} {Physical Review B}\ }\textbf {\bibinfo
  {volume} {50}},\ \bibinfo {pages} {17953} (\bibinfo {year}
  {1994})}\BibitemShut {NoStop}%
\bibitem [{\citenamefont {Kresse}\ and\ \citenamefont
  {Joubert}(1999)}]{Kresse1999a}%
  \BibitemOpen
  \bibfield  {author} {\bibinfo {author} {\bibfnamefont {G.}~\bibnamefont
  {Kresse}}\ and\ \bibinfo {author} {\bibfnamefont {D.}~\bibnamefont
  {Joubert}},\ }\href {\doibase 10.1103/physrevb.59.1758} {\bibfield  {journal}
  {\bibinfo  {journal} {Physical Review B}\ }\textbf {\bibinfo {volume} {59}},\
  \bibinfo {pages} {1758} (\bibinfo {year} {1999})}\BibitemShut {NoStop}%
\bibitem [{\citenamefont {Perdew}\ \emph {et~al.}(1996)\citenamefont {Perdew},
  \citenamefont {Burke},\ and\ \citenamefont {Ernzerhof}}]{Perdew1996a}%
  \BibitemOpen
  \bibfield  {author} {\bibinfo {author} {\bibfnamefont {J.~P.}\ \bibnamefont
  {Perdew}}, \bibinfo {author} {\bibfnamefont {K.}~\bibnamefont {Burke}}, \
  and\ \bibinfo {author} {\bibfnamefont {M.}~\bibnamefont {Ernzerhof}},\ }\href
  {\doibase 10.1103/physrevlett.77.3865} {\bibfield  {journal} {\bibinfo
  {journal} {Physical Review Letters}\ }\textbf {\bibinfo {volume} {77}},\
  \bibinfo {pages} {3865} (\bibinfo {year} {1996})}\BibitemShut {NoStop}%
\bibitem [{\citenamefont {Wang}\ \emph {et~al.}(1994)\citenamefont {Wang},
  \citenamefont {Hubbard}, \citenamefont {Alexander}, \citenamefont {Becher},
  \citenamefont {Fernandez-Baca},\ and\ \citenamefont {Spooner}}]{Wang1994a}%
  \BibitemOpen
  \bibfield  {author} {\bibinfo {author} {\bibfnamefont {X.-L.}\ \bibnamefont
  {Wang}}, \bibinfo {author} {\bibfnamefont {C.~R.}\ \bibnamefont {Hubbard}},
  \bibinfo {author} {\bibfnamefont {K.~B.}\ \bibnamefont {Alexander}}, \bibinfo
  {author} {\bibfnamefont {P.~F.}\ \bibnamefont {Becher}}, \bibinfo {author}
  {\bibfnamefont {J.~A.}\ \bibnamefont {Fernandez-Baca}}, \ and\ \bibinfo
  {author} {\bibfnamefont {S.}~\bibnamefont {Spooner}},\ }\href {\doibase
  10.1111/j.1151-2916.1994.tb09758.x} {\bibfield  {journal} {\bibinfo
  {journal} {Journal of the American Ceramic Society}\ }\textbf {\bibinfo
  {volume} {77}},\ \bibinfo {pages} {1569} (\bibinfo {year}
  {1994})}\BibitemShut {NoStop}%
\bibitem [{\citenamefont {French}(1990)}]{French1990a}%
  \BibitemOpen
  \bibfield  {author} {\bibinfo {author} {\bibfnamefont {R.~H.}\ \bibnamefont
  {French}},\ }\href {\doibase 10.1111/j.1151-2916.1990.tb06541.x} {\bibfield
  {journal} {\bibinfo  {journal} {Journal of the American Ceramic Society}\
  }\textbf {\bibinfo {volume} {73}},\ \bibinfo {pages} {477} (\bibinfo {year}
  {1990})}\BibitemShut {NoStop}%
\bibitem [{\citenamefont {Zemann}(1965)}]{Zemann1965a}%
  \BibitemOpen
  \bibfield  {author} {\bibinfo {author} {\bibfnamefont {J.}~\bibnamefont
  {Zemann}},\ }\href {\doibase 10.1107/s0365110x65000361} {\bibfield  {journal}
  {\bibinfo  {journal} {Acta Crystallographica}\ }\textbf {\bibinfo {volume}
  {18}},\ \bibinfo {pages} {139} (\bibinfo {year} {1965})}\BibitemShut
  {NoStop}%
\bibitem [{\citenamefont {Yim}\ \emph {et~al.}(1973)\citenamefont {Yim},
  \citenamefont {Stofko}, \citenamefont {Zanzucchi}, \citenamefont {Pankove},
  \citenamefont {Ettenberg},\ and\ \citenamefont {Gilbert}}]{Yim1973a}%
  \BibitemOpen
  \bibfield  {author} {\bibinfo {author} {\bibfnamefont {W.~M.}\ \bibnamefont
  {Yim}}, \bibinfo {author} {\bibfnamefont {E.~J.}\ \bibnamefont {Stofko}},
  \bibinfo {author} {\bibfnamefont {P.~J.}\ \bibnamefont {Zanzucchi}}, \bibinfo
  {author} {\bibfnamefont {J.~I.}\ \bibnamefont {Pankove}}, \bibinfo {author}
  {\bibfnamefont {M.}~\bibnamefont {Ettenberg}}, \ and\ \bibinfo {author}
  {\bibfnamefont {S.~L.}\ \bibnamefont {Gilbert}},\ }\href {\doibase
  10.1063/1.1661876} {\bibfield  {journal} {\bibinfo  {journal} {Journal of
  Applied Physics}\ }\textbf {\bibinfo {volume} {44}},\ \bibinfo {pages} {292}
  (\bibinfo {year} {1973})}\BibitemShut {NoStop}%
\bibitem [{\citenamefont {Speziale}\ and\ \citenamefont
  {Duffy}(2002)}]{Speziale2002a}%
  \BibitemOpen
  \bibfield  {author} {\bibinfo {author} {\bibfnamefont {S.}~\bibnamefont
  {Speziale}}\ and\ \bibinfo {author} {\bibfnamefont {T.~S.}\ \bibnamefont
  {Duffy}},\ }\href {\doibase 10.1007/s00269-002-0250-x} {\bibfield  {journal}
  {\bibinfo  {journal} {Physics and Chemistry of Minerals}\ }\textbf {\bibinfo
  {volume} {29}},\ \bibinfo {pages} {465} (\bibinfo {year} {2002})}\BibitemShut
  {NoStop}%
\bibitem [{\citenamefont {Rubloff}(1972)}]{Rubloff1972a}%
  \BibitemOpen
  \bibfield  {author} {\bibinfo {author} {\bibfnamefont {G.~W.}\ \bibnamefont
  {Rubloff}},\ }\href {\doibase 10.1103/physrevb.5.662} {\bibfield  {journal}
  {\bibinfo  {journal} {Physical Review B}\ }\textbf {\bibinfo {volume} {5}},\
  \bibinfo {pages} {662} (\bibinfo {year} {1972})}\BibitemShut {NoStop}%
\bibitem [{\citenamefont {Senyshyn}\ \emph {et~al.}(2004)\citenamefont
  {Senyshyn}, \citenamefont {Kraus}, \citenamefont {Mikhailik},\ and\
  \citenamefont {Yakovyna}}]{Senyshyn2004a}%
  \BibitemOpen
  \bibfield  {author} {\bibinfo {author} {\bibfnamefont {A.}~\bibnamefont
  {Senyshyn}}, \bibinfo {author} {\bibfnamefont {H.}~\bibnamefont {Kraus}},
  \bibinfo {author} {\bibfnamefont {V.~B.}\ \bibnamefont {Mikhailik}}, \ and\
  \bibinfo {author} {\bibfnamefont {V.}~\bibnamefont {Yakovyna}},\ }\href
  {\doibase 10.1103/physrevb.70.214306} {\bibfield  {journal} {\bibinfo
  {journal} {Physical Review B}\ }\textbf {\bibinfo {volume} {70}} (\bibinfo
  {year} {2004}),\ 10.1103/physrevb.70.214306}\BibitemShut {NoStop}%
\bibitem [{\citenamefont {Mikhailik}\ \emph {et~al.}(2004)\citenamefont
  {Mikhailik}, \citenamefont {Kraus}, \citenamefont {Wahl}, \citenamefont
  {Itoh}, \citenamefont {Koike},\ and\ \citenamefont
  {Bailiff}}]{Mikhailik2004a}%
  \BibitemOpen
  \bibfield  {author} {\bibinfo {author} {\bibfnamefont {V.~B.}\ \bibnamefont
  {Mikhailik}}, \bibinfo {author} {\bibfnamefont {H.}~\bibnamefont {Kraus}},
  \bibinfo {author} {\bibfnamefont {D.}~\bibnamefont {Wahl}}, \bibinfo {author}
  {\bibfnamefont {M.}~\bibnamefont {Itoh}}, \bibinfo {author} {\bibfnamefont
  {M.}~\bibnamefont {Koike}}, \ and\ \bibinfo {author} {\bibfnamefont {I.~K.}\
  \bibnamefont {Bailiff}},\ }\href {\doibase 10.1103/physrevb.69.205110}
  {\bibfield  {journal} {\bibinfo  {journal} {Physical Review B}\ }\textbf
  {\bibinfo {volume} {69}} (\bibinfo {year} {2004}),\
  10.1103/physrevb.69.205110}\BibitemShut {NoStop}%
\bibitem [{\citenamefont {Lipp}\ \emph {et~al.}(2006)\citenamefont {Lipp},
  \citenamefont {Yoo}, \citenamefont {Strachan},\ and\ \citenamefont
  {Daniels}}]{Lipp2006a}%
  \BibitemOpen
  \bibfield  {author} {\bibinfo {author} {\bibfnamefont {M.~J.}\ \bibnamefont
  {Lipp}}, \bibinfo {author} {\bibfnamefont {C.~H.}\ \bibnamefont {Yoo}},
  \bibinfo {author} {\bibfnamefont {D.}~\bibnamefont {Strachan}}, \ and\
  \bibinfo {author} {\bibfnamefont {W.~B.}\ \bibnamefont {Daniels}},\ }\href
  {\doibase 10.1103/physrevb.73.085121} {\bibfield  {journal} {\bibinfo
  {journal} {Physical Review B}\ }\textbf {\bibinfo {volume} {73}} (\bibinfo
  {year} {2006}),\ 10.1103/physrevb.73.085121}\BibitemShut {NoStop}%
\bibitem [{\citenamefont {C.~D.~Clark}\ and\ \citenamefont
  {Harris}(1964)}]{Clark1964a}%
  \BibitemOpen
  \bibfield  {author} {\bibinfo {author} {\bibfnamefont {P.~J.~D.}\
  \bibnamefont {C.~D.~Clark}}\ and\ \bibinfo {author} {\bibfnamefont {P.~V.}\
  \bibnamefont {Harris}},\ }\href {\doibase 10.1098/rspa.1964.0025} {\bibfield
  {journal} {\bibinfo  {journal} {Proceedings of the Royal Society of London.
  Series A. Mathematical and Physical Sciences}\ }\textbf {\bibinfo {volume}
  {277}},\ \bibinfo {pages} {312} (\bibinfo {year} {1964})}\BibitemShut
  {NoStop}%
\bibitem [{\citenamefont {Levinshtein}\ \emph {et~al.}(1996)\citenamefont
  {Levinshtein}, \citenamefont {Rumyantsev},\ and\ \citenamefont
  {Shur}}]{Levinshtein1996a}%
  \BibitemOpen
  \bibfield  {author} {\bibinfo {author} {\bibfnamefont {M.}~\bibnamefont
  {Levinshtein}}, \bibinfo {author} {\bibfnamefont {S.}~\bibnamefont
  {Rumyantsev}}, \ and\ \bibinfo {author} {\bibfnamefont {M.}~\bibnamefont
  {Shur}},\ }\href {\doibase 10.1142/2046-vol1} {\emph {\bibinfo {title}
  {Handbook Series on Semiconductor Parameters}}}\ (\bibinfo  {publisher}
  {{WORLD} {SCIENTIFIC}},\ \bibinfo {year} {1996})\BibitemShut {NoStop}%
\bibitem [{\citenamefont {Paszkowicz}\ \emph {et~al.}(2004)\citenamefont
  {Paszkowicz}, \citenamefont {Podsiad{\l}o},\ and\ \citenamefont
  {Minikayev}}]{Paszkowicz2004a}%
  \BibitemOpen
  \bibfield  {author} {\bibinfo {author} {\bibfnamefont {W.}~\bibnamefont
  {Paszkowicz}}, \bibinfo {author} {\bibfnamefont {S.}~\bibnamefont
  {Podsiad{\l}o}}, \ and\ \bibinfo {author} {\bibfnamefont {R.}~\bibnamefont
  {Minikayev}},\ }\href {\doibase 10.1016/j.jallcom.2004.05.036} {\bibfield
  {journal} {\bibinfo  {journal} {Journal of Alloys and Compounds}\ }\textbf
  {\bibinfo {volume} {382}},\ \bibinfo {pages} {100} (\bibinfo {year}
  {2004})}\BibitemShut {NoStop}%
\bibitem [{\citenamefont {Teisseyre}\ \emph {et~al.}(1994)\citenamefont
  {Teisseyre}, \citenamefont {Perlin}, \citenamefont {Suski}, \citenamefont
  {Grzegory}, \citenamefont {Porowski}, \citenamefont {Jun}, \citenamefont
  {Pietraszko},\ and\ \citenamefont {Moustakas}}]{Teisseyre1994a}%
  \BibitemOpen
  \bibfield  {author} {\bibinfo {author} {\bibfnamefont {H.}~\bibnamefont
  {Teisseyre}}, \bibinfo {author} {\bibfnamefont {P.}~\bibnamefont {Perlin}},
  \bibinfo {author} {\bibfnamefont {T.}~\bibnamefont {Suski}}, \bibinfo
  {author} {\bibfnamefont {I.}~\bibnamefont {Grzegory}}, \bibinfo {author}
  {\bibfnamefont {S.}~\bibnamefont {Porowski}}, \bibinfo {author}
  {\bibfnamefont {J.}~\bibnamefont {Jun}}, \bibinfo {author} {\bibfnamefont
  {A.}~\bibnamefont {Pietraszko}}, \ and\ \bibinfo {author} {\bibfnamefont
  {T.~D.}\ \bibnamefont {Moustakas}},\ }\href {\doibase 10.1063/1.357592}
  {\bibfield  {journal} {\bibinfo  {journal} {Journal of Applied Physics}\
  }\textbf {\bibinfo {volume} {76}},\ \bibinfo {pages} {2429} (\bibinfo {year}
  {1994})}\BibitemShut {NoStop}%
\bibitem [{\citenamefont {Adachi}(1989)}]{Adachi1989a}%
  \BibitemOpen
  \bibfield  {author} {\bibinfo {author} {\bibfnamefont {S.}~\bibnamefont
  {Adachi}},\ }\href {\doibase 10.1063/1.343580} {\bibfield  {journal}
  {\bibinfo  {journal} {Journal of Applied Physics}\ }\textbf {\bibinfo
  {volume} {66}},\ \bibinfo {pages} {6030} (\bibinfo {year}
  {1989})}\BibitemShut {NoStop}%
\bibitem [{\citenamefont {Sze}\ and\ \citenamefont {Irvin}(1968)}]{Sze1968a}%
  \BibitemOpen
  \bibfield  {author} {\bibinfo {author} {\bibfnamefont {S.}~\bibnamefont
  {Sze}}\ and\ \bibinfo {author} {\bibfnamefont {J.}~\bibnamefont {Irvin}},\
  }\href {\doibase 10.1016/0038-1101(68)90012-9} {\bibfield  {journal}
  {\bibinfo  {journal} {Solid-State Electronics}\ }\textbf {\bibinfo {volume}
  {11}},\ \bibinfo {pages} {599} (\bibinfo {year} {1968})}\BibitemShut
  {NoStop}%
\bibitem [{\citenamefont {Littler}\ and\ \citenamefont
  {Seiler}(1985)}]{Littler1985a}%
  \BibitemOpen
  \bibfield  {author} {\bibinfo {author} {\bibfnamefont {C.~L.}\ \bibnamefont
  {Littler}}\ and\ \bibinfo {author} {\bibfnamefont {D.~G.}\ \bibnamefont
  {Seiler}},\ }\href {\doibase 10.1063/1.95789} {\bibfield  {journal} {\bibinfo
   {journal} {Applied Physics Letters}\ }\textbf {\bibinfo {volume} {46}},\
  \bibinfo {pages} {986} (\bibinfo {year} {1985})}\BibitemShut {NoStop}%
\bibitem [{\citenamefont {Ullrich}\ \emph {et~al.}(1992)\citenamefont
  {Ullrich}, \citenamefont {Uhlig}, \citenamefont {Geise}, \citenamefont
  {Horn},\ and\ \citenamefont {Waltinger}}]{Ullrich1992a}%
  \BibitemOpen
  \bibfield  {author} {\bibinfo {author} {\bibfnamefont {H.-J.}\ \bibnamefont
  {Ullrich}}, \bibinfo {author} {\bibfnamefont {A.}~\bibnamefont {Uhlig}},
  \bibinfo {author} {\bibfnamefont {G.}~\bibnamefont {Geise}}, \bibinfo
  {author} {\bibfnamefont {H.}~\bibnamefont {Horn}}, \ and\ \bibinfo {author}
  {\bibfnamefont {H.}~\bibnamefont {Waltinger}},\ }\href {\doibase
  10.1007/bf01244483} {\bibfield  {journal} {\bibinfo  {journal} {Mikrochimica
  Acta}\ }\textbf {\bibinfo {volume} {107}},\ \bibinfo {pages} {283} (\bibinfo
  {year} {1992})}\BibitemShut {NoStop}%
\bibitem [{\citenamefont {Singh}\ and\ \citenamefont
  {Gallon}(1984)}]{Singh1984a}%
  \BibitemOpen
  \bibfield  {author} {\bibinfo {author} {\bibfnamefont {G.}~\bibnamefont
  {Singh}}\ and\ \bibinfo {author} {\bibfnamefont {T.}~\bibnamefont {Gallon}},\
  }\href {\doibase 10.1016/0038-1098(84)90687-2} {\bibfield  {journal}
  {\bibinfo  {journal} {Solid State Communications}\ }\textbf {\bibinfo
  {volume} {51}},\ \bibinfo {pages} {281} (\bibinfo {year} {1984})}\BibitemShut
  {NoStop}%
\bibitem [{\citenamefont {Thomas}\ \emph {et~al.}(1973)\citenamefont {Thomas},
  \citenamefont {Stephan}, \citenamefont {Lemonnier}, \citenamefont {Nisar},\
  and\ \citenamefont {Robin}}]{Thomas1973a}%
  \BibitemOpen
  \bibfield  {author} {\bibinfo {author} {\bibfnamefont {J.}~\bibnamefont
  {Thomas}}, \bibinfo {author} {\bibfnamefont {G.}~\bibnamefont {Stephan}},
  \bibinfo {author} {\bibfnamefont {J.~C.}\ \bibnamefont {Lemonnier}}, \bibinfo
  {author} {\bibfnamefont {M.}~\bibnamefont {Nisar}}, \ and\ \bibinfo {author}
  {\bibfnamefont {S.}~\bibnamefont {Robin}},\ }\href {\doibase
  10.1002/pssb.2220560115} {\bibfield  {journal} {\bibinfo  {journal} {Physica
  Status Solidi (b)}\ }\textbf {\bibinfo {volume} {56}},\ \bibinfo {pages}
  {163} (\bibinfo {year} {1973})}\BibitemShut {NoStop}%
\bibitem [{MgO()}]{MgO}%
  \BibitemOpen
  in\ \href {\doibase 10.1007/10681719_206} {\emph {\bibinfo {booktitle}
  {{II}-{VI} and I-{VII} Compounds; Semimagnetic Compounds}}}\ (\bibinfo
  {publisher} {Springer-Verlag})\ pp.\ \bibinfo {pages} {1--6}\BibitemShut
  {NoStop}%
\bibitem [{\citenamefont {Whited}\ \emph {et~al.}(1973)\citenamefont {Whited},
  \citenamefont {Flaten},\ and\ \citenamefont {Walker}}]{Whited1973a}%
  \BibitemOpen
  \bibfield  {author} {\bibinfo {author} {\bibfnamefont {R.}~\bibnamefont
  {Whited}}, \bibinfo {author} {\bibfnamefont {C.~J.}\ \bibnamefont {Flaten}},
  \ and\ \bibinfo {author} {\bibfnamefont {W.}~\bibnamefont {Walker}},\ }\href
  {\doibase 10.1016/0038-1098(73)90754-0} {\bibfield  {journal} {\bibinfo
  {journal} {Solid State Communications}\ }\textbf {\bibinfo {volume} {13}},\
  \bibinfo {pages} {1903} (\bibinfo {year} {1973})}\BibitemShut {NoStop}%
\bibitem [{\citenamefont {Brown}\ \emph {et~al.}(1970)\citenamefont {Brown},
  \citenamefont {G{\"a}hwiller}, \citenamefont {Fujita}, \citenamefont {Kunz},
  \citenamefont {Scheifley},\ and\ \citenamefont {Carrera}}]{Brown1970a}%
  \BibitemOpen
  \bibfield  {author} {\bibinfo {author} {\bibfnamefont {F.~C.}\ \bibnamefont
  {Brown}}, \bibinfo {author} {\bibfnamefont {C.}~\bibnamefont
  {G{\"a}hwiller}}, \bibinfo {author} {\bibfnamefont {H.}~\bibnamefont
  {Fujita}}, \bibinfo {author} {\bibfnamefont {A.~B.}\ \bibnamefont {Kunz}},
  \bibinfo {author} {\bibfnamefont {W.}~\bibnamefont {Scheifley}}, \ and\
  \bibinfo {author} {\bibfnamefont {N.}~\bibnamefont {Carrera}},\ }\href
  {\doibase 10.1103/physrevb.2.2126} {\bibfield  {journal} {\bibinfo  {journal}
  {Physical Review B}\ }\textbf {\bibinfo {volume} {2}},\ \bibinfo {pages}
  {2126} (\bibinfo {year} {1970})}\BibitemShut {NoStop}%
\bibitem [{\citenamefont {Deshpande}(1961)}]{Deshpande1961a}%
  \BibitemOpen
  \bibfield  {author} {\bibinfo {author} {\bibfnamefont {V.~T.}\ \bibnamefont
  {Deshpande}},\ }\href {\doibase 10.1107/s0365110x61002357} {\bibfield
  {journal} {\bibinfo  {journal} {Acta Crystallographica}\ }\textbf {\bibinfo
  {volume} {14}},\ \bibinfo {pages} {794} (\bibinfo {year} {1961})}\BibitemShut
  {NoStop}%
\bibitem [{\citenamefont {Roy}\ \emph {et~al.}(1985)\citenamefont {Roy},
  \citenamefont {Singh},\ and\ \citenamefont {Gallon}}]{Roy1985a}%
  \BibitemOpen
  \bibfield  {author} {\bibinfo {author} {\bibfnamefont {G.}~\bibnamefont
  {Roy}}, \bibinfo {author} {\bibfnamefont {G.}~\bibnamefont {Singh}}, \ and\
  \bibinfo {author} {\bibfnamefont {T.}~\bibnamefont {Gallon}},\ }\href
  {\doibase 10.1016/0039-6028(85)90519-9} {\bibfield  {journal} {\bibinfo
  {journal} {Surface Science}\ }\textbf {\bibinfo {volume} {152-153}},\
  \bibinfo {pages} {1042} (\bibinfo {year} {1985})}\BibitemShut {NoStop}%
\bibitem [{\citenamefont {Rodnyi}(1997)}]{Rodnyi1997a}%
  \BibitemOpen
  \bibfield  {author} {\bibinfo {author} {\bibfnamefont {P.~A.}\ \bibnamefont
  {Rodnyi}},\ }\href
  {https://www.amazon.com/Physical-Processes-Inorganic-Scintillators-Technology/dp/0849337887?SubscriptionId=AKIAIOBINVZYXZQZ2U3A&tag=chimbori05-20&linkCode=xm2&camp=2025&creative=165953&creativeASIN=0849337887}
  {\emph {\bibinfo {title} {Physical Processes in Inorganic Scintillators
  (Laser \& Optical Science \& Technology)}}}\ (\bibinfo  {publisher} {CRC
  Press},\ \bibinfo {year} {1997})\BibitemShut {NoStop}%
\bibitem [{\citenamefont {Strehlow}\ and\ \citenamefont
  {Cook}(1973)}]{Strehlow1973a}%
  \BibitemOpen
  \bibfield  {author} {\bibinfo {author} {\bibfnamefont {W.~H.}\ \bibnamefont
  {Strehlow}}\ and\ \bibinfo {author} {\bibfnamefont {E.~L.}\ \bibnamefont
  {Cook}},\ }\href {\doibase 10.1063/1.3253115} {\bibfield  {journal} {\bibinfo
   {journal} {Journal of Physical and Chemical Reference Data}\ }\textbf
  {\bibinfo {volume} {2}},\ \bibinfo {pages} {163} (\bibinfo {year}
  {1973})}\BibitemShut {NoStop}%
\bibitem [{\citenamefont {Ahrenkiel}\ \emph {et~al.}(2007)\citenamefont
  {Ahrenkiel}, \citenamefont {Johnston}, \citenamefont {Metzger},\ and\
  \citenamefont {Dippo}}]{Ahrenkiel2007a}%
  \BibitemOpen
  \bibfield  {author} {\bibinfo {author} {\bibfnamefont {R.}~\bibnamefont
  {Ahrenkiel}}, \bibinfo {author} {\bibfnamefont {S.}~\bibnamefont {Johnston}},
  \bibinfo {author} {\bibfnamefont {W.}~\bibnamefont {Metzger}}, \ and\
  \bibinfo {author} {\bibfnamefont {P.}~\bibnamefont {Dippo}},\ }\href
  {\doibase 10.1007/s11664-007-0325-z} {\bibfield  {journal} {\bibinfo
  {journal} {Journal of Electronic Materials}\ }\textbf {\bibinfo {volume}
  {37}},\ \bibinfo {pages} {396} (\bibinfo {year} {2007})}\BibitemShut
  {NoStop}%
\bibitem [{\citenamefont {Page}\ and\ \citenamefont
  {Donnay}(1976)}]{Page1976a}%
  \BibitemOpen
  \bibfield  {author} {\bibinfo {author} {\bibfnamefont {Y.~L.}\ \bibnamefont
  {Page}}\ and\ \bibinfo {author} {\bibfnamefont {G.}~\bibnamefont {Donnay}},\
  }\href {\doibase 10.1107/s0567740876007966} {\bibfield  {journal} {\bibinfo
  {journal} {Acta Crystallographica Section B Structural Crystallography and
  Crystal Chemistry}\ }\textbf {\bibinfo {volume} {32}},\ \bibinfo {pages}
  {2456} (\bibinfo {year} {1976})}\BibitemShut {NoStop}%
\bibitem [{\citenamefont {Sato}\ \emph {et~al.}(2015)\citenamefont {Sato},
  \citenamefont {Yabana}, \citenamefont {Shinohara}, \citenamefont {Otobe},
  \citenamefont {Lee},\ and\ \citenamefont {Bertsch}}]{Sato2015a}%
  \BibitemOpen
  \bibfield  {author} {\bibinfo {author} {\bibfnamefont {S.~A.}\ \bibnamefont
  {Sato}}, \bibinfo {author} {\bibfnamefont {K.}~\bibnamefont {Yabana}},
  \bibinfo {author} {\bibfnamefont {Y.}~\bibnamefont {Shinohara}}, \bibinfo
  {author} {\bibfnamefont {T.}~\bibnamefont {Otobe}}, \bibinfo {author}
  {\bibfnamefont {K.-M.}\ \bibnamefont {Lee}}, \ and\ \bibinfo {author}
  {\bibfnamefont {G.~F.}\ \bibnamefont {Bertsch}},\ }\href {\doibase
  10.1103/physrevb.92.205413} {\bibfield  {journal} {\bibinfo  {journal}
  {Physical Review B}\ }\textbf {\bibinfo {volume} {92}} (\bibinfo {year}
  {2015}),\ 10.1103/physrevb.92.205413}\BibitemShut {NoStop}%
\bibitem [{\citenamefont {Srikant}\ and\ \citenamefont
  {Clarke}(1998)}]{Srikant1998a}%
  \BibitemOpen
  \bibfield  {author} {\bibinfo {author} {\bibfnamefont {V.}~\bibnamefont
  {Srikant}}\ and\ \bibinfo {author} {\bibfnamefont {D.~R.}\ \bibnamefont
  {Clarke}},\ }\href {\doibase 10.1063/1.367375} {\bibfield  {journal}
  {\bibinfo  {journal} {Journal of Applied Physics}\ }\textbf {\bibinfo
  {volume} {83}},\ \bibinfo {pages} {5447} (\bibinfo {year}
  {1998})}\BibitemShut {NoStop}%
\bibitem [{\citenamefont {Jumpertz}(1955)}]{Jumpertz1955}%
  \BibitemOpen
  \bibfield  {author} {\bibinfo {author} {\bibfnamefont {E.~A.}\ \bibnamefont
  {Jumpertz}},\ }\href@noop {} {\bibfield  {journal} {\bibinfo  {journal}
  {Zeitschrift f{\"{u}}r Elektrochemie und angewandte physikalische Chemie}\
  }\textbf {\bibinfo {volume} {59}},\ \bibinfo {pages} {419} (\bibinfo {year}
  {1955})}\BibitemShut {NoStop}%
\bibitem [{\citenamefont {Khamala}\ \emph {et~al.}(2016)\citenamefont
  {Khamala}, \citenamefont {Franklin}, \citenamefont {Malozovsky},
  \citenamefont {Stewart}, \citenamefont {Saleem},\ and\ \citenamefont
  {Bagayoko}}]{Khamala2016a}%
  \BibitemOpen
  \bibfield  {author} {\bibinfo {author} {\bibfnamefont {B.}~\bibnamefont
  {Khamala}}, \bibinfo {author} {\bibfnamefont {L.}~\bibnamefont {Franklin}},
  \bibinfo {author} {\bibfnamefont {Y.}~\bibnamefont {Malozovsky}}, \bibinfo
  {author} {\bibfnamefont {A.}~\bibnamefont {Stewart}}, \bibinfo {author}
  {\bibfnamefont {H.}~\bibnamefont {Saleem}}, \ and\ \bibinfo {author}
  {\bibfnamefont {D.}~\bibnamefont {Bagayoko}},\ }\href {\doibase
  10.1016/j.cocom.2015.12.001} {\bibfield  {journal} {\bibinfo  {journal}
  {Computational Condensed Matter}\ }\textbf {\bibinfo {volume} {6}},\ \bibinfo
  {pages} {18} (\bibinfo {year} {2016})}\BibitemShut {NoStop}%
\bibitem [{\citenamefont {Bystrom}\ \emph {et~al.}()\citenamefont {Bystrom},
  \citenamefont {Broberg}, \citenamefont {Dwaraknath}, \citenamefont
  {Persson},\ and\ \citenamefont {Asta}}]{Bystrom2019a}%
  \BibitemOpen
  \bibfield  {author} {\bibinfo {author} {\bibfnamefont {K.}~\bibnamefont
  {Bystrom}}, \bibinfo {author} {\bibfnamefont {D.}~\bibnamefont {Broberg}},
  \bibinfo {author} {\bibfnamefont {S.}~\bibnamefont {Dwaraknath}}, \bibinfo
  {author} {\bibfnamefont {K.~A.}\ \bibnamefont {Persson}}, \ and\ \bibinfo
  {author} {\bibfnamefont {M.}~\bibnamefont {Asta}},\ }\href@noop {} {\
  }\Eprint {http://arxiv.org/abs/http://arxiv.org/abs/1904.11572v1}
  {http://arxiv.org/abs/1904.11572v1} \BibitemShut {NoStop}%
\bibitem [{\citenamefont {Krukau}\ \emph {et~al.}(2006)\citenamefont {Krukau},
  \citenamefont {Vydrov}, \citenamefont {Izmaylov},\ and\ \citenamefont
  {Scuseria}}]{Krukau2006a}%
  \BibitemOpen
  \bibfield  {author} {\bibinfo {author} {\bibfnamefont {A.~V.}\ \bibnamefont
  {Krukau}}, \bibinfo {author} {\bibfnamefont {O.~A.}\ \bibnamefont {Vydrov}},
  \bibinfo {author} {\bibfnamefont {A.~F.}\ \bibnamefont {Izmaylov}}, \ and\
  \bibinfo {author} {\bibfnamefont {G.~E.}\ \bibnamefont {Scuseria}},\ }\href
  {\doibase 10.1063/1.2404663} {\bibfield  {journal} {\bibinfo  {journal} {The
  Journal of Chemical Physics}\ }\textbf {\bibinfo {volume} {125}},\ \bibinfo
  {pages} {224106} (\bibinfo {year} {2006})}\BibitemShut {NoStop}%
\bibitem [{\citenamefont {Togo}\ and\ \citenamefont {Tanaka}(2015)}]{phonopy}%
  \BibitemOpen
  \bibfield  {author} {\bibinfo {author} {\bibfnamefont {A.}~\bibnamefont
  {Togo}}\ and\ \bibinfo {author} {\bibfnamefont {I.}~\bibnamefont {Tanaka}},\
  }\href {\doibase 10.1016/j.scriptamat.2015.07.021} {\bibfield  {journal}
  {\bibinfo  {journal} {Scripta Materialia}\ }\textbf {\bibinfo {volume}
  {108}},\ \bibinfo {pages} {1} (\bibinfo {year} {2015})}\BibitemShut {NoStop}%
\bibitem [{\citenamefont {Hinuma}\ \emph {et~al.}()\citenamefont {Hinuma},
  \citenamefont {Pizzi}, \citenamefont {Kumagai}, \citenamefont {Oba},\ and\
  \citenamefont {Tanaka}}]{Hinuma:aa}%
  \BibitemOpen
  \bibfield  {author} {\bibinfo {author} {\bibfnamefont {Y.}~\bibnamefont
  {Hinuma}}, \bibinfo {author} {\bibfnamefont {G.}~\bibnamefont {Pizzi}},
  \bibinfo {author} {\bibfnamefont {Y.}~\bibnamefont {Kumagai}}, \bibinfo
  {author} {\bibfnamefont {F.}~\bibnamefont {Oba}}, \ and\ \bibinfo {author}
  {\bibfnamefont {I.}~\bibnamefont {Tanaka}},\ }\href
  {http://arxiv.org/abs/1602.06402} {\ }\Eprint
  {http://arxiv.org/abs/1602.06402} {arXiv:1602.06402} \BibitemShut {NoStop}%
\end{thebibliography}%

\end{document}